# A Generative Deep Learning and Explainable Machine Learning Framework for Heat Transfer Prediction and Analysis in Porous Structures with Oscillatory Flows

**Lichang Zhu[a], Laura Schaefer[b], Leitao Chen[c], Ben Xu[a]***

[a] Department of Mechanical and Aerospace Engineering, University of Houston, Houston, TX 77204

[b] Department of Mechanical Engineering, Rice University, Houston, TX 77005

[c] Department of Mechanical Engineering, Embry-Riddle Aeronautical University, Daytona Beach, FL 32114

***Corresponding Author**

Ben Xu, Email: bxu12@central.uh.edu

## Abstract

Predicting and interpreting thermal performance under oscillating flow in porous structures remains a critical challenge due to the complex coupling between fluid dynamics and geometric features. This study introduces a data-driven framework that integrates generative deep learning, numerical simulation based on the lattice Boltzmann method (LBM), and interpretable machine learning to predict and explain the thermal behavior in such systems. A wide range of porous structures with diverse topologies was synthesized using a Wasserstein generative adversarial network with gradient penalty (wGAN-GP), significantly expanding the design space. High-fidelity thermal data were then generated through LBM simulations across various Reynolds ($Re$) and Strouhal numbers ($St$). Among ten machine learning models evaluated via nested cross-validation (Nested_CV) and Bayesian optimization, Extreme Gradient Boosting (XGBoost) achieved the best predictive performance for the average Nusselt number ($\overline{Nu}$ ) ($R^2 = 0.9981$). Furthermore, model interpretation using SHapley Additive exPlanations (SHAP) identified the Reynolds number, Strouhal number, porosity, specific surface area, and pore size dispersion as the most influential predictors, while also revealing synergistic interactions among them. For example, SHAP-derived interactive thresholds, including $Re$ > 75 and porosity > 0.6256, provide practical guidance for enhancing convective heat transfer. This data-driven framework novelly integrates a hybrid approach to predict thermal performance in porous media under oscillatory flow, with the implementation of explainable machine learning, delivering both quantitative predictive accuracy and physical interpretability, offering guidelines for identifying and designing favorable oscillatory flow and structural conditions that enhance thermal performance in complex porous media.

## 1 Introduction

Recently, numerous advanced energy technologies have emerged in which oscillating flow and heat transfer in porous structures play a critical role in improving system thermal efficiency. Examples include the use of phase change materials (PCMs) as the heat transfer medium in solar thermal energy storage (TES) systems for concentrated solar power (CSP) plants [1–3], flow and condensation processes in the regenerators of thermoacoustic coolers using air as the working fluid [4–6], and the operation of enhanced geothermal systems (EGS) that extract energy from fractured rock formations [7,8].

Despite their widespread application, the thermal performance of oscillatory flows in porous media remains poorly understood and difficult to predict reliably. Although numerous experimental and numerical investigations have been conducted, the findings in the literature often show inconsistent or even contradictory trends, making it challenging to develop a unified predictive framework. For instance, Fu et al. [9] conducted experimental tests using aluminum foams with a high porosity of 0.9, subjecting them to oscillating airflow in the frequency range of 2–8 Hz. They observed that oscillatory flow significantly improved the average Nusselt number ($\overline{Nu}$) compared to a steady flow. Moreover, they found that $\overline{Nu}$ positively correlated with both the dimensionless amplitude and frequency. In a related experimental study, Bayomy and Saghir [10] explored oscillating flow in aluminum foam radiators at frequencies of 0.04–0.1 Hz and dimensionless amplitudes of 297–1353. Their results revealed a 14% improvement in $\overline{Nu}$ relative to stable flow. Additionally, they reported improved internal temperature uniformity across the system.

However, such enhancement is not universally observed. Paek et al. [11], using experimental bead-packed channels, showed that at small oscillation amplitudes, the improvement in heat transfer was negligible across a frequency range of 0.5–3 Hz. In another study, Huang and Yang [12], who carried out numerical simulations in a channel with porous-block heat sources at the Strouhal numbers ($St$) range of 0.5–3. They identified a critical $St$, above or below which the heat transfer benefit weakened. This non-monotonic trend was also confirmed by Al-Sumaily and Thompson [13], who reported numerical simulations that $\overline{Nu}$ initially increased but then decreased with increasing $St$ between 0 and 2, suggesting the presence of an optimal frequency beyond which the thermal benefit diminishes

Conversely, Forooghi et al. [14] conducted a numerical study on both steady and oscillatory flows in a channel that was partially filled with a two-layer porous medium (porosity = 0.5) under a constant wall heat flux. Their simulation results demonstrated that increasing the oscillation amplitude (from 0 to 1.8) enhanced the average Nusselt number. However, the influence of oscillation frequency, characterized by the Womersley number ($Wo$), ranging from 0 to 20, exhibited a non-monotonic trend. Specifically, the average Nusselt number initially decreased with increasing $Wo$, reaching a minimum at an intermediate frequency, and then increased again at higher $Wo$ values.

These discrepancies underscore that the behavior of $\overline{Nu}$ is highly dependent on the porous medium's structures and the nature of the oscillations such as frequency and amplitude [15]. Oscillating flow can affect $\overline{Nu}$ through various ways, such as boundary layer disruption, enhanced flow mixing, and the induction of nonlinear velocity profiles [16]. Meanwhile, the heterogeneity and anisotropy of porous structures further complicate predictive modeling [17]. To date, achieving a unified methodology or conclusion using traditional experimental or numerical approaches remains extremely challenging; the specific obstacles are outlined below.

To develop a method capable of accurately predicting and interpreting the thermal performance under oscillating conditions in porous structures, the foremost requirement is that the porous structures studied possess high topological diversity to capture inherent heterogeneity and anisotropy. However, most existing studies employ regular, simplified, or idealized porous geometries [18–23]. While such approaches provide valuable insights for specific scenarios, they fail to represent the complexity of real-world porous structures, such as rock formations in EGS [8], thereby limiting generalizability. A second major challenge is that achieving high topological diversity in experimental or computational models can significantly increase fabrication and computational costs. Furthermore, conventional predictive pipelines and classical regression models may prove insufficient even when large and diverse structural datasets are obtained. While traditional regressors are often valued for their physical interpretability, they frequently exhibit relatively large prediction errors [24].

To address these challenges, recent advancements in artificial intelligence (AI) offer promising solutions. For instance, machine learning (ML) models can perform regression analysis and make predictions based on large datasets at a lower computational cost and prediction errors [25]. ML models have been widely adopted for regression and prediction tasks in thermal-fluid applications. For example, Imran et al. [26] used Artificial Neural Networks (ANNs) to predict flow and temperature fields for fluid along a shrinking sheet using an engine oil-based fluid emulsified with nanoparticles. Fowler et al. [27] employed Physics-Informed Neural Networks (PINNs) to perform a forward simulation without leveraging known data, enabling higher-order parameter space solutions to be feasible. Cheng et al. [28] used Extreme Gradient Boosting (XGBoost) to predict heat and mass transfer in phase change systems. Furthermore, generative AI approaches, such as deep learning (DL) models, can create new porous structures with diverse topologies. In particular, Generative Adversarial Networks (GANs) have been increasingly used for this purpose due to their ability to learn from training data and generate realistic synthetic samples [29]. Tan et al. [30] developed a hybrid Convolutional Neural Network (CNN) - GAN approach to generate microstructures and predict their thermal performance. Similarly, the Wasserstein GAN with Gradient Penalty (wGAN-GP) [31] has been employed to design porous structures for enhanced heat transfer [32], and CNN-wGAN models have been used to design high-performance heat sinks [33].

By leveraging these capabilities, we integrate generative and predictive ML with physics-based simulation to estimate and elucidate oscillatory flow in porous media with heat transfer. A wGAN-GP model, trained on micro-computed tomography (micro-CT) images of natural coral rock, which is chosen for its natural exposure to oscillating ocean currents and thermal gradients, generates high-diversity porous structures suitable for biomimetic design. The lattice Boltzmann method (LBM) [34] is then used to evaluate flow and heat transfer through each generated geometry and to label the data with $\overline{Nu}$ under prescribed oscillation conditions (varying Reynolds number, $Re$, and Strouhal number, $St$). This produces a dataset that links thermal performance to oscillatory parameters and topological descriptors (Table 1). Ten predictive ML models (see Sec. 2.3) are trained and compared for $\overline{Nu}$ prediction using nested cross-validation (Nested_CV) [35], with hyperparameters tuned via Bayesian Optimization (BO) [36]. The best-performing model is subsequently retrained on the full dataset and assessed on a final hold-out set for unbiased performance estimation. Beyond prediction, we apply SHapley Additive exPlanations (SHAP) [37] to interpret and quantify the influence of oscillation parameters and structural features on $\overline{Nu}$, providing insights into the key factors governing thermal performance.

The novelty of this study lies in its methodological integration and in the first systematic application of such a hybrid approach to oscillatory flow in porous media with heat transfer, in order to predict the thermal performance with enhanced structural patterns across a broad design space. Additionally, explainable AI, SHAP, generates quantitatively ranked hypotheses about influential features and identifies the mechanism of synergistic interactions among them.

## 2 Methodology

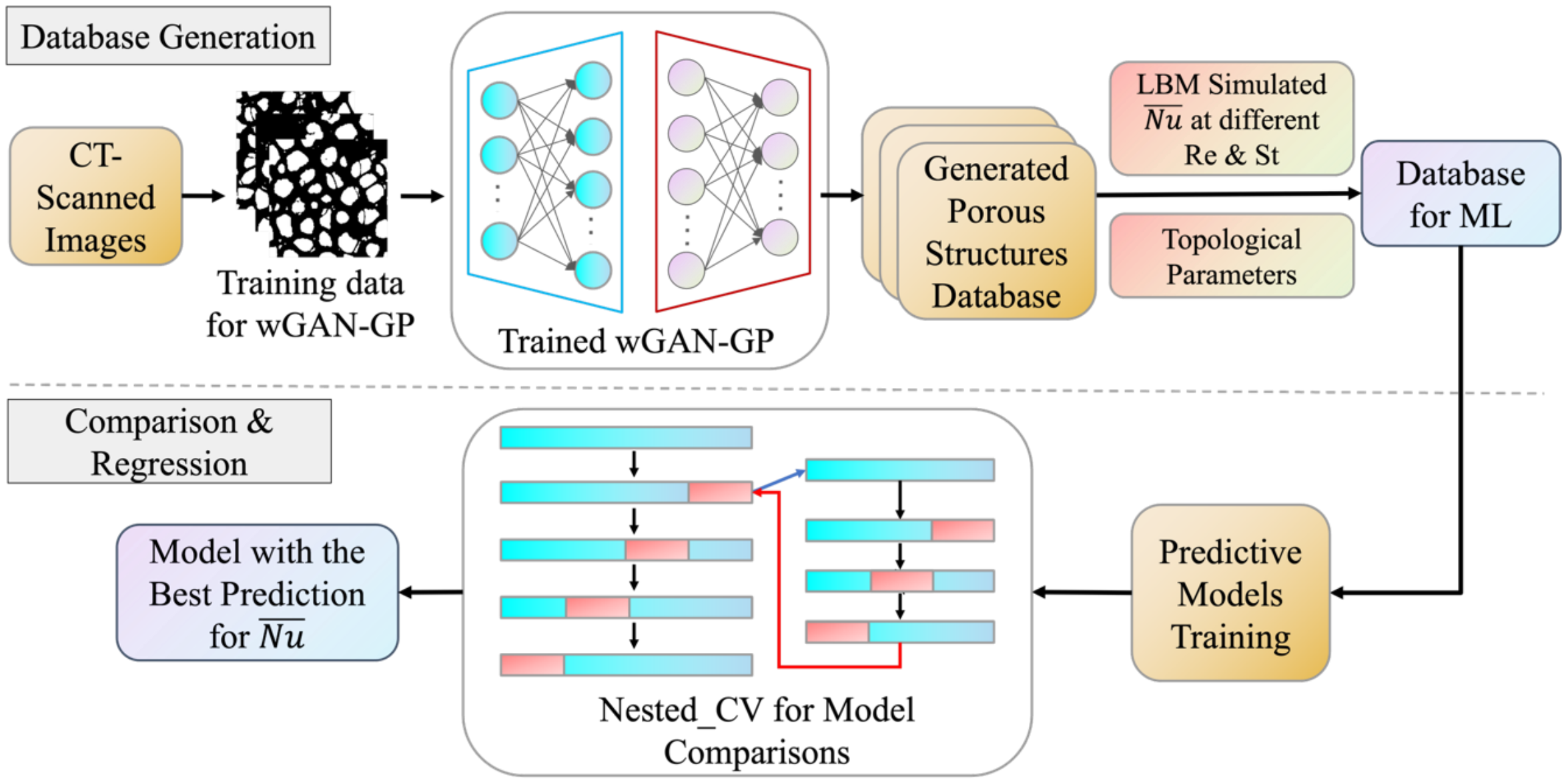

***Fig. 1*** *Flowchart of the proposed wGAN-LBM-Nested_CV architecture*

This section outlines the detailed implementation process of the proposed wGAN-LBM-Nested_CV architecture, as illustrated in Fig. 1. It consists of two major phases: 1) data generation; and 2) model comparison and regression. Fig. 1 shows that two databases are required in the data generation phase. The first one consists of porous structure images obtained from micro-CT scans of natural coral rock, which are used to train the wGAN-GP model (described in Sec. 2.1). The second database, described in detail in Sec. 2.2, contains $\overline{Nu}$ values computed using LBM simulations based on the porous structures generated by the trained wGAN-GP. This database comprises thermal performance metrics ($\overline{Nu}$) as the target variable and oscillatory flow parameters ($Re$ and $St$), along with topological features as the feature variables. It is then used to train and evaluate machine learning models, with the model evaluation and comparison techniques discussed in Sec. 3.3.

### *2.1 CT-scanned porous structures post-processing for wGAN-GP training*

The training dataset for the wGAN-GP was obtained non-destructively via X-ray micro-CT. A natural coral rock sample (Fig. 2(A)) was scanned using a SkyScan 2214 system with a sub-500 nm spot size at the Advanced Manufacturing Institute of the University of Houston. This process generated a sequence of 2D cross-sectional images at a spatial resolution of 27 μm. The resulting image stack was then reconstructed into a 3D visualization of the coral structure using Autodesk 3ds MAX 2024, as shown in Fig. 2(B).

The 2D slices were extracted from the reconstructed 3D structure for the subsequent deep learning tasks. A representative slice is shown in Fig. 2(C), where the white regions correspond to the solid matrix and the black regions to voids. A region of interest (ROI) measuring 0.2 in × 0.2 in with a pixel resolution of 187 × 187 is highlighted in red in Fig. 2(C). These ROIs were used as training data for the wGAN-GP so that the generated images could serve as inputs for LBM simulations. However, the original void regions from the CT scans appear disconnected, potentially blocking flow channels. To address this, the binary profiles of the selected ROIs were inverted so that previously connected solid regions became voids, creating flow domains more suitable for simulations. This approach preserves the coral's structural characteristics while adapting them for engineered flow studies. An example of an inverted ROI is shown in Fig. 2(D), where the voids appear black and the solid phase appears white.

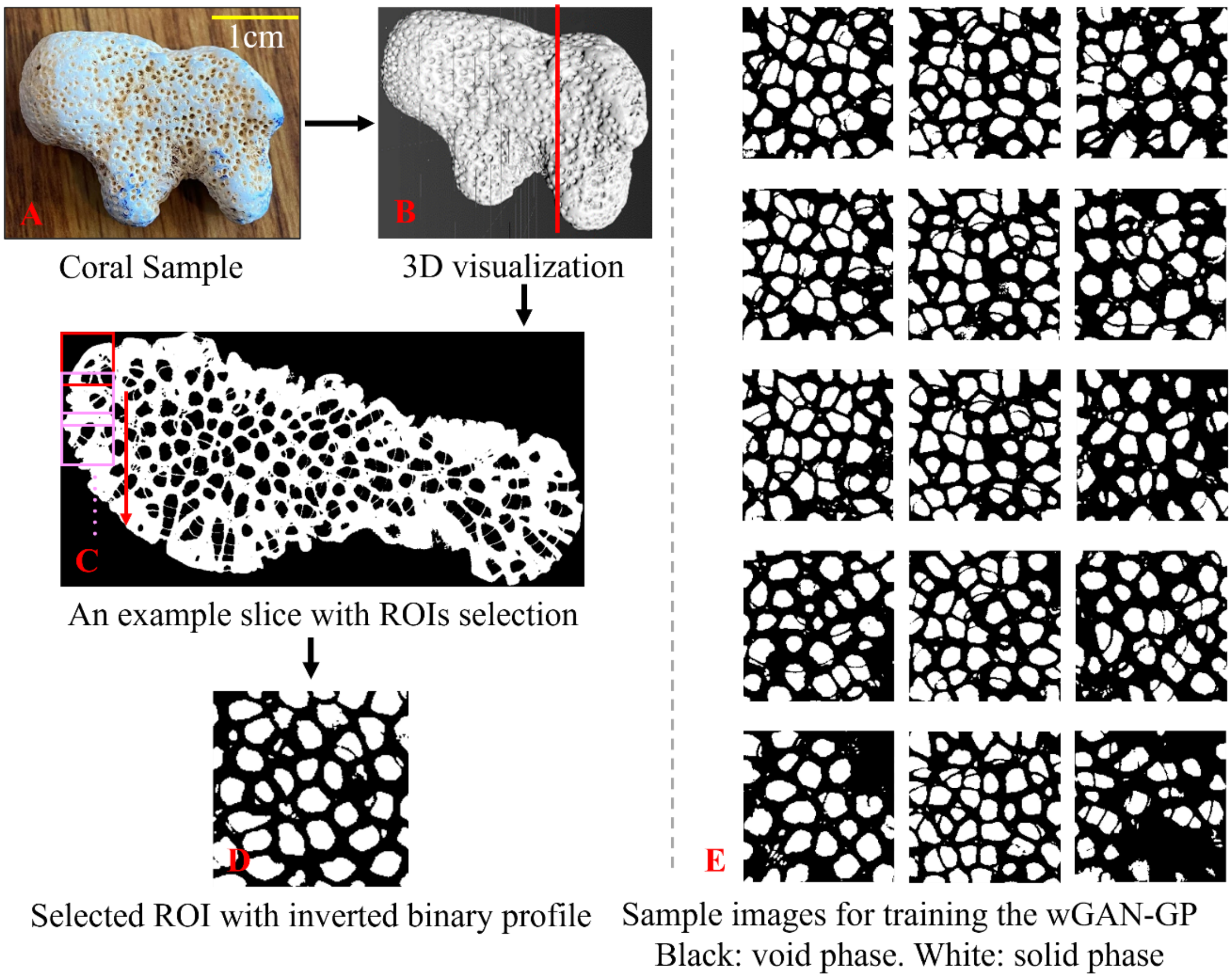

***Fig. 2** Workflow for generating the 2D porous structure dataset from CT-scanned coral rock*

To generate a large dataset of these 187 × 187 2D porous structures, a batch-processing workflow was created using the open-source platform FIJI [32,38]. The porosity (∅) of each structure was calculated using Eq. (1) [39]:

$$\emptyset = \frac{Area\ of\ voids}{Total\ area} \tag{1}$$

which is implemented via `NumPy` in Python 3.12.

An in-house macro, developed in Java within the FIJI platform, was employed to automate the extraction of two-dimensional square ROIs. As illustrated by the pink boxes in Fig. 2(C), ROIs were systematically sampled across the image slices at horizontal and vertical intervals of 40 pixels and 20 pixels, respectively. This procedure yielded a dataset comprising 20,000 unique porous structures, with porosity values ranging from 47 % to 84 %. Representative examples are shown in Fig. 2(E), where the black pixels represent the void space and the white pixels denote the solid matrix.

*Table 1 Topological parameters to characterize the porous structures*

| Parameters | Definition | Major Python Library |
|---|---|---|
| Connectivity | The number of disconnected blocks | `scikit-image` |
| Porosity | The area of voids over the total area | `NumPy` |
| Percolation Strength | Ratio of the largest connected pore area to the total pore area | `scikit-image` |
| Tortuosity | Ratio of actual path length to straight path | `SciPy` |
| Specific Surface Area (SSA) | Ratio of interface area to the total image area | `SciPy` |
| Mean Pore Diameter | Average pore diameter (note: pore diameter is the maximum inscribed circle diameter) | `SciPy` |
| Pore Size Dispersion | Standard deviation of the pore diameters | `SciPy` |
| Euler Number | Euler number = number of connected components - number of unconnected components | `scikit-image` |
| Network Connectivity Index (NCI) | The normalized average node degree ($\bar{k}$) of the skeleton graph (note: $NCI = \bar{k}/8$) | `NetworkX` |

The CT-derived dataset was used to train wGAN-GP, enabling it to generate synthetic porous structures with high topological diversity for LBM simulation. All topological parameters used in this work are listed in Table 1 [40], and their statistical representations, computed using `R version 4.5.0`, are summarized in Table 2.

***Table 2** Statistics of topological parameters for the 20,000 CT-scanned porous structures*

| | Porosity | Connec-tivity | Percolation Strength | Tortuosity | SSA | Mean Pore Diamet-er | Pore Size Dispers-ion | Euler Number | NCI |
|---|---|---|---|---|---|---|---|---|---|
| Min. | 0.4789 | 1 | 0.5943 | 1.165 | 0.0652 | 5.662 | 3.165 | -85 | 0.5089 |
| Median | 0.6085 | 6 | 0.9992 | 1.358 | 0.1155 | 9.540 | 6.516 | -42 | 0.5201 |
| Mean | 0.6120 | 6.442 | 0.9984 | 1.354 | 0.1165 | 10.475 | 7.586 | -40.6 | 0.5199 |
| Max. | 0.8397 | 32 | 1.0000 | 1.407 | 0.1521 | 37.581 | 38.527 | 3 | 0.5308 |

The wGAN-GP is a generative deep learning model belonging to the family of Generative Adversarial Networks (GANs). Unlike conventional GANs, which rely on cross-entropy loss and are often prone to training instabilities, wGAN-GP employs the Wasserstein distance as a metric to quantify the difference between generated and real data distributions [41]. This approach provides a more robust and stable training process. To ensure the validity of the Wasserstein distance, the Critic (C) network in wGAN-GP must satisfy the 1-Lipschitz continuity condition. This requirement is enforced by introducing a gradient penalty term into the Critic's loss function, substantially enhancing both training stability and the fidelity of generated samples [31].

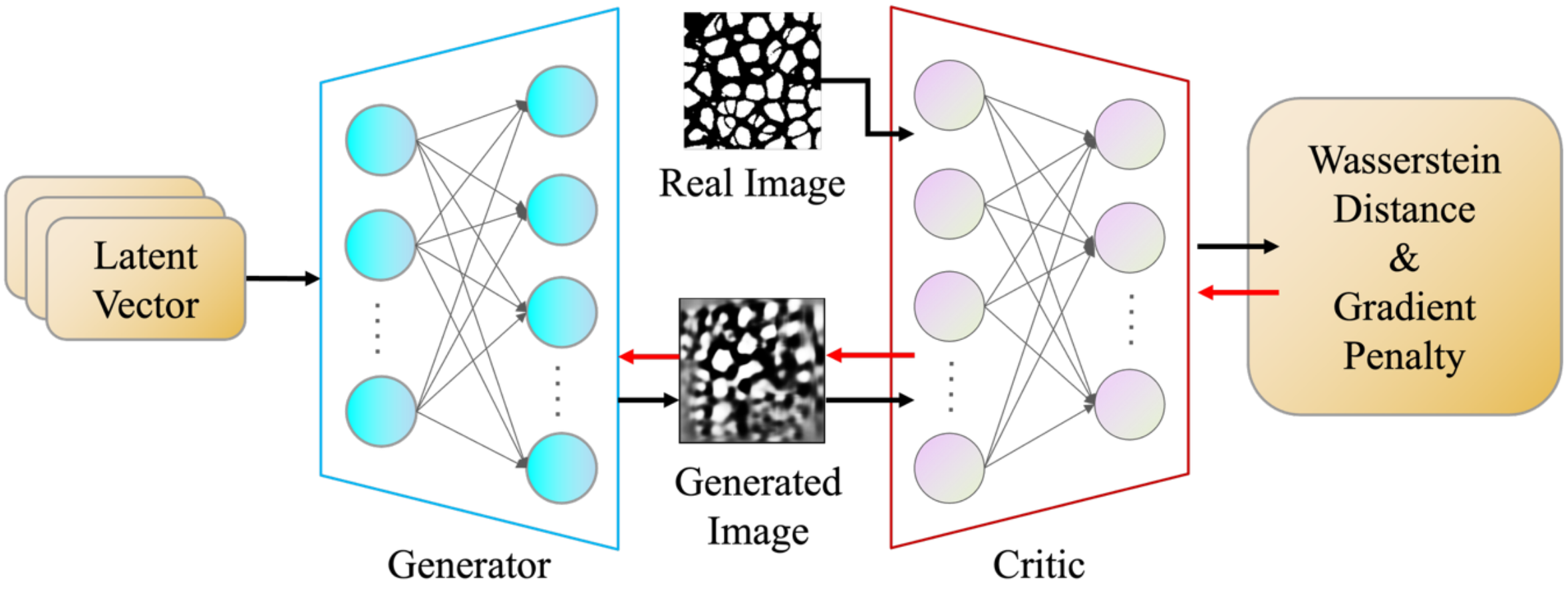

***Fig. 3** Schematic of the wGAN-GP model*

Fig. 3 shows the schematic of the wGAN-GP framework implemented in this study. A random noise vector (latent vector) is first fed into the Generator (G), which produces a synthetic porous structure image. The Critic then evaluates the Wasserstein distance between the generated and real images. This distance is used to compute the Generator's loss, providing a feedback signal that iteratively improves the Generator's ability to produce highly realistic porous structures.

By leveraging the combination and interpolation capabilities of wGAN-GP in latent space, porous structures with enhanced topological diversity can be generated. These images are subsequently employed as computational domains in LBM simulations to evaluate the target variable, $\overline{Nu}$. The topological parameters extracted from the generated images constitute part of the feature set for the predictive model, while additional features related to oscillatory flow parameters are introduced in the following section.

*2.2 LBM simulations*

*2.2.1 LBM fundamentals*

The lattice Boltzmann method (LBM) [34] was applied in this work, as it is a robust numerical simulation approach for modeling flow in porous structure. The LBM is a discretization of the Boltzmann equation where a discrete number of particles (groupings of matter) collide in space and stream over time based on a defined lattice. The LBM momentum equation with the commonly used Bhatnagar–Gross–Krook (BGK) collision model is given in Eq. (2):

$$f_i(x + e_i\Delta t,\ t + \Delta t) - f_i(x,\ t) = -\frac{1}{\tau}\left(f_i(x,\ t) - f_i^{eq}(x,t)\right) \tag{2}$$

The LBM heat equation with BGK collision is:

$$g_i(x + e_i\Delta t,\ t + \Delta t) - g_i(x,\ t) = -\frac{1}{\tau_g}\left(g_i(x,\ t) - g_i^{eq}(x,t)\right) \tag{3}$$

where $f_i$ and $g_i$ are the distribution functions for fluid and thermal particle fields, $e_i$ is the discrete velocity, and $\tau$ and $\tau_g$ are the dimensionless momentum relaxation and thermal relaxation time, respectively.

The equilibrium and heat equilibrium distribution functions at $x, t$ are defined as Eq. (4) and (5), where $\rho$, T, and $w_i$ are density, temperature, and lattice weight, respectively, $c_s$ is the speed of sound, and $u$ is velocity.

$$f_i^{eq}(x,t) = w_i\rho(1 + \frac{u \cdot e_i}{c_s^2} + \frac{(u \cdot e_i)^2}{2c_s^2} - \frac{u \cdot u}{2c_s^2}) \tag{4}$$

$$g_i^{eq}(x,t) = w_i T(1 + \frac{u \cdot e_i}{c_s^2} + \frac{(u \cdot e_i)^2}{2c_s^2} - \frac{u \cdot u}{2c_s^2}) \tag{5}$$

The relaxation times $\tau$ and $\tau_g$ can be expressed as equations (6) and (7):

$$\tau = \frac{\nu}{c_s^2 dt} + 0.5 \tag{6}$$

$$\tau_g = \frac{\alpha}{c_s^2 dt} + 0.5 \tag{7}$$

where $\nu$ and $\alpha$ are kinematic viscosity and thermal diffusivity, and $dx$ and $dt$ are the lattice length and the time step.

In this work, the D2Q9 lattice model was used for the velocity field, and the D2Q5 model for the temperature field, as D2Q5 reduces computational cost without compromising physical accuracy. The corresponding lattice schemes are illustrated in Fig. 4. Therefore, for D2Q9, the lattice weights are $w_i$ ={4/9, 1/9, 1/9, 1/9, 1/9, 1/36, 1/36, 1/36, 1/36}, where $i = 0{\sim}8$. Similarly, for D2Q5, $w_i$ ={1/3, 1/6, 1/6, 1/6, 1/6}, where $i = 0{\sim}4$.

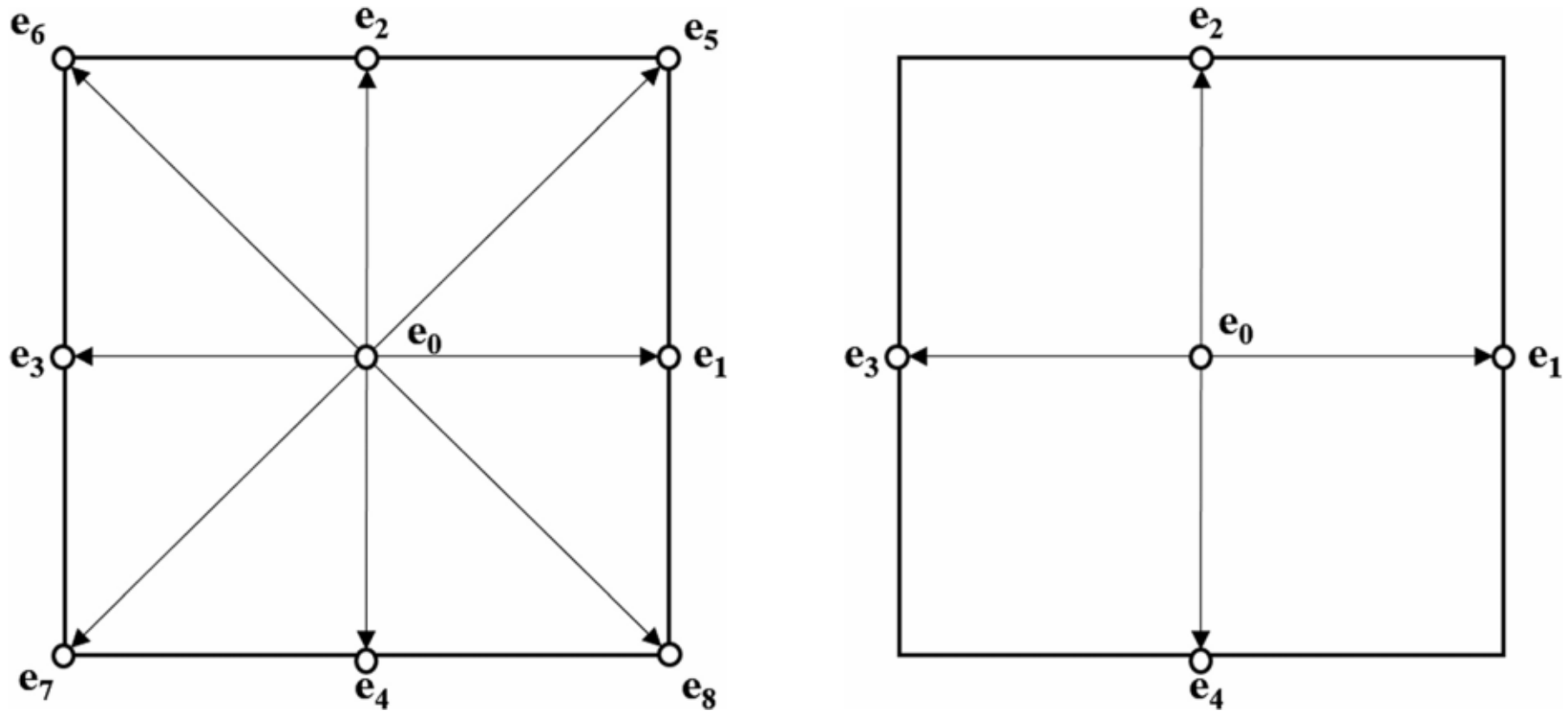

***Fig. 4*** *D2Q9 (left) and D2Q5 (right) lattice schemes*

*2.2.2 Simulation setup and boundary conditions*

*Palabos*, an open-source LBM solver, was chosen for the LBM simulations [42]. The setup is illustrated in Fig. 5, where single-phase, low-temperature water flows through heated porous structures subjected to an oscillating inlet velocity imposed at the left boundary of the porous region. All the flow domains for the batch processing have the same dimensions of $l_x$ = $l_y$ = 0.2 in = 5.08 mm = 187 pixels, consistent with the wGAN-generated structures. Periodic boundary conditions are applied at the top and bottom, while the outlet is set to be adiabatic with a constant pressure of 1 atm. The initial fluid temperature is $T_{cold} = T_{in} = 300$ K, and the solid matrix is maintained at $T_{hot} = T_{solid} = 350$ K. Oscillatory inlet velocity is defined as:

$$U_{in} = A\, sin(2\pi f t + \varphi) \tag{8}$$

where $A$ is amplitude (also is $U_{in,max}$), $f$ is frequency, $\varphi$ is the phase angle that is defined as 0 in this work, and $t$ is time.

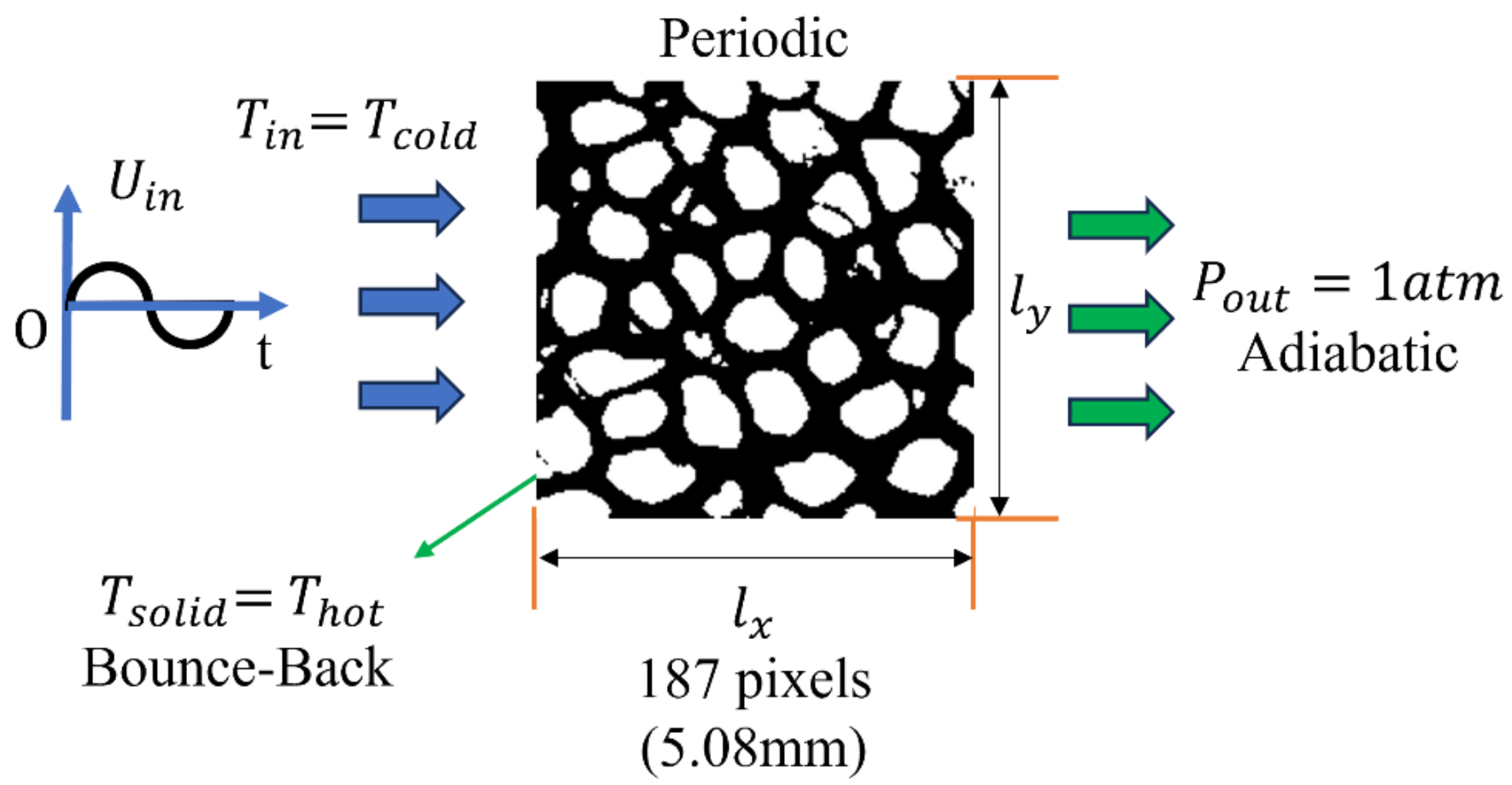

***Fig. 5** Dimensions and boundary conditions for the 2D LBM simulations*

Relevant dimensionless numbers include:

$$Re = \frac{U_{in,max} \cdot l_y}{\nu}, \qquad St = \frac{f \cdot l_x}{U_{in,max}}, \qquad Pr = \frac{\nu}{\alpha} \tag{9, 10, 11}$$

The Prandtl number (*Pr*), determined by fluid properties and an average temperature of 325K, was set to $Pr = 3.5$, giving $\nu = 0.53\ mm^2/s$. The target variable of this work, the average Nusselt number, for the entire domain, was calculated from $Nu = hl/k$, where $h = q''/\Delta T$, giving $Nu = q''l/\Delta Tk$. Accordingly, the average Nusselt number for the whole domain $\overline{Nu}$ is defined in Eq. (12), where $q''$, $S$, and $k$ are heat flux, domain area, and thermal conductivity, respectively.

$$\overline{Nu} = \frac{1}{S} \iint \frac{q''l}{\Delta Tk} dxdy \tag{12}$$

A total of 56 simulation groups were conducted, covering all combinations of 8 Reynolds numbers and 7 Strouhal numbers, as defined in Eq. (13).

$$\begin{cases} Re = 15, 30, 45, 60, 75, 90, 105, 120 \\ St = 1, 100, 200, 400, 600, 800, 1000 \end{cases} \tag{13}$$

Since *Pr* was fixed in this study, *Re* and *St* were selected as the feature variables representing the oscillation conditions. It is worth noting that approximating a non-oscillating condition without imposing a completely static flow ($U_{in} = 0$ when $St = 0$), the lowest Strouhal number was set to $St = 1$ in this work.

*2.3 ML model comparison using nested cross-validation with Bayesian Optimization*

The database was constructed using $Re$, $St$, and the topological parameters listed in Table 1, resulting in a total of 11 feature variables, with $\overline{Nu}$ as the target variable. This database was then used to evaluate the performance of the selected ML models.

This study compared the predictive performance of 10 widely used ML models for estimating fluid and thermal properties [43–47]. These models fall into five main categories: support vector machines (SVM) [48], tree-based methods [49–51], neural networks [52], k-nearest neighbors (k-NN) [53], and linear models [54,55]. A complete list of the models, their classifications, and the Python libraries used is provided in Table A.1 of Appendix A.

To obtain an unbiased and comprehensive estimate of model performance, Nested_CV was employed, with Bayesian Optimization (BO) used for hyperparameter tuning. The outer loop of the Nested_CV performs classic $k$-fold validation [35]. Model performance for each validation is evaluated using Root Mean Squared Error (RMSE):

$$RMSE = \sqrt{\frac{1}{n}\sum_{i=1}^{n}(a_i - \widehat{a_i})^2} \quad (14)$$

where $a_i$ is the actual value of $i - th$ data point while $\widehat{a_i}$ is the predicted value.

This procedure is repeated $k$ times, such that each of the $k$ subsets serves as the testing set. The cross-validated performance is then calculated as the average of the performance metrics obtained from the $k$ individual trials. The upside of Fig. 6 shows this process when outer_fold = 4, while the right side depicts the inner loop when inner_fold = 3. For each outer loop iteration, a separate internal CV is performed solely on the training data of that outer fold. The purpose of this inner loop is to determine the optimal hyperparameters. The optimizer proposes various hyperparameter configurations, and their performance, which is, for example, measured by average RMSE, is evaluated. The set of hyperparameters yielding the best performance (i.e., the best RMSE_tuned in Fig. 6) is then passed to the outer loop as the optimal configuration for the current fold. The model is trained on the complete outer training set using these hyperparameters and subsequently evaluated on the hold-out test set. This process is repeated for all $k$ outer folds, and the average of the resulting test scores constitutes the final unbiased performance estimate for the model (i.e., RMSE_Model). This metric is used to compare the different ML models, with the best-performing model corresponding to the lowest RMSE_Model value.

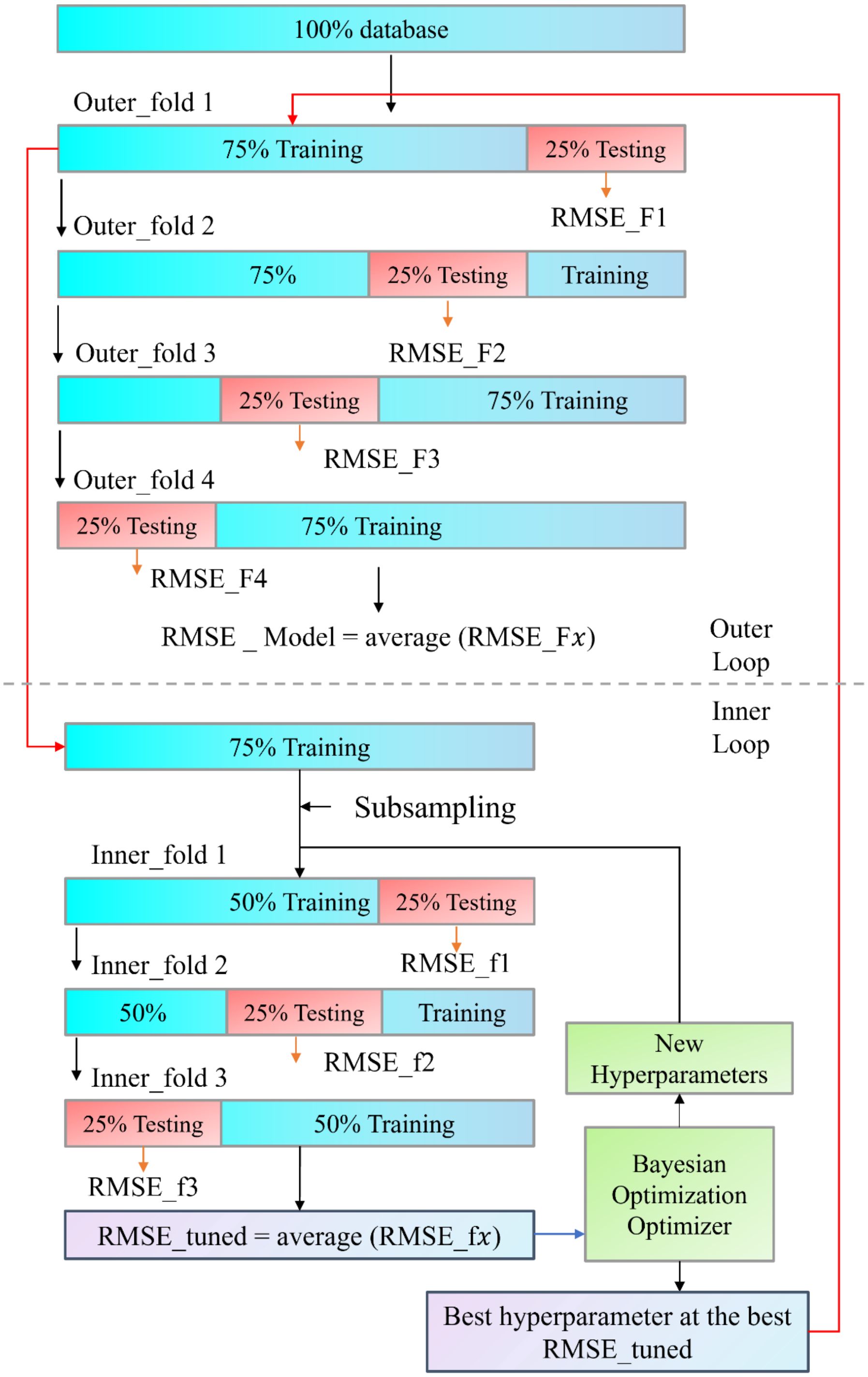

***Fig. 6*** *Schematic of the Nested_CV process (using outer_fold = 4 and inner_fold = 3 as an example)*

To reduce computational cost during the inner loop, a subsampled portion of the training set may be used for hyperparameter tuning. Although this can slightly alter the performance value for a specific hyperparameter set, it does not affect the overall ranking of models.

Hyperparameter optimization is a critical step in maximizing model performance [56]. While methods such as Grid Search or Random Search are widely applied [57], this work utilized Bayesian Optimization (BO) for its computational efficiency. BO balances exploration and exploitation through an acquisition function, with the generalization performance of the learning algorithm modeled via a Gaussian process (GP) surrogate [58]. This approach typically achieves better results with fewer evaluations. In this study, BO was implemented using the Python library `bayes_opt`.

In addition to the RMSE, the model performance was also assessed using the Mean Absolute Error (MAE) and the coefficient of determination ( $R^2$ ). The mathematical formulations for these metrics are provided in Eqs. (15-16), where SSE represents the Sum of Squared Errors, SST is the Total Sum of Squares, and $\bar{a}$ denotes the mean of the actual values.

$$R^2 = 1 - \frac{SSE}{SST} = 1 - \frac{\sum_{i=1}^{n}(a_i - \hat{a}_i)^2}{\sum_{i=1}^{n}(a_i - \bar{a})^2} \tag{15}$$

$$MAE = \frac{1}{n}\sum_{i=1}^{n}|a_i - \hat{a}_i| \tag{16}$$

The value of RMSE, MAE, and $R^2$, are computed using the `mean_squared_error`, `mean_absolute_error`, and `r2_score` functions, respectively, from the `sklearn.metrics` library in Python.

The Nested_CV framework enables a comprehensive comparison of the selected ML models. Once the best-performing model is identified, it is subsequently employed for further hold-out implementation. Model interpretation is then conducted using the SHAP method to provide physical insights into oscillating flow and heat transfer phenomena in porous structures. The comparison results of the ML models are presented in Sec. 3.3, the hold-out test results are detailed in Sec. 3.4, and the SHAP-based interpretation is discussed in Sec. 4.

## 3 Results

### *3.1 wGAN-GP*

#### 3.1.1 Performance of the wGAN-GP training

Training of the wGAN-GP model was performed for 200 epochs on an NVIDIA RTX 6000

Ada GPU, requiring approximately 24 hours. The training dynamics were monitored using the Generator and Critic losses, as shown in Fig. 7.

Analysis of the loss functions indicates that the wGAN-GP model exhibits stable and effective learning behavior. The Generator loss shows a gradual upward trend, which in the context of the wGAN-GP framework, does not indicate degraded performance. Instead, it reflects the Generator's continuous adaptation in response to a Critic of increasing capability. A transient plateau observed between epochs ~125–155 may represent a temporary equilibrium, where the model has already captured the primary structural features of the training data. Beyond this point, training appears to shift towards learning finer structural details. However, the plateau may also suggest over-optimization of the Critic, which would potentially reduce the quality of generated porous structure images.

Correspondingly, the Critic loss remains stable at approximately –2.5 without significant oscillations. This stability indicates that the Critic is reliable in estimating the Wasserstein distance between real and generated images, while the gradient penalty term effectively enforces the Lipschitz constraint, thereby ensuring a consistent and stable gradient flow to the Generator.

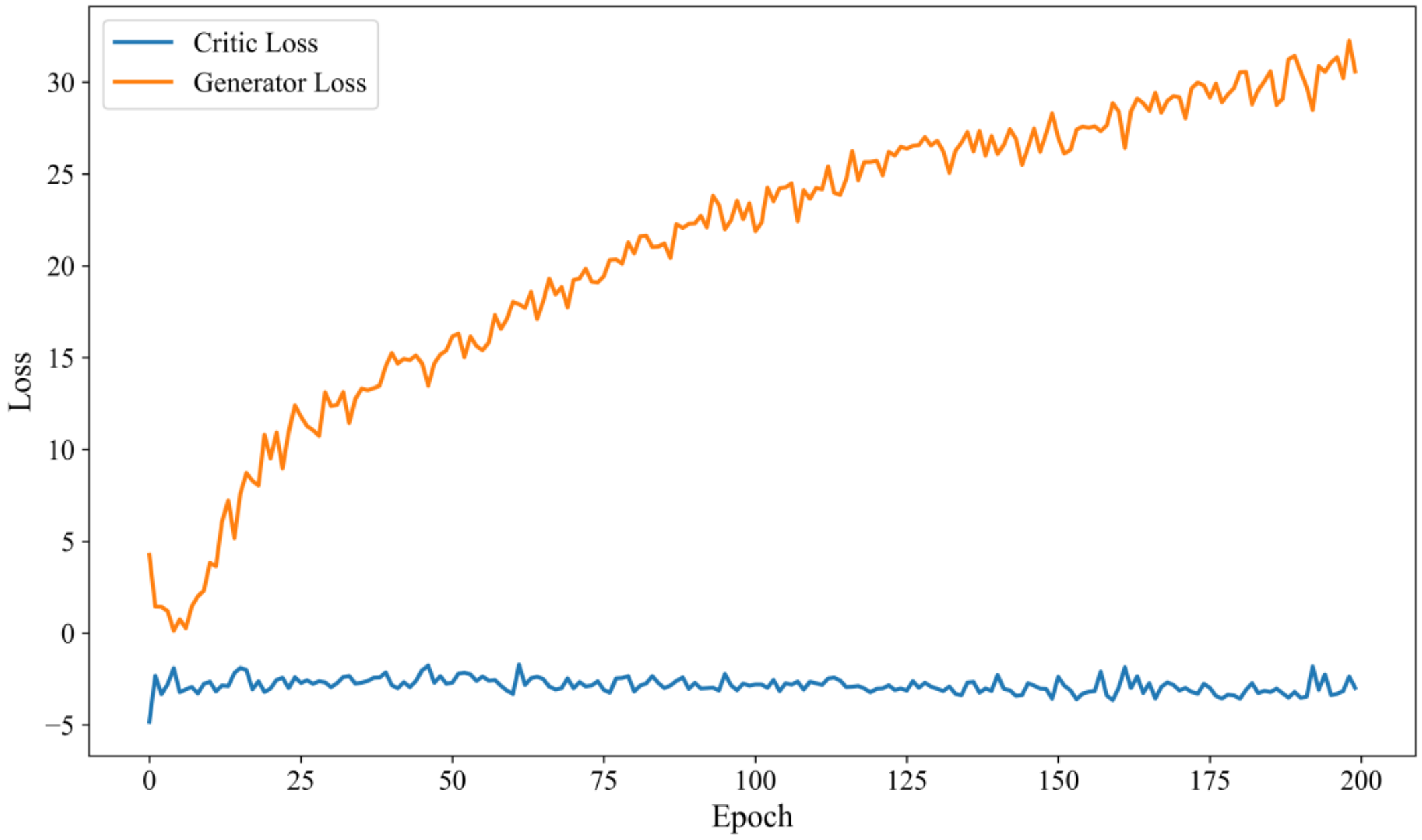

**Fig. 7** Training loss curves of the wGAN-GP model (Generator loss and Critic loss)

In summary, the behavior of both loss functions suggests a stable and robust training process, effectively avoiding common GAN-related issues such as mode collapse. The temporary stabilization of the Generator loss between epochs 125 and 155 suggests that near-optimal performance may occur within this range. However, the results may also imply that training for 200 epochs may not be sufficient for the wGAN-GP to fully converge, and extended training could further enhance performance.

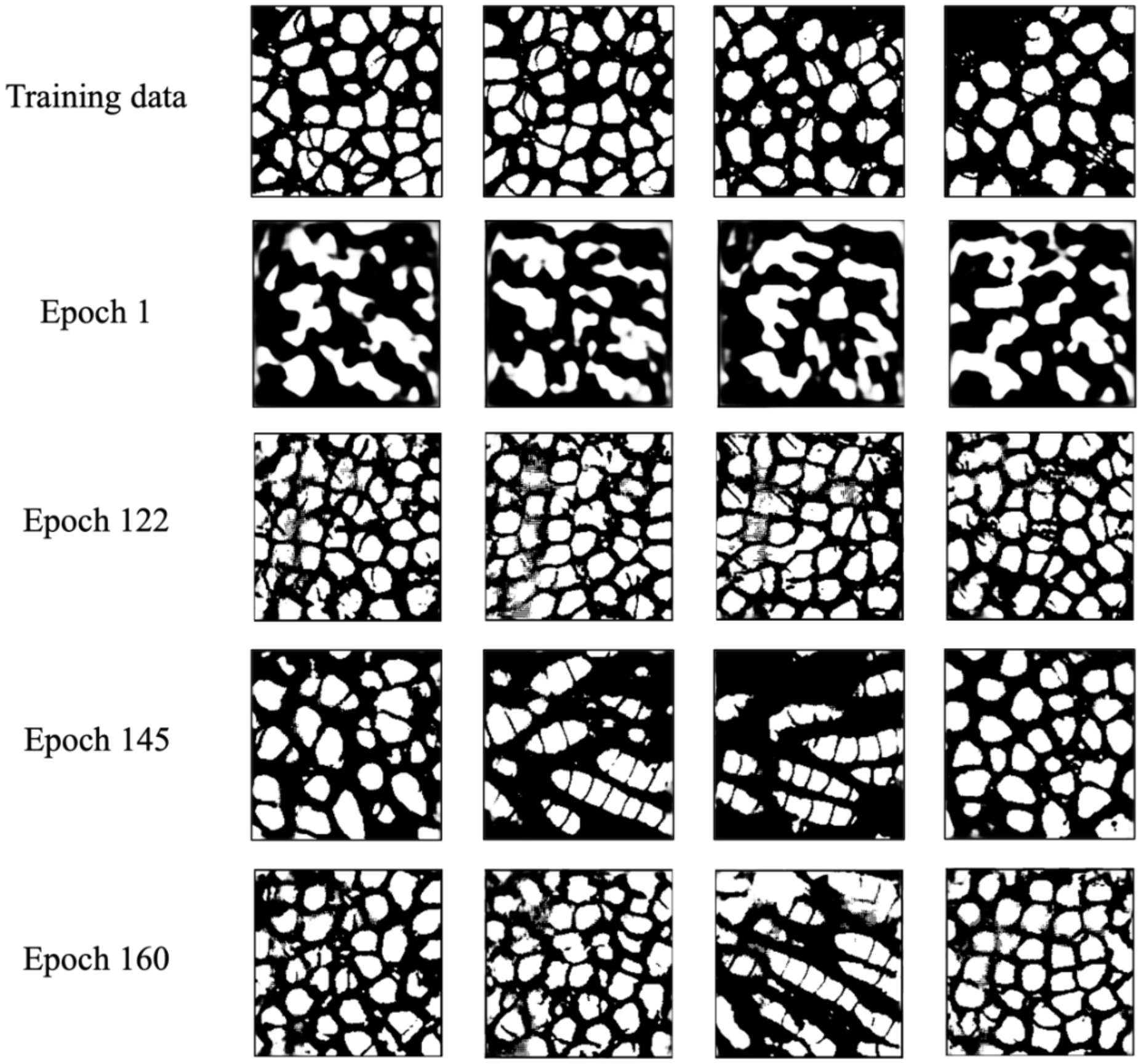

***Fig. 8*** *Comparison between wGAN-GP generated images (epochs 1, 122, 145, and 160) and training images. Images at Epoch 145 exhibit fewer gray, blurry, and unclear regions compared with Epochs 122 and 160.*

To further assess model quality, visual inspection was performed with particular attention to outputs at epochs 125 and 155. As shown in Fig. 8, random images generated by the wGAN-GP model at epochs 1, 122, 145, and 160 were compared against samples from the training dataset. A satisfactory model should produce images with a distinct separation between solid and void phases and exhibit geometric characteristics closely matching the training data.

At epoch 1, corresponding to the initial stage of training, the generated images were dominated by noise, which is a common outcome when the network has only begun learning the data distribution. Outputs from epochs 122 and 160 contained gray, blurred, and indistinct regions, where those from epoch 145 displayed sharper boundaries and more realistic features. This aligns with the plateau in the Generator loss curve, suggesting that training had temporarily converged during this period, with performance degrading due to under- or over-optimization before and after.

Based on these observations, the model at epoch 145 was selected for subsequent simulations, as it produced porous geometries with well-defined phase boundaries and strong visual fidelity to the training data.

*3.1.2 Comparison of Topological diversity for generated porous structure images*

A detailed comparative analysis was conducted to evaluate the topological characteristics of porous structures generated by the wGAN-GP model relative to those from the original CT-scanned dataset. To ensure consistency, 1,000 samples were selected from each dataset (both original and generated) based on matched porosity distributions. The resulting porosity distributions of the selected samples are shown in Fig. 9, ensuring that both datasets were compared under equivalent sampling criteria.

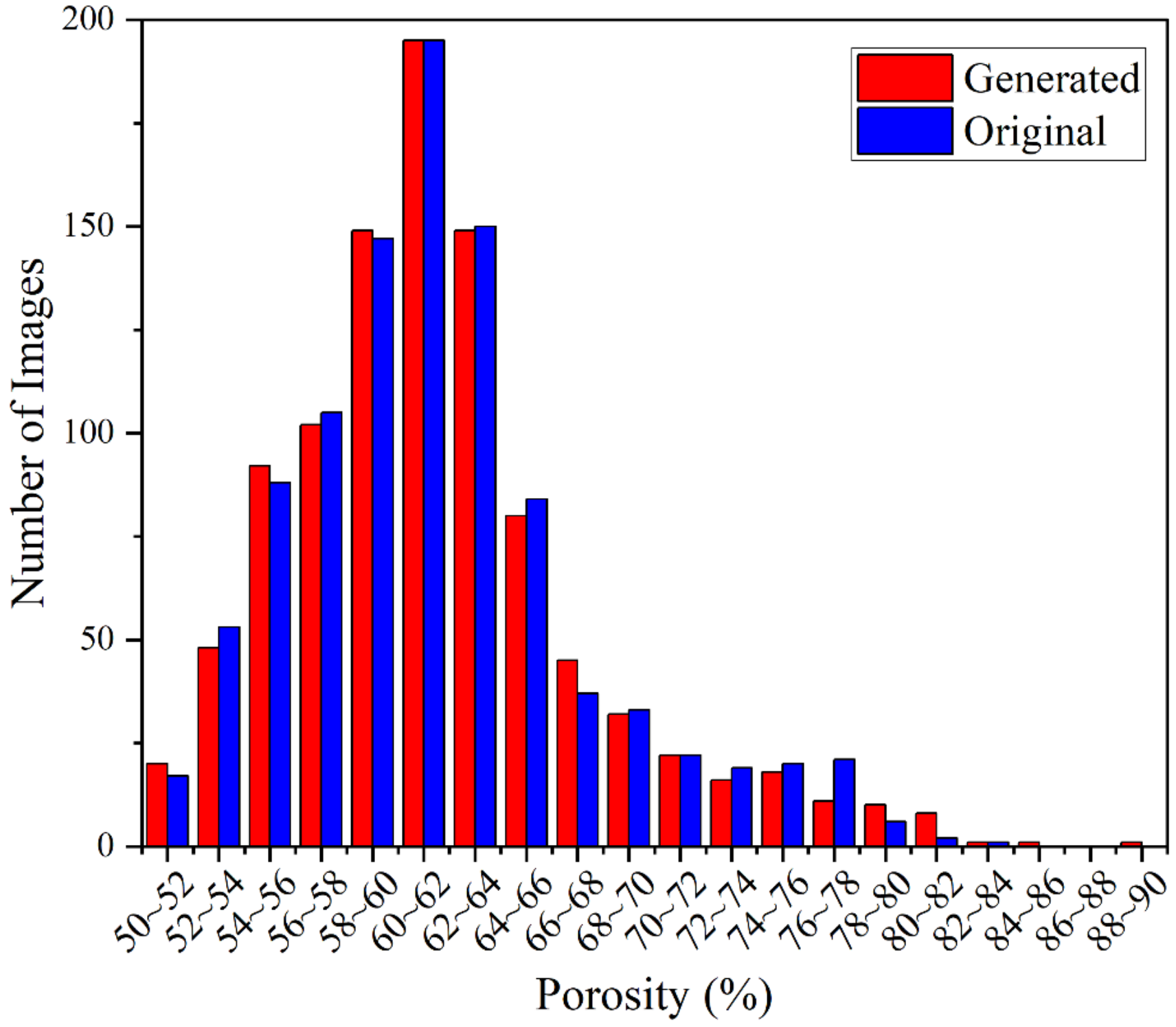

**Fig. 9** Porosity distributions of selected generated and original CT-scanned structures

Three metrics were employed to assess topological diversity: Mean Pairwise Distance (MPD), Information Entropy, and Principal Component Analysis (PCA). These metrics were applied to the 9 topological parameter features listed in Table 1. MPD quantifies the average Euclidean distance between all image pairs in the feature space, providing a measure of diversity—larger values indicate greater variation. Eq. (17) defines this calculation, where $x_i$ and $x_j$ represent 9-dimensional topological feature vectors for the *i-th* and *j-th* image, with $x_i \neq x_j$, and $dist$ denotes the calculation of Euclidean distance.

$$\textit{Mean Pairwise Distance (image diversity)} = \frac{1}{n(n-1)}\sum dist(x_i, x_j) \tag{17}$$

Information Entropy, shown in Eq. (18), is computed by normalizing each feature, discretizing it into histogram bins, calculating the frequency distribution $p(x)$, and then computing entropy. Higher entropy values indicate broader distributions, suggesting greater topological variability.

$$\text{Information Entropy} = -\sum p(x) log p(x) \tag{18}$$

PCA can be employed to reduce the 9-dimensional feature space to 2 dimensions while preserving maximal variance, enabling visual inspection of data spread and clustering behavior.

The calculations of MPD, Information Entropy, and PCA were performed using the Python libraries `sklearn.metrics.pairwise_distances`, `scipy.stats.entropy`, and `sklearn.decomposition.PCA`, respectively.

**Table 3** Comparison of topological metrics between 1,000 generated and 1,000 selected original structures

| Items | Images type | Values |
|---|---|---|
| Mean Pairwise Distance (image diversity) | Original structures | 17.1697 |
| | Generated structures | 25.1092 |
| Information Entropy | Original structures | 2.2803 |
| | Generated structures | 2.1011 |

Table 3 summarizes the diversity and entropy results for the 1,000 original and 1,000

generated structures. The generated structures exhibit a substantially higher MPD, indicating greater overall diversity, while their slightly lower entropy values suggest reduced dispersion within individual features. The PCA results shown in Fig. 10 further support this conclusion, as the generated images occupy a broader region in the projected feature space, demonstrating enhanced topological diversity.

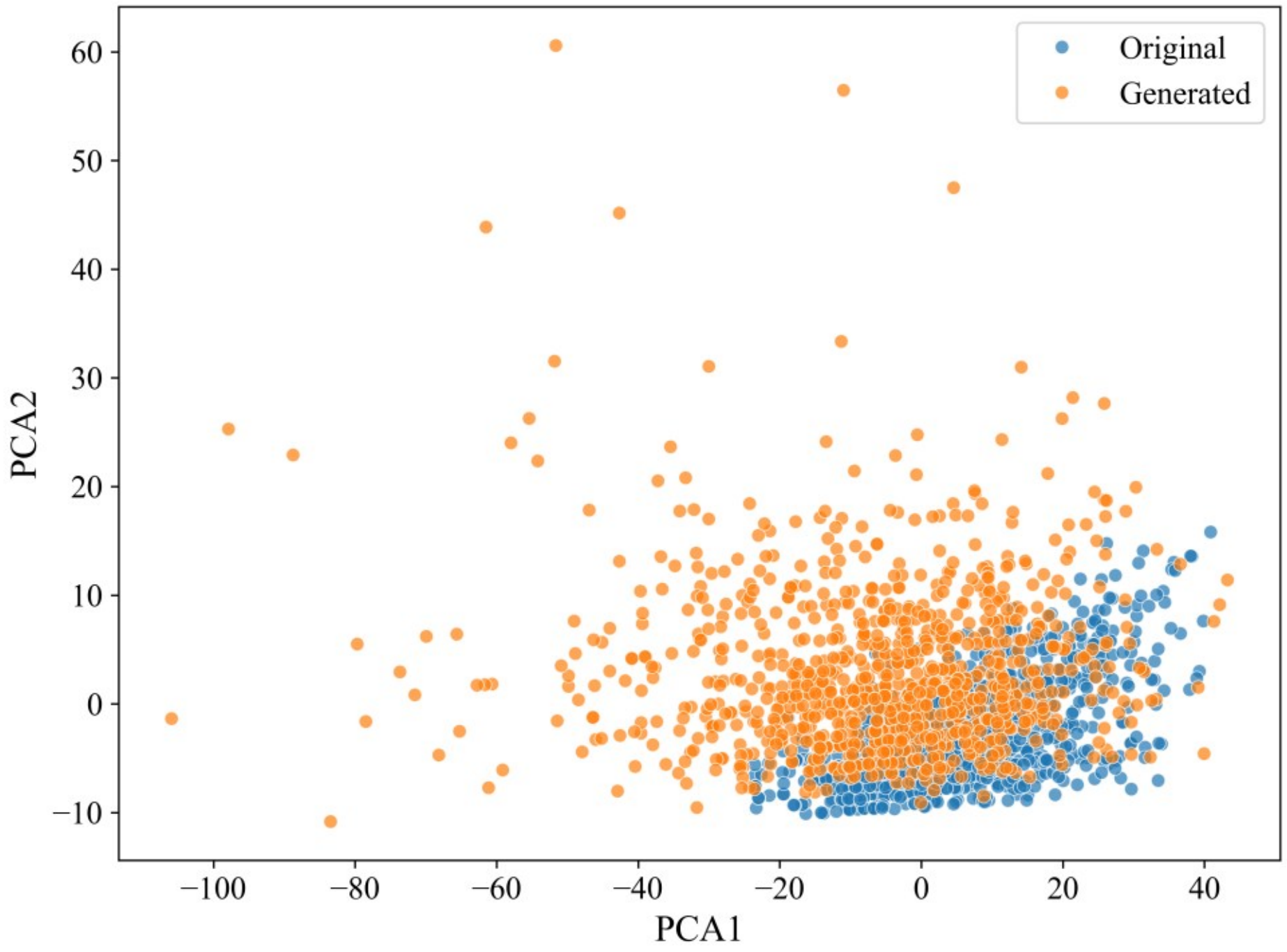

**Fig. 10** PCA projection of topological features for 1,000 generated and 1,000 selected original structures

To minimize potential selection bias in the original dataset, a more comprehensive comparison was performed: the 1,000 generated structures were retained, and their metrics were compared against all 20,000 original CT-scanned structures. The updated results are reported in Table 4, with the corresponding PCA visualization shown in Fig. 11. The findings confirm that the generated structures surpass the original dataset in both feature diversity and dispersion. In PCA space, the original structures form a dense cluster, whereas the generated structures are more widely distributed, consistent with the goal of producing high-diversity porous structures.

***Table 4*** *Comparison of topological metrics between 1,000 generated and 20,000 CT-scanned structures*

| Items | Images type | Values |
|---|---|---|
| Mean Pairwise Distance (image diversity) | Original structures | 16.2960 |
| | Generated structures | 25.1092 |
| Information Entropy | Original structures | 2.0319 |
| | Generated structures | 2.1011 |

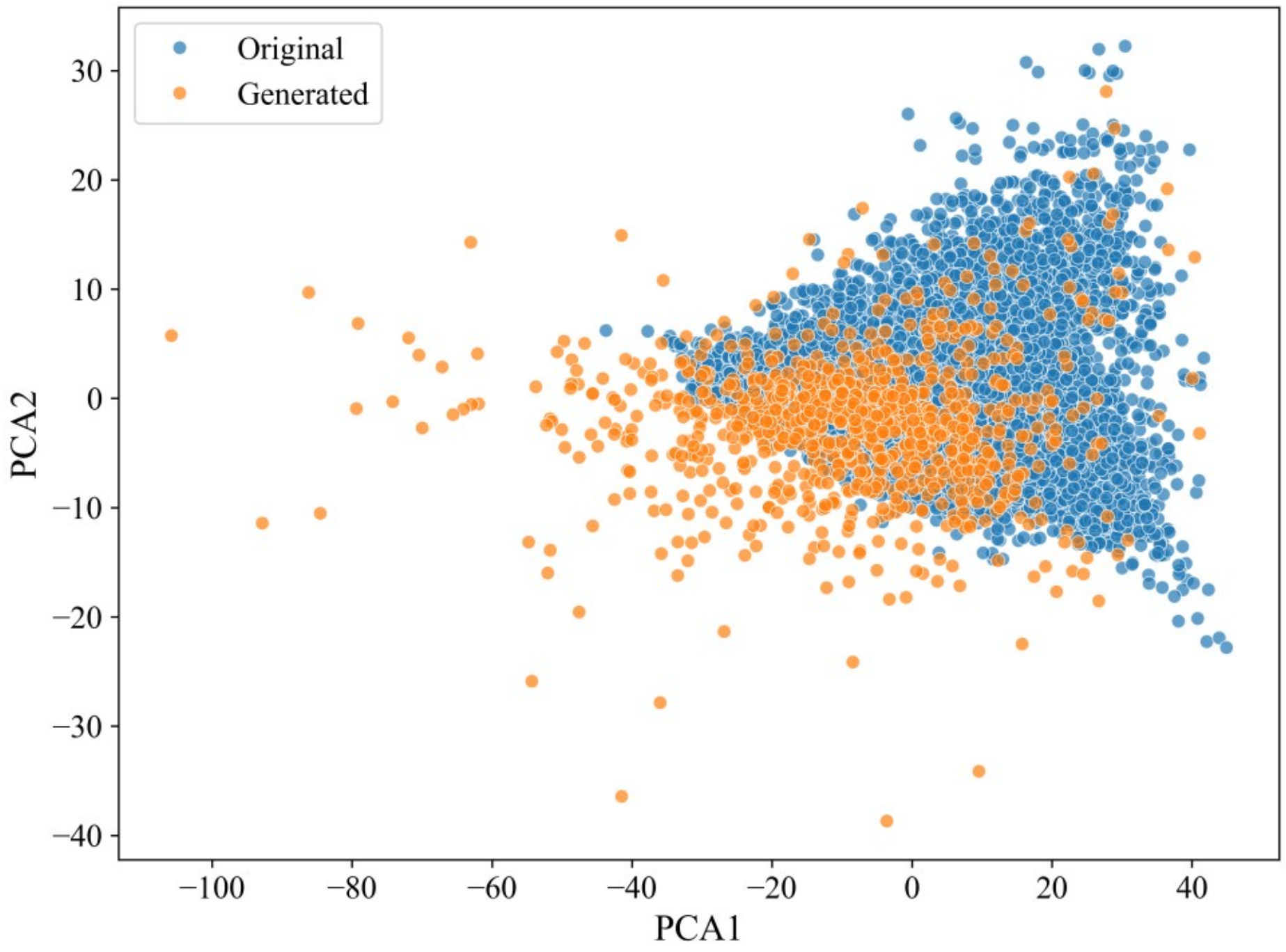

***Fig. 11*** *PCA projection of topological features for 1,000 generated and 20,000 CT-scanned structures*

These results demonstrate that the wGAN-GP effectively generates porous structures with greater topological diversity than the original input structures. This capability arises from the generative nature of GANs, which can interpolate and combine features from the training data to produce novel designs. In the context of oscillating flow and heat transfer in porous structures, such diversity enables broader parametric investigations and supports more generalizable conclusions.

The final set of 1,000 generated images was selected for subsequent LBM simulations and machine learning analyses. Their statistical characteristics with respect to topological parameters are summarized in Table 5.

***Table 5*** *Statistical representation of topological parameters for 1,000 wGAN-GP generated porous structures*

| | Porosity | Connec-tivity | Percolation Strength | Tortuosity | SSA | Mean Pore Diamet-er | Pore Size Dispers-ion | Euler Number | NCI |
|---|---|---|---|---|---|---|---|---|---|
| Min. | 0.4933 | 1 | 0.9654 | 1.177 | 0.0515 | 5.511 | 3.157 | -152 | 0.5120 |
| Median | 0.6092 | 12 | 0.9979 | 1.358 | 0.1185 | 9.341 | 6.295 | -50 | 0.5269 |
| Mean | 0.6161 | 13.51 | 0.9970 | 1.351 | 0.1193 | 10.515 | 7.639 | -51.37 | 0.5282 |
| Max. | 0.8921 | 74 | 1.0000 | 1.407 | 0.1723 | 42.168 | 40.542 | -3 | 0.5668 |

## *3.2 Results of LBM simulations*

### *3.2.1 Lattice resolution independence study*

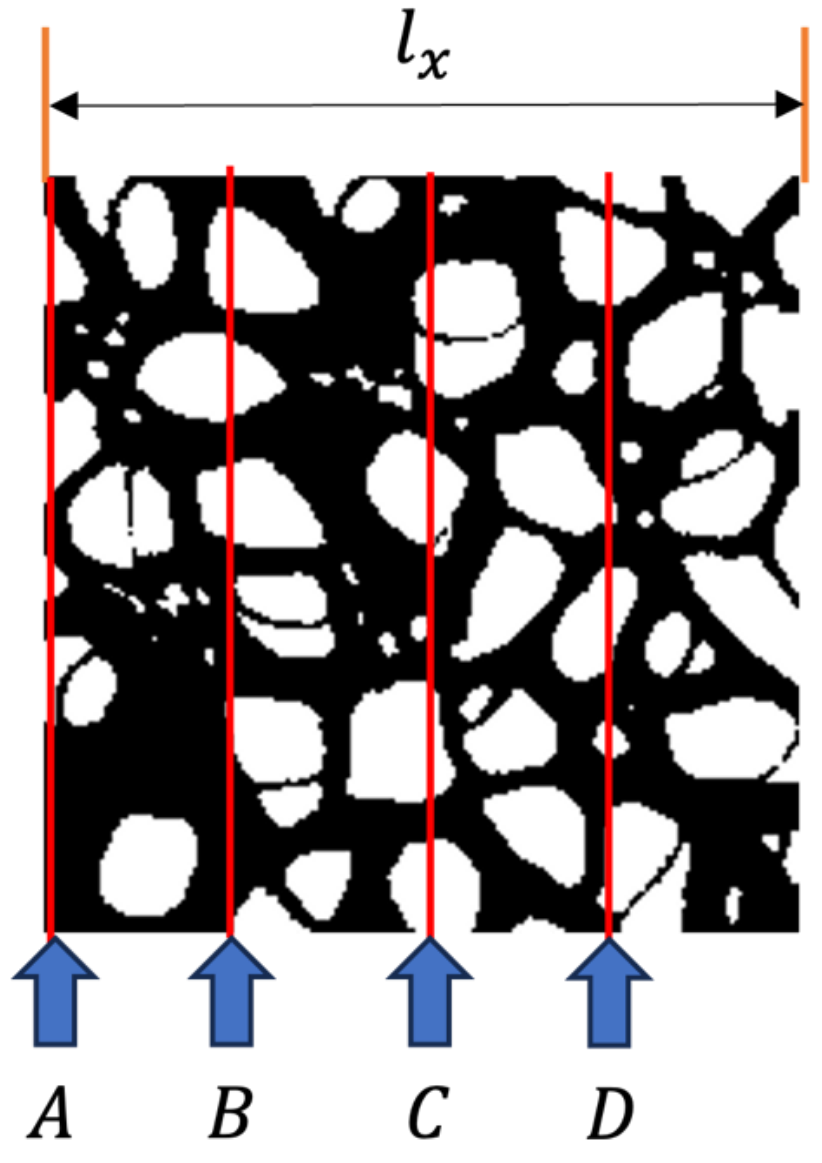

***Fig. 12*** *Locations for the resolution independent study: A is at inlet, B is at* $\frac{1}{4}l_x$, *C is at* $\frac{1}{2}l_x$, *and D is at* $\frac{3}{4}l_x$.

A lattice resolution independent study was conducted to ensure the reliability of the LBM simulations. A porous sample, with a porosity of 58.85%, was selected from the dataset. The physical dimensions were $l_x$ = $l_y$ = 0.2 in = 5.08 mm. The average temperature difference ($\Delta T$) and the average velocity difference ($\Delta U$) were computed at four locations along the flow direction: the inlet (or $0l_x$), $\frac{1}{4}l_x$, $\frac{1}{2}l_x$, and $\frac{3}{4}l_x$, as illustrated in Fig. 12. These two variables were calculated under different resolutions $N_i$, defined in Eqs. (19-21).

$$N_i = 100,\ 300,\ 500,\ 700,\ \ldots,\ 2700,\ 2900, 3100 \tag{19}$$

$$\Delta T = \bar{T}_{N_{i+1}} - \bar{T}_{N_i} \tag{20}$$

$$\Delta U = \bar{U}_{N_{i+1}} - \bar{U}_{N_i} \tag{21}$$

The simulation setup followed the conditions described in Section 2.2.2, except that the inlet velocity was fixed at $U_{in} = 2mm/s$, corresponding to a Reynolds number of *Re* = 19.17 and a Strouhal number of *St* = 0. A constant velocity inlet was adopted to simplify the analysis relative to oscillatory conditions.

The results of the lattice resolution independent study are presented in Fig. 13. Once the resolution exceeded 1900, variations in $\Delta T$ and $\Delta U$ became negligible, with $\Delta T < 0.02\,K$, $\Delta U < 0.05mm/s$. Accordingly, a resolution of 2200 was selected to ensure accuracy while maintaining computational efficiency.

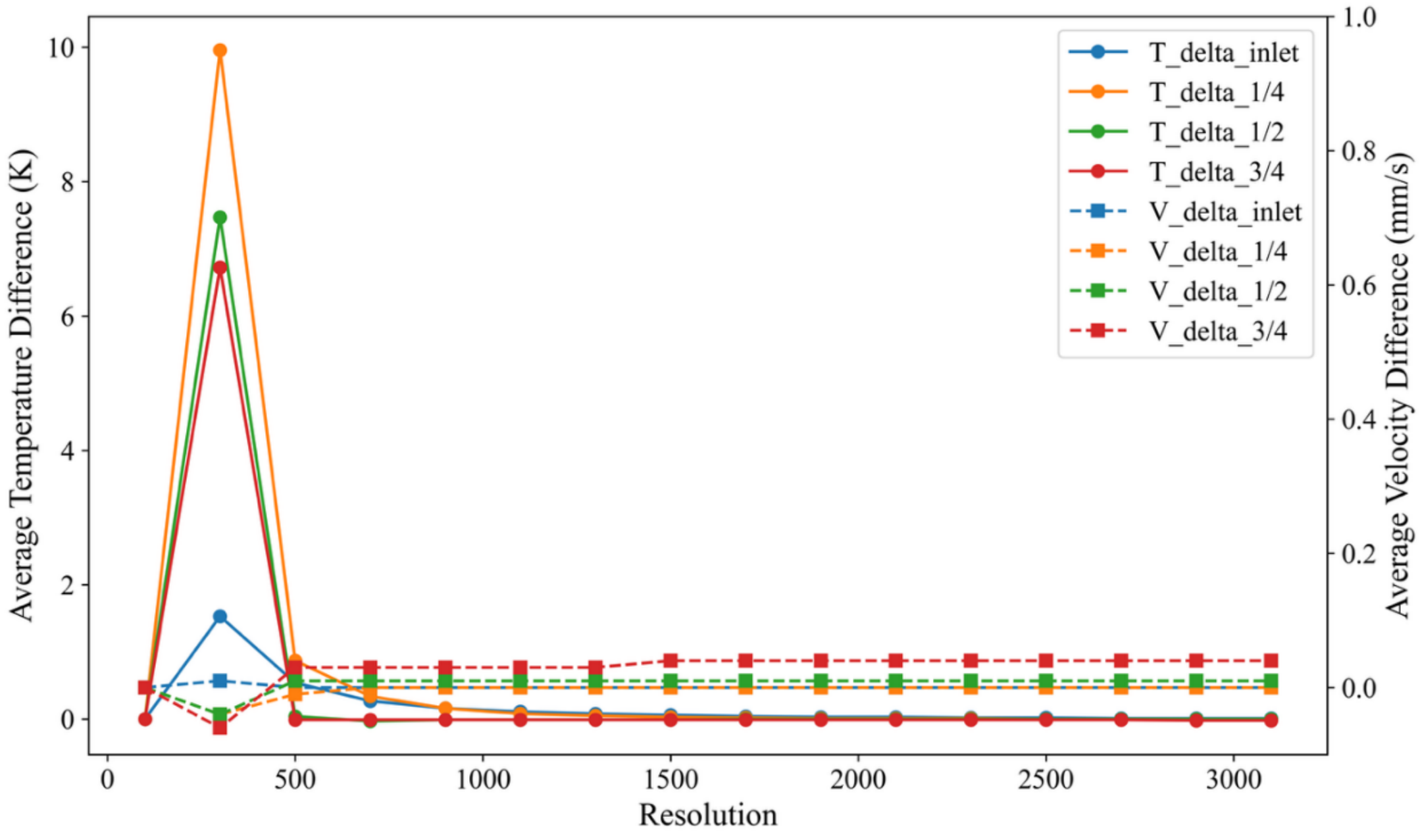

***Fig. 13*** *Lattice resolution independent study*

### <u>3.2.2 $\overline{Nu}$ data generation using LBM</u>

LBM simulations were conducted in batch mode using the 1,000 porous structures generated by the wGAN-GP model. Following the 56 combinations of $Re$ and $St$ numbers defined in Eq. (13), 56 groups of simulations were performed, each comprising 1,000 distinct porous domains (statistical characteristics summarized in Table 5). This results in a total of 56,000 simulations and corresponding average Nusselt number values. All simulations were executed with *Palabos* on an Intel i7-12700 CPU with 32 GB of RAM, requiring approximately one month to complete. For each oscillatory flow simulation, $\overline{Nu}$ denotes the cycle-averaged Nusselt number over one oscillation period.

Fig. 14 and 15 illustrate the distribution of the averaged $\overline{Nu}$ as a function of $Re$ and $St$. Fig. 14 presents a 3D scatter plot, while Fig. 15 shows the corresponding 2D colormap scatter plot highlighting the relationships among *Re*, *St*, and $\overline{Nu}$. Each vertical column in Fig. 14 corresponds to 1,000 $\overline{Nu}$ values obtained from the wGAN-GP-generated porous structures. These values are then projected onto a 2D plane to produce the colormap in Fig. 15.

Figs. 14 and 15 reveal that $\overline{Nu}$ is primarily governed by the *Re*, consistent with physical intuition: higher inlet velocity $U_{in}$ enhances convective heat transfer. Although $\overline{Nu}$ also increases with *St*, its effect is secondary. Notably, at *St* = 1, $\overline{Nu}$ values are significantly lower because *St* is two orders of magnitude smaller than in other cases, leading to a reduced inlet velocity and weaker convective strength. A detailed analysis of the effects of *Re*, *St*, and the topological parameters on $\overline{Nu}$ predictions is presented in Sec. 4.

In addition to the $\overline{Nu}$ distribution, supplementary LBM simulation results, such as the temperature and velocity fields of representative porous structures, are provided in Appendix B. The topological characteristics of these representative structures are further discussed in Sec. 4.3.

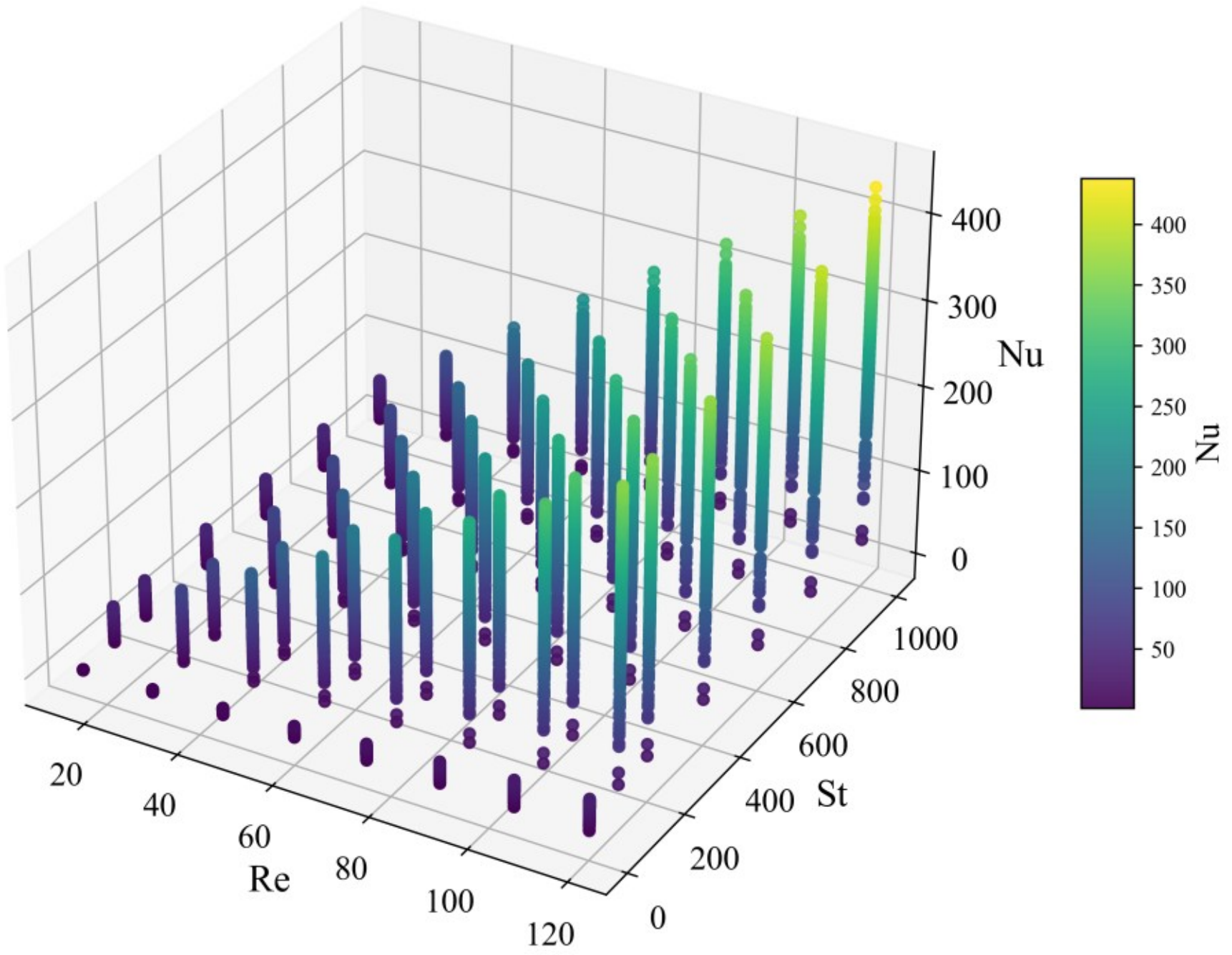

***Fig. 14*** *3D scatter plot of Re, St, and generated $\overline{Nu}$ data*

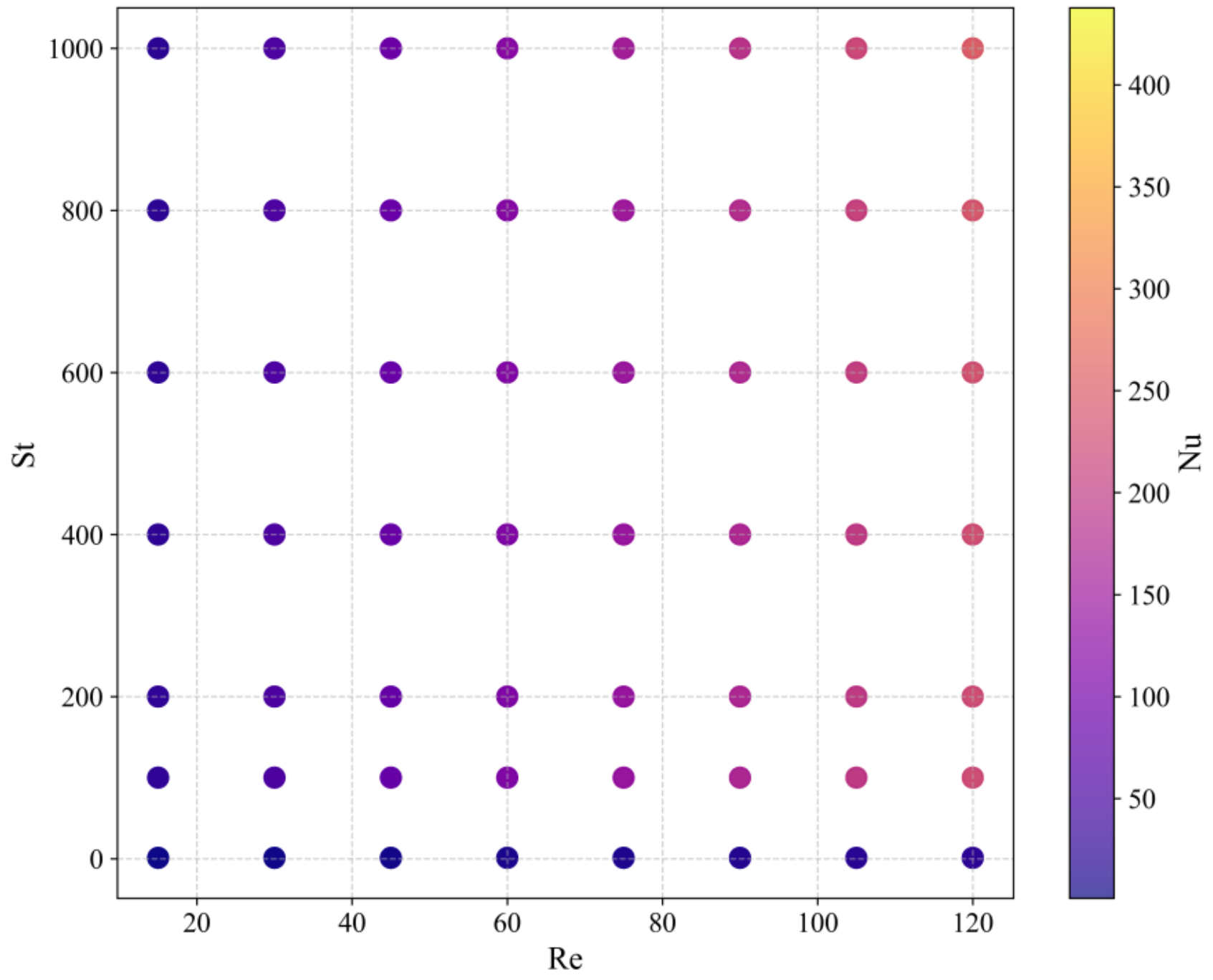

***Fig. 15*** *Color mapped scatter plot of Re, St, and generated $\overline{Nu}$ data*

*3.3 ML model comparisons*

The performance of 10 machine learning models was evaluated using Nested_CV. Both the outer and inner folds were set to 5, and a subsampling factor of 0.25 was applied to reduce computational cost. Hyperparameter tuning was performed using Bayesian optimization (BO), with five initial exploration points followed by 20 iterations. The computation was executed on an Intel i7-13700F CPU, requiring about 3.5 hours to complete.

Table 6 summarizes the model performance, while Fig. 16 presents the mean RMSE values with standard deviations error bars. The Ridge regression models (PolyRidge with D = 2, 3) correspond to a second- and third-order polynomial regressions with Ridge regularization. Among all models, tree-based models, such as XGBoost and Random Forest, significantly outperformed the others. XGBoost achieved the best overall performance, with an average $R^2 = 0.9853$ and the lowest average RMSE and MAE values. The ANN model also demonstrated strong predictive ability ($R^2 > 0.90$), suggesting that advanced neural network architectures may yield further improvements. In contrast, linear- and SVR-based models consistently underperformed, with average $R^2$ values below 0.90. Overall, the Nested_CV results indicate that XGBoost offers the most accurate predictive performance among all evaluated models. Further implementation and analysis of XGBoost are presented in the following section.

***Table 6** Performance comparison of machine learning models based on RMSE, MAE, and $R^2$*

| Models | Mean_RMSE | Std_RMSE | Mean_MAE | Mean_$R^2$ |
|---|---|---|---|---|
| XGBoost | 10.4063 | 0.6246 | 6.3686 | 0.9853 |
| Random Forest | 18.3266 | 0.6016 | 11.3231 | 0.9545 |
| ANN (MLP) | 24.5367 | 0.3645 | 16.6713 | 0.9185 |
| CART | 27.5974 | 0.4916 | 17.5683 | 0.8969 |
| PolyRidge (D =3) | 33.9604 | 0.2225 | 25.8702 | 0.8440 |
| Radial SVR | 41.9355 | 0.4184 | 28.8344 | 0.7621 |
| PolyRidge (D =2) | 42.5107 | 0.2221 | 32.2726 | 0.7555 |
| k-NN | 45.9306 | 0.5524 | 32.7630 | 0.7146 |
| ElasticNet | 50.1074 | 0.4121 | 37.4067 | 0.6603 |
| Linear SVR | 53.4388 | 0.3868 | 35.5206 | 0.6136 |
| Sigmoid SVR | 54.4512 | 0.3371 | 39.0374 | 0.5989 |

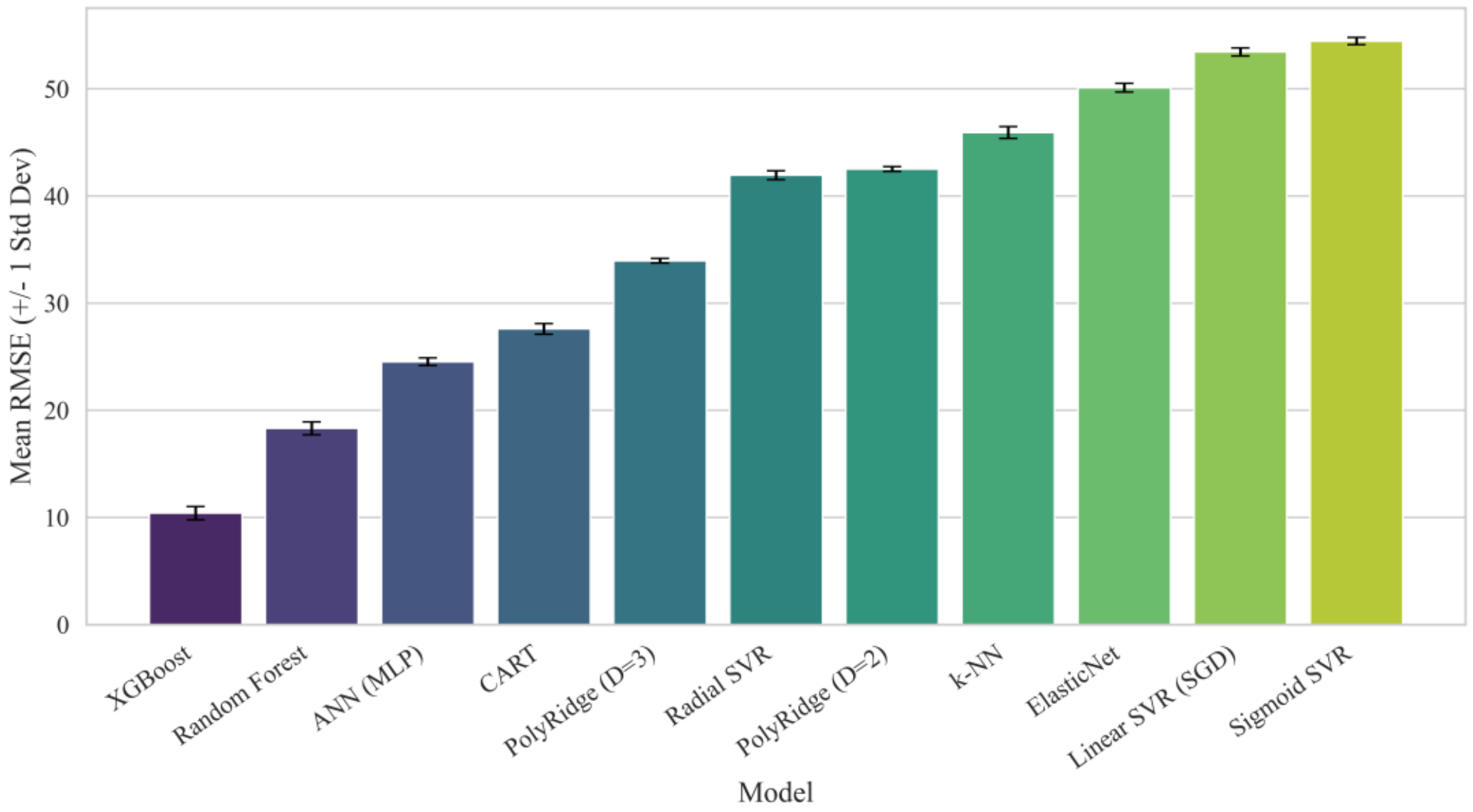

***Fig. 16*** *Comparison of model performance using mean RMSE with standard deviations shown as error bars*

*3.4 Final implementation of XGBoost*

Based on the Nested_CV comparisons, XGBoost was identified as the best-performing model. Since subsampling was applied during Nested_CV to reduce computational cost, the final training was performed on the full dataset without subsampling to obtain a more robust performance evaluation. The complete dataset of 56,000 samples was split into a training set (80%) and a hold-out testing set (20%). BO was then applied to the training set to determine the optimal hyperparameters, following the same iteration scheme as in Nested_CV. The tuned hyperparameters are summarized in Table 7.

***Table*** *7 Optimal hyperparameters for the XGBoost model*

| Hyperparameter | n_estimators | max_depth | learning_rate | gamma | subsample |
|---|---|---|---|---|---|
| Values | 208 | 8 | 0.22009204420958423 | 0.6110235106775829 | 0.8612216912851107 |

The XGBoost model was trained with the optimal hyperparameters on the full training set and subsequently evaluated on the hold-out test set. The results, summarized in Table 8, demonstrate excellent predictive accuracy, with a coefficient of determination of $R^2 = 0.9981$.

***Table 8*** *Performance of the trained XGBoost model on the hold-out test set*

| | |
|---|---|
| RMSE | 3.7347 |
| MAE | 2.4703 |
| $R^2$ | 0.9981 |

Fig. 17 presents a scatter plot comparing predicted and actual values for the test data. The close alignment of data points along the $y = x$ line demonstrates strong agreement between predictions and observations. Fig. 18 shows the residual plot, which exhibits no funnel-shaped pattern, supporting the assumption of homoscedasticity (constant variance) and thereby increasing confidence in the regression results.

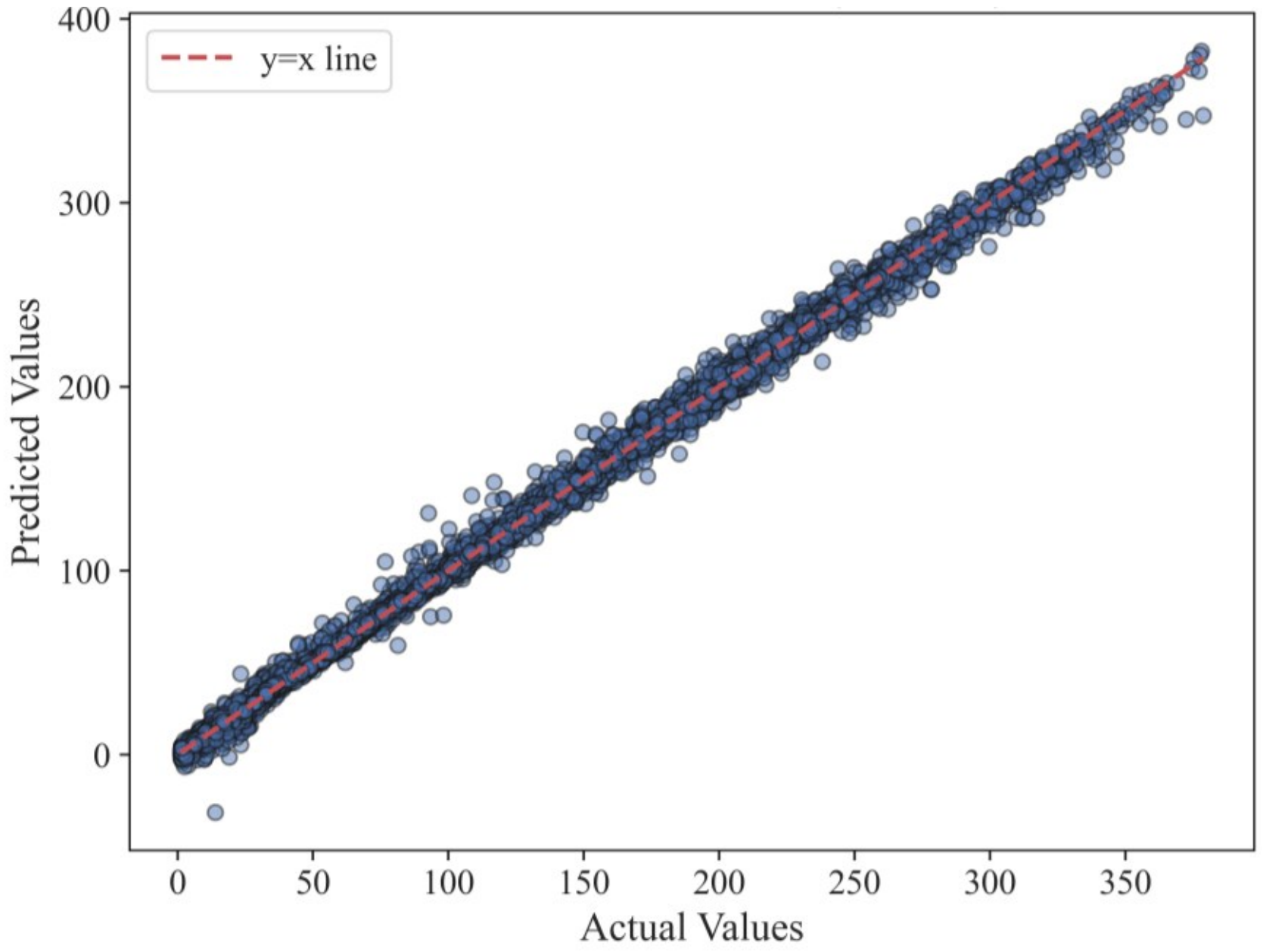

***Fig. 17*** *Scatter plot of predicted versus actual values for the XGBoost model*

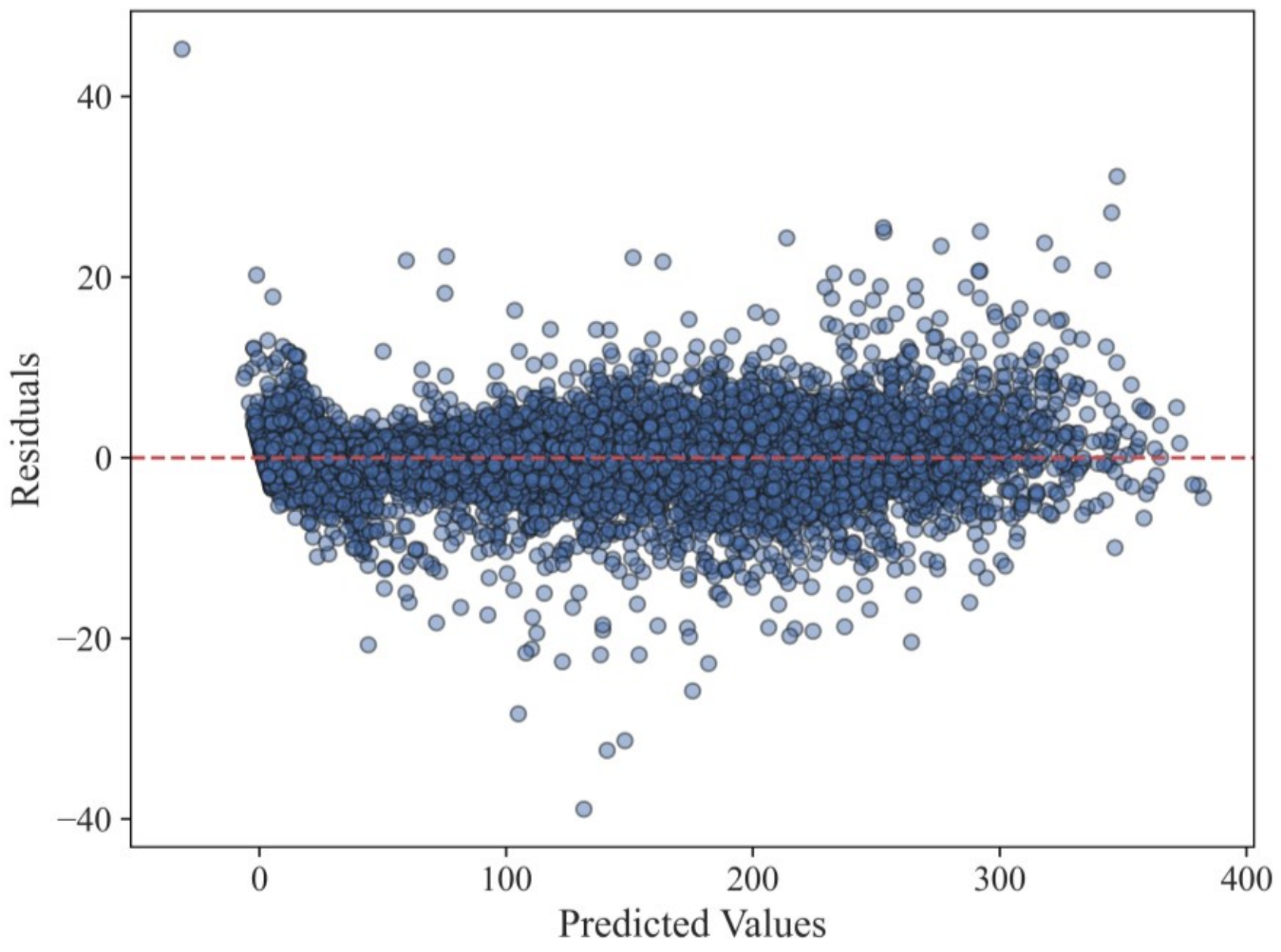

***Fig. 18*** *Residual plot for the XGBoost predictions*

Fig. 19 illustrates the histogram of prediction errors, which is approximately symmetric and centered at zero. Most errors fall within ±5 units, indicating low bias and supporting the validity of RMSE and MAE as performance metrics. The central peak at zero further confirms the model's unbiasedness. Fig. 20 provides the Q-Q plot of prediction errors. The central portion of the curve closely aligns with the theoretical normal distribution line, while deviations at the lower-left and upper-right tails indicate heavy-tailed behavior, which means extreme errors occur more frequently than expected under a standard normal distribution. This pattern is commonly observed in predictive ML models and is generally considered acceptable, as capturing the central tendency of the residuals is more critical for predictive accuracy than perfectly matching the tails of the normal distribution.

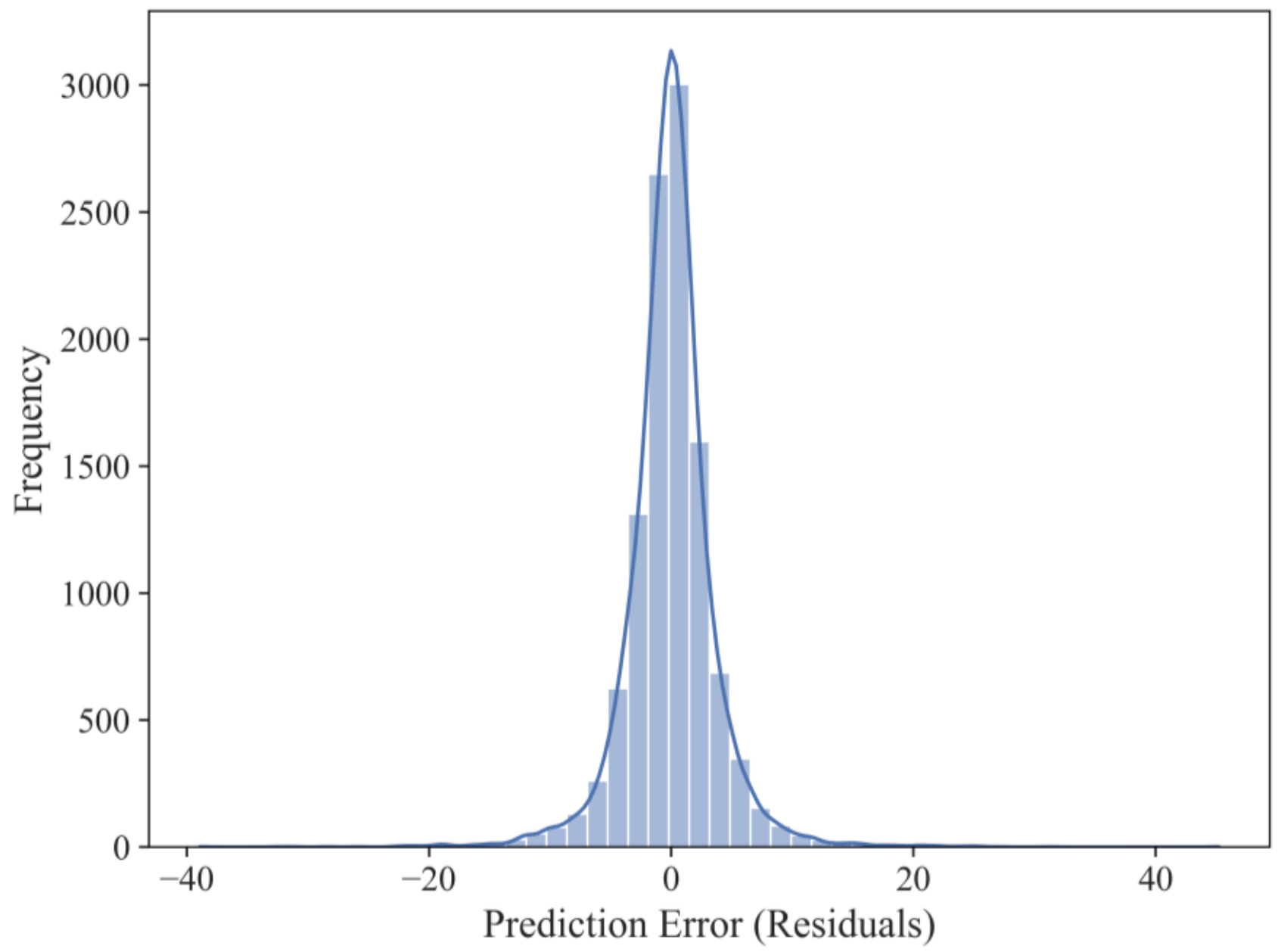

***Fig. 19*** *Histogram of prediction errors for the XGBoost*

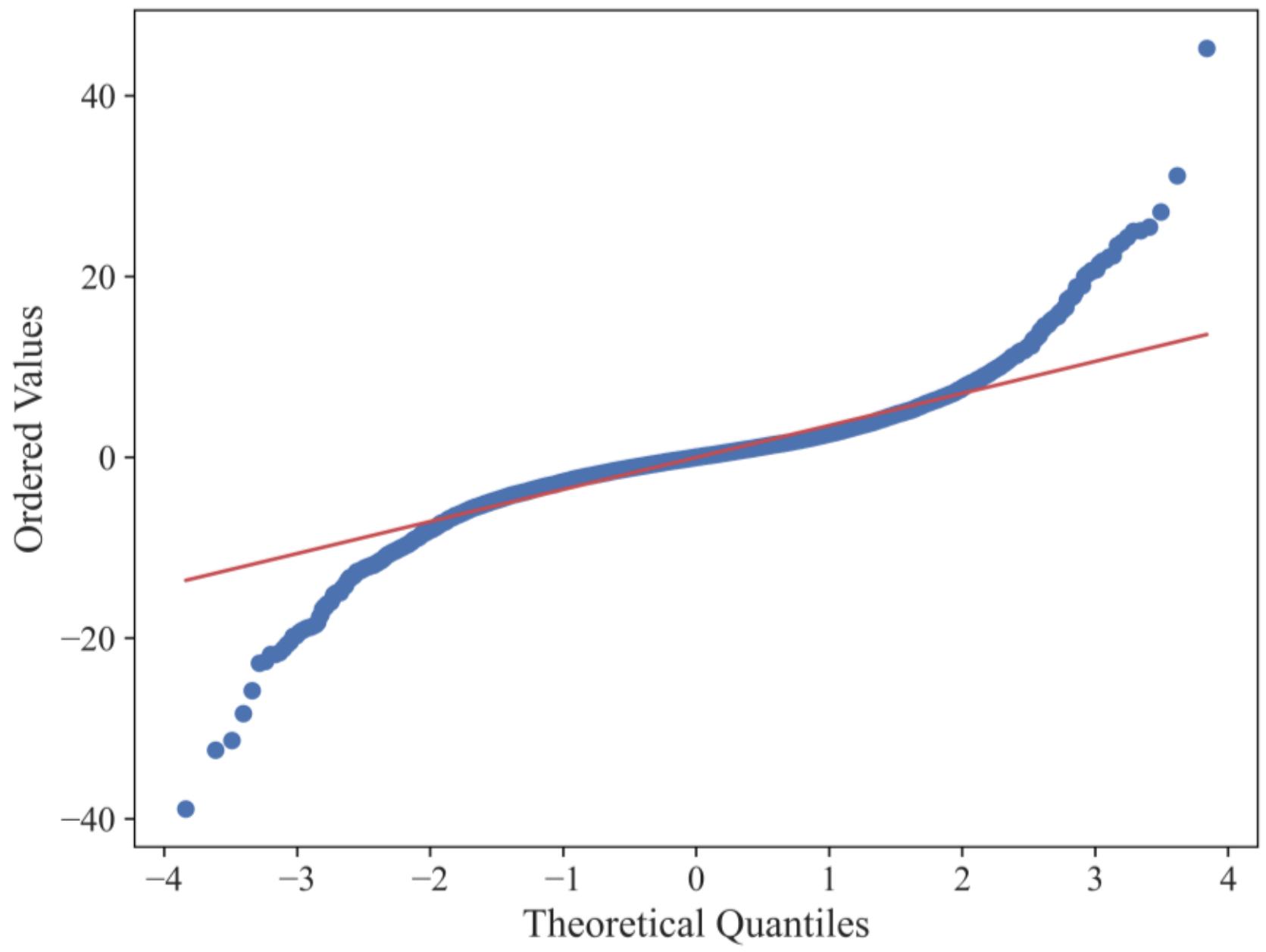

***Fig. 20*** *Q–Q plot of residuals for the XGBoost predictions*

Overall, the results confirm that the XGBoost model achieves high predictive accuracy with small, unbiased errors in most cases. The presence of a limited number of outliers ─ likely associated with extreme topological parameter values ─ does not significantly influence the model's overall reliability.

## 4 XGBoost prediction and interpretations

### *4.1 The implementation of SHAP analysis*

Understanding the underlying factors driving the predictions made by XGBoost is a key objective of this study. To this end, SHapley Additive exPlanations (SHAP) analysis was employed to interpret and explain the outputs of the XGBoost. Prior studies have successfully applied SHAP to extract physical insights from ML predictions in engineering contexts [59–61]. SHAP provides a unified framework for interpreting complex model predictions by assigning each feature a Shapley value–based importance score for individual predictions [24].

Fig. 21 presents the SHAP summary plot, which illustrates both the relative importance of features and their directional impact on model's prediction outcomes. Features are ranked from top to bottom by their mean absolute SHAP values, reflecting their relative influence in the predictive model. The *x*-axis represents SHAP values, quantifying the positive or negative contribution of each feature to the model output for a given sample. The *y*-axis lists the feature names, and the color encodes feature magnitude (red = high, blue = low).

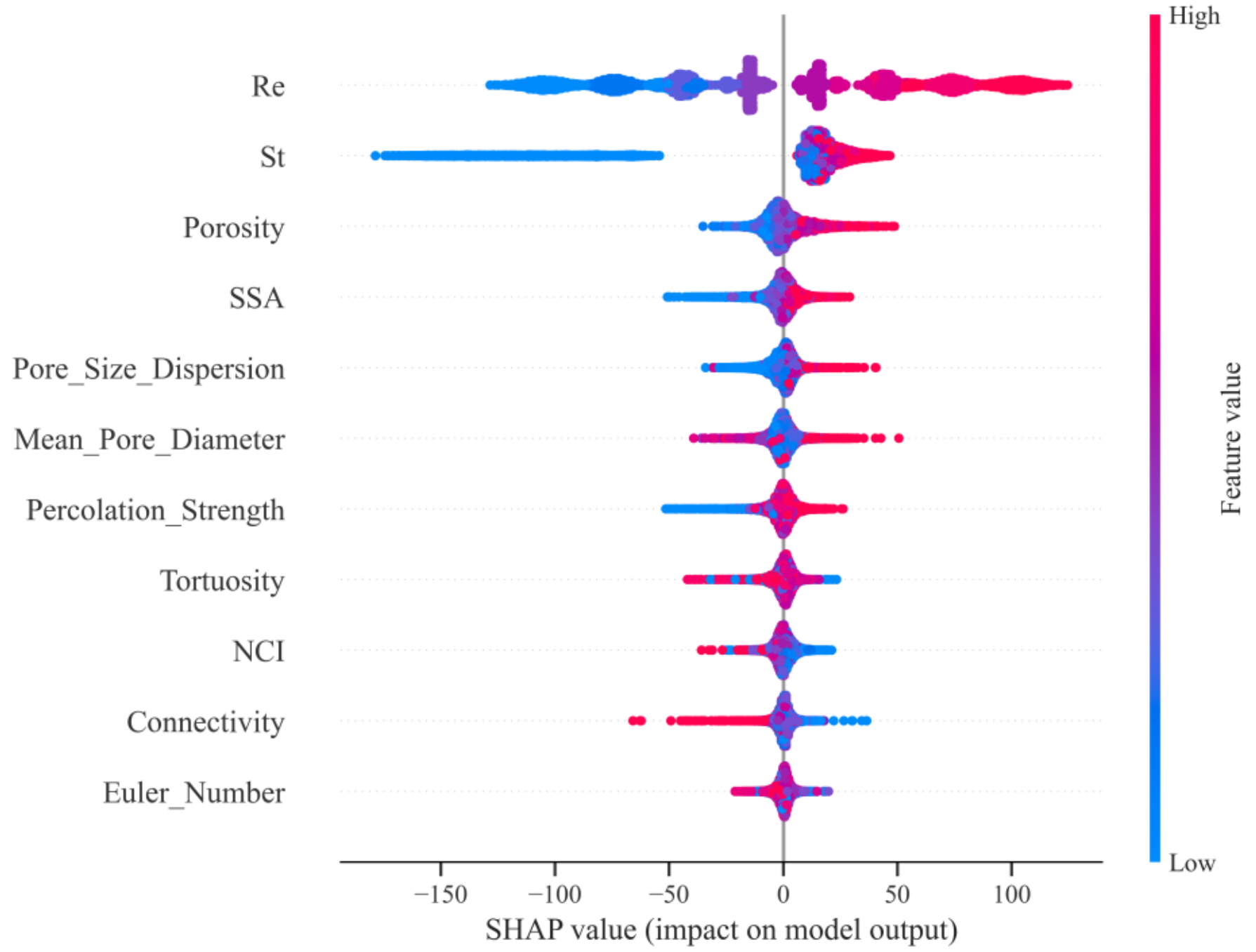

***Fig. 21*** *SHAP summary plot of the XGBoost model. Features are ranked from top to bottom by their mean absolute SHAP values, indicating their relative importance, while the horizontal axis represents their positive or negative contribution to the prediction.*

The SHAP results indicate that the Reynolds number ($Re$) and Strouhal number ($St$) are the most influential predictors. High-$Re$ samples (red) predominantly appear in the positive SHAP region, confirming that under oscillatory flow conditions, higher flow velocities enhance convective heat transfer and increase the predicted thermal performance $\overline{Nu}$. In contrast, low-$Re$ cases (blue) are associated with negative SHAP values, suggesting reduced flow strength and weaker heat transfer. For $St$, low values correspond strongly to negative SHAP values, suggesting that low oscillation frequencies limit fluid–solid thermal interaction and thus diminish heat transfer. Higher $St$ values generally contribute positively, though their effect depends on interaction with $Re$.

Among all the topological parameters, porosity, specific surface area (SSA), and pore size dispersion cluster in the positive SHAP region, indicating that a larger connected channel areas and greater effective heat exchange surfaces can strongly enhance heat transfer. Percolation strength follows a similar pattern, with high values contributing positively and low values negatively. mean pore diameter exhibits both positive and negative SHAP values at high magnitudes, implying that larger pores can either facilitate or hinder heat transfer depending on other conditions, while low values cluster near zero, indicating minimum influence.

Connectivity-related parameters, including tortuosity, network connectivity index (NCI), and connectivity, tend to show low values in the positive SHAP region and high values in the negative region, suggesting that reduced flow resistance and more direct pathways enhance heat transfer. The Euler number, by contrast, shows a balanced red–blue distribution across positive and negative SHAP regions and contributes the lowest overall, suggesting a weak direct influence on thermal performance. Therefore, the Euler number could potentially be excluded in future development of predictive models or correlation formulations.

*4.2 Analysis of SHAP features' dependence*

While Fig. 21 provides an overview of feature importance and directional influence, a quantitative comparison of feature thresholds can be obtained through SHAP feature dependence analysis. Thresholds (defined as the feature values corresponding to a SHAP value of 0) were calculated for all features and are listed in Table 9. For example, $Re$ is the most influential predictor, with a threshold at $Re$ = 75. This indicates that $Re$ values greater than 75 positively contribute to $\overline{Nu}$, thereby enhancing heat transfer performance. Similarly, the threshold for porosity is 0.6256, meaning porosity above this value tends to promote heat transfer, while lower values may hinder it.

***Table 9*** *Feature thresholds derived from SHAP dependence analysis*

| Parameters | Threshold |
|---|---|
| *Re* | 75 |
| *St* | 100 |
| Porosity | 0.6256 |
| Specific Surface Area (SSA) | 0.1175 |
| Pore Size Dispersion | 5.5705 |
| Mean Pore Diameter | 8.0283 |
| Percolation Strength | 0.9976 |
| Tortuosity | 1.3201 |
| NCI | 0.5187 |
| Connectivity | 1 |
| Euler Number | -152 |

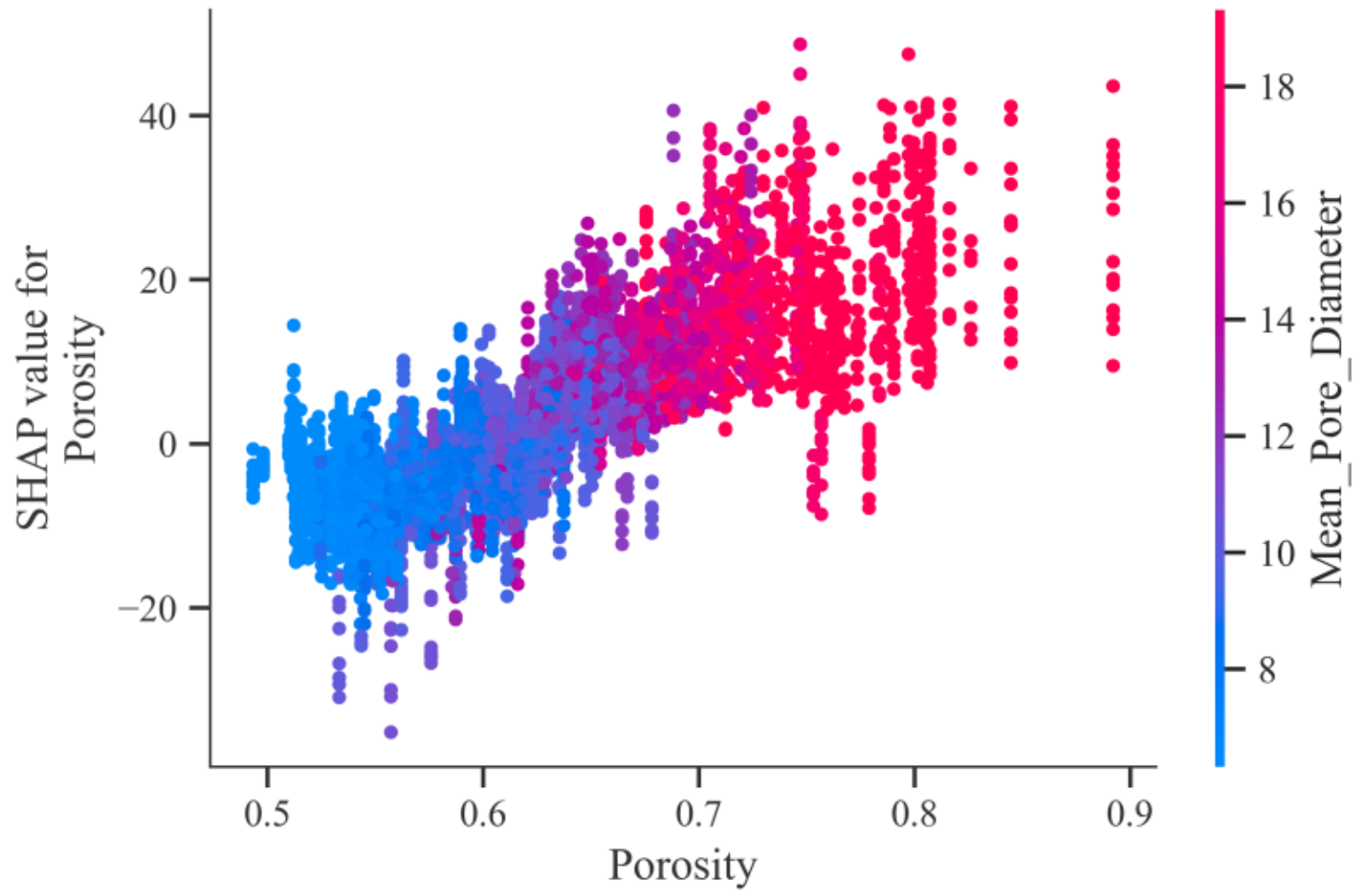

***Fig. 22*** *SHAP dependence plot for Porosity*

To investigate feature interactions, SHAP dependence plots were generated. Unlike summary plots, which provide an overall view of feature importance, dependence plots reveal the relationship between a feature's value and its corresponding SHAP value. The color scale encodes the value of the most interactive feature, thereby highlighting potential interaction effects. Among the porous structure parameters, porosity is the most influential factor, as shown in Fig. 21, with mean pore diameter identified as its strongest interacting feature. Fig. 22 shows that when porosity exceeds its threshold (0.6256 from Table 9), its SHAP value becomes predominantly positive, indicating a favorable impact on $\overline{Nu}$. This effect is further amplified when the mean pore diameter exceeds its threshold of 8.0283, suggesting that the combination of high porosity and large pore diameter could enhance heat transfer.

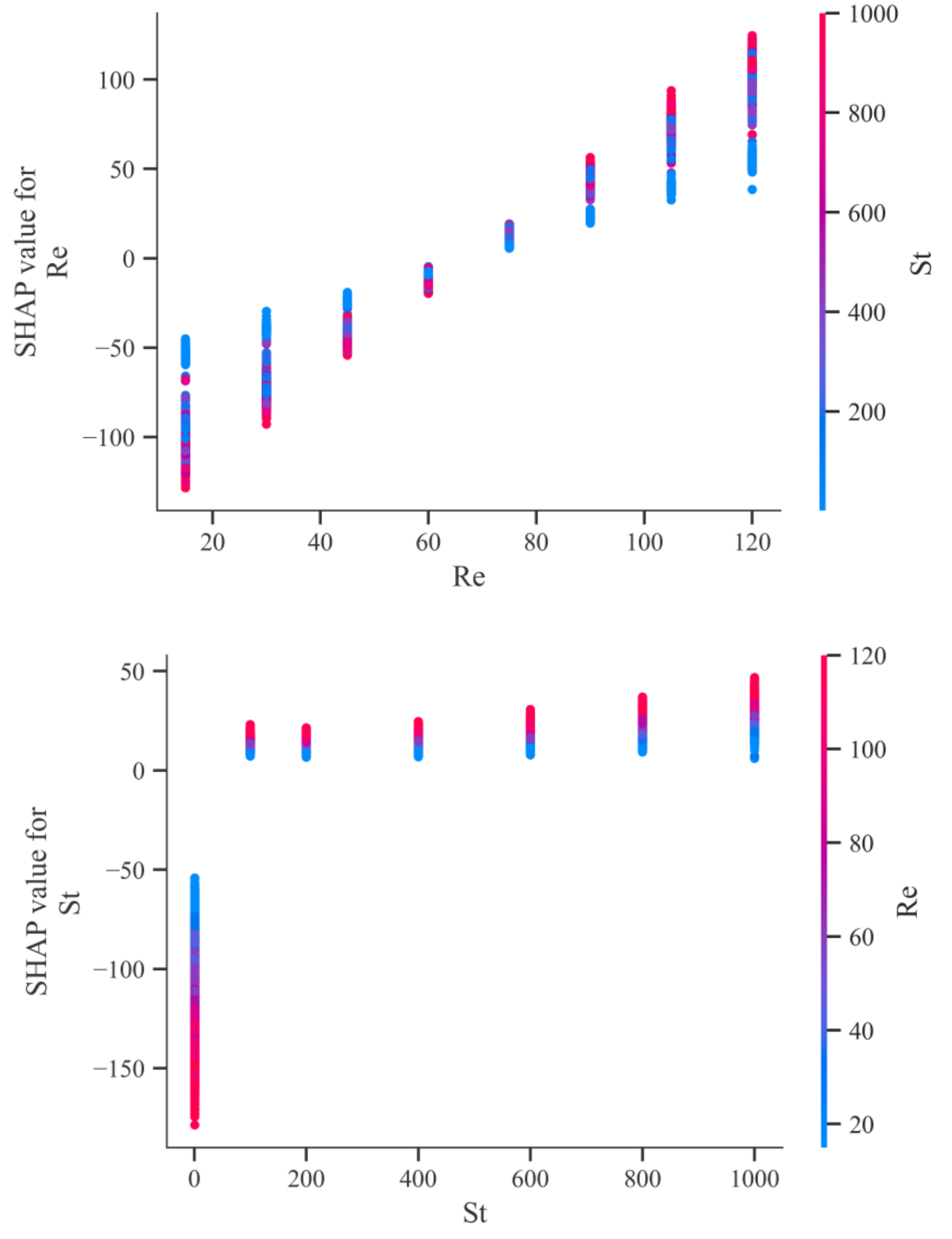

***Fig. 23** SHAP dependence plots for oscillatory flow parameters (Re and St)*

Fig. 23 presents the SHAP dependence plots for $Re$ and $St$. These two parameters show a strong correlation with heat transfer performance. When $Re$ >75, SHAP values become strongly positive, indicating a marked increase in $\overline{Nu}$. Similarly, $St$ values above approximately 100 contribute positively, while lower values lead to a pronounced reduction. The color bar patterns reveal that high $Re$ combined with moderate-to-high $St$ (>100) maximizes convective heat transfer.

Figs. C.1–C.8 in Appendix C illustrate the SHAP dependence plots for the remaining topological features with their most relevant synergistic patterns. Although these features are not among the most dominant predictors, the thresholds summarized in Table 9 highlight several notable interaction patterns that still influence heat transfer performance. For example, higher SSA, particularly above 0.1175, exerts a strong positive influence when pore size is small. Pore size dispersion values above 5.5705 enhance heat transfer under high Reynolds number conditions; however, when pore size dispersion is low, a high $Re$ may instead hinder heat transfer performance. A mean pore diameter above 8.0283 contributes positively under high-porosity conditions. High percolation strength (near 1.00), coupled with low tortuosity (<1.3201), reduces flow resistance and promotes heat transfer. Lower NCI values (around 0.5187) generally benefit $\overline{Nu}$, while excessive connectivity may hinder performance unless accompanied by strong percolation. The Euler Number shows minimal influence and no distinct interaction patterns.

Based on the SHAP interpretation of the XGBoost model, optimal heat transfer performance in porous media under oscillating flow requires coupling favorable flow conditions with optimized structural characteristics. For flow parameters, $Re$ > 75 and $St$ > 100 are necessary to ensure strong convective heat transfer, with their interaction significantly boosting $\overline{Nu}$. For structural properties, porosity should exceed 0.6256 and be paired with a mean pore diameter greater than 8.0283 to balance permeability and effective heat exchange area. SSA should be at least 0.1175, and pore size dispersion above 5.57 further promotes heat transfer under high $Re$. Percolation strength close to 1.00 combined with tortuosity below 1.32 minimizes flow resistance and enhances the transport. NCI values should remain below 0.5187 to avoid excessive connectivity that can cause short-circuiting flow.

These findings highlight the coupled relationship between velocity, frequency, and optimized pore structures, offering and elucidating quantitative guidance and analysis for designing porous media with enhanced thermal performance under oscillating flow conditions from a data-driven perspective.

*4.3 Implementation of SHAP conclusions*

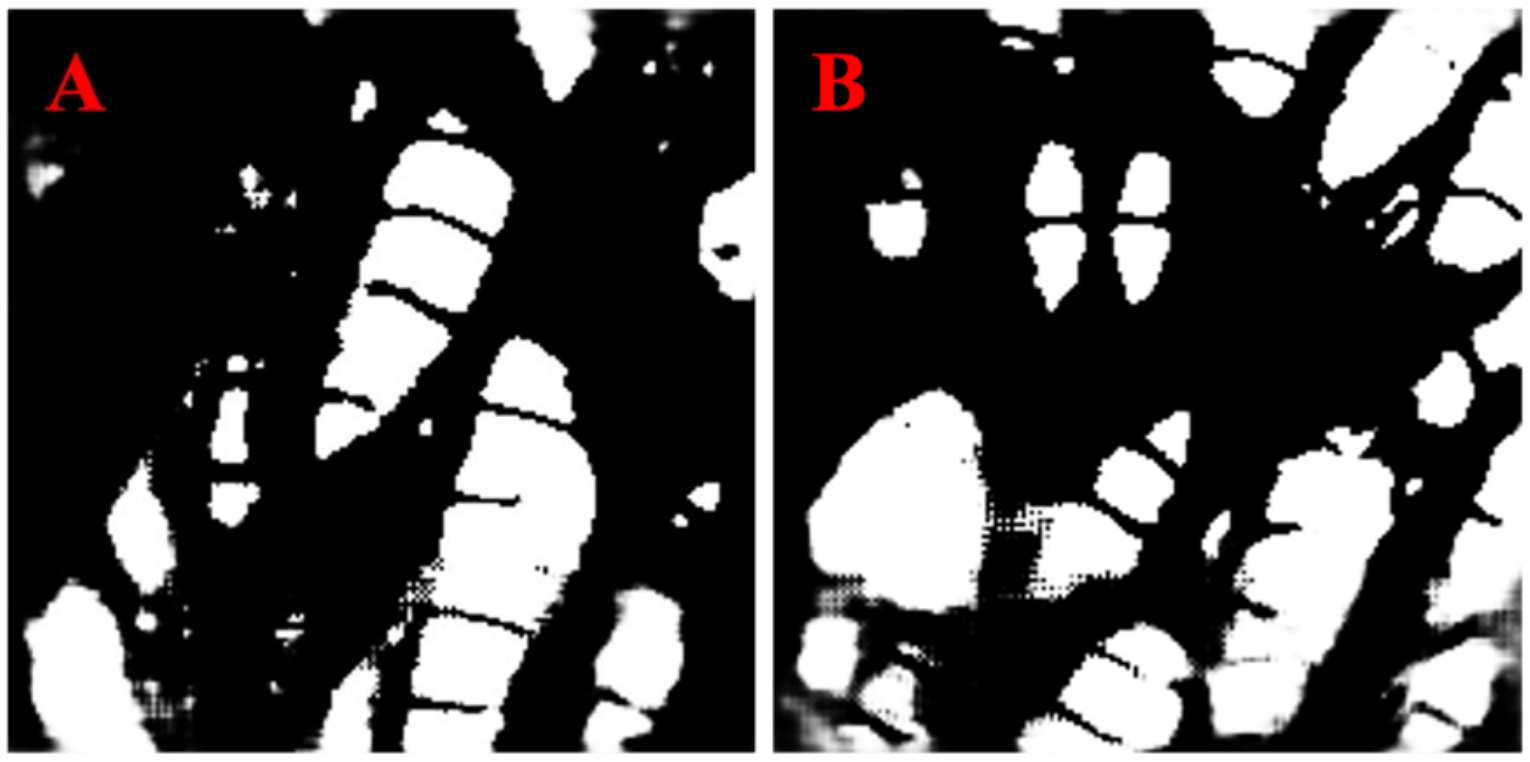

***Fig. 24*** *wGAN-GP generated typical high $\overline{Nu}$ porous structures A and B*

***Table 10*** *Feature thresholds derived from SHAP dependence analysis*

| Parameters | Structure A | Structure B |
|---|---|---|
| *Re* | 120 | 120 |
| *St* | 1000 | 1000 |
| Porosity | 0.7744 | 0.6848 |
| Specific Surface Area (SSA) | 0.0839 | 0.0959 |
| Pore Size Dispersion | 16.4832 | 15.5621 |
| Mean Pore Diameter | 20.7650 | 18.1038 |
| Percolation Strength | 0.9991 | 0.9988 |
| Tortuosity | 1.3311 | 1.3689 |
| NCI | 0.5280 | 0.5232 |
| Connectivity | 5 | 26 |
| Euler Number | -60 | -20 |
| $\overline{Nu}$ | 400.7818 | 399.5285 |

To further demonstrate the practical relevance of the SHAP-based interpretation, two representative porous structures were selected from the dataset, as shown in Fig. 24. These samples exhibit exceptionally high average Nusselt numbers, with $\overline{Nu} = 400.7818$ and $399.5285$, respectively. The corresponding structural and flow parameters are summarized in Table 10.

Both structures exhibit thermal performance consistent with the SHAP-identified enhancement patterns. Their Reynolds and Strouhal numbers are set at $Re = 120$ and $St = 1000$, well above the enhancement thresholds ($Re > 75$, $St > 100$), indicating strong convective effects under oscillatory flow. Structurally, both samples feature high porosity (> 0.6256), large mean pore diameters, and high pore size dispersion—all positively contributing to $\overline{Nu}$ as identified by SHAP. In addition, their percolation strength values exceed 0.9976, reflecting well-connected internal flow pathways.

Although some features deviate from the SHAP-defined optimal thresholds—namely, specific surface area < 0.1175, tortuosity > 1.32, and network connectivity index slightly > 0.5187—their overall thermal performance remains excellent. This highlights the importance of synergistic interactions between parameters. SHAP ranks SSA as the second most influential topological variable, whereas tortuosity and NCI rank significantly lower. Hence, suboptimal values in the latter can be offset by strong performance in more influential parameters such as porosity and pore size dispersion, particularly under favorable flow conditions. The LBM-simulated velocity streamlines and temperature contours for structures A and B, as demonstrated in Fig. 24, are presented in Appendix B.

## 5 Conclusions and future work

### *5.1 Conclusions*

This study proposed a data-driven wGAN–LBM–Nested_CV framework that novelly integrates generative modeling (wGAN-GP), physics-based simulation (LBM), and comprehensive comparison for 10 predictive models (Nested_CV), and finally adopts XGBoost to predict the average Nusselt number subject to oscillatory flow in porous media, with the application of explainable AI, further improving the physical interpretability of the machine learning model.

The key contributions and findings are summarized as follows:

1) The wGAN-GP model successfully generated porous structures with substantially

greater topological diversity than the original CT-scanned samples. This diversity enabled broader and more representative exploration of oscillatory flow and heat transfer behaviors.

2) Among the ten machine learning models evaluated by leveraging the dataset of wGAN-GP generated porous structures, XGBoost achieved the best performance in predicting the average Nusselt number, reaching an $R^2$ of 0.9981. The integration of generative deep learning with predictive ML models offers both reliable performance estimation and generalizable insights for heat transfer in porous media systems.
3) SHAP interpretation of the XGBoost model identified the Reynolds number, Strouhal number, porosity, specific surface area, and pore size dispersion as the most influential predictors of the average Nusselt number. Quantitative threshold analysis revealed optimal synergistic interactive enhancement ranges, including $Re > 75$, $St > 100$, porosity > 0.6256, SSA > 0.1175, and pore size dispersion > 5.57. Dependence plots highlighted synergistic effects, such as high porosity combined with large mean pore diameter and high Reynolds number, coupled with moderate-to-high Strouhal number, that maximize convective heat transfer performance. The SHAP-derived analysis provides guidance for further porous structural heat exchangers design and elucidates how oscillation and structural patterns interactively affect the thermal performance from a data-driven perspective.

*5.2 Limitations and suggestions for future work*

While the proposed framework demonstrates strong predictive accuracy and interpretability, the current findings remain data-driven rather than derived from physical principles. The SHAP-based insights reflect statistical associations learned by XGBoost within the simulated dataset and thus are valid only within the tested range of parameters and flow conditions. Moreover, the LBM simulations used to generate the training database inevitably simplify real-world phenomena by neglecting, for instance, surface energy effects, wall wettability, and multiphase interactions, which may influence boundary-layer behavior and convective transport. The framework also focuses solely on a two-dimensional flow domain, which limits experimental validation and ignores the fabrication constraints, such as that melt-pool phenomena may change the geometry of small pores during 3D printing.

Future work will extend the range of the dataset to expand the prediction capabilities and extend this methodology toward a three-dimensional generative–simulation–learning pipeline that incorporates additional physics and manufacturability, enabling experimental validation of predicted structures. Such integration of modeling and experimentation will strengthen the physical grounding of the framework and enhance its applicability to real porous media and engineered thermal systems.

**Acknowledgements**

The study was supported by the National Science Foundation (NSF), under grant No. 2318107. Any opinions, findings, and conclusions or recommendations expressed in this material are those of the authors and do not necessarily reflect the views of the NSF.

**Nomenclature**

| | |
|---|---|
| $A$ | amplitude |
| $\widehat{a_i}$ | the predicted value of i-th data point |
| $a_i$ | the actual value of i-th data point |
| $\bar{a}$ | the mean of the actual values |
| $c_s$ | speed of sound |
| $e_i$ | discrete velocity |
| $f$ | frequency |
| $f_i$ | distribution functions for fluid particle fields |
| $f_i^{eq}$ | equilibrium distribution function |
| $g_i^{eq}$ | heat equilibrium distribution function |
| $g_i$ | distribution functions for thermal particle fields |
| $k$ | thermal conductivity |
| $\bar{k}$ | normalized average node degree |
| $l$ | length |
| $n$ | sample size |
| $N$ | resolution |
| $Nu$ | Nusselt numbers |
| $\overline{Nu}$ | average Nusselt numbers |
| $P$ | pressure |
| $p$ | frequency distribution function |
| $Pr$ | Prandtl number |
| $q''$ | heat flux |
| $Re$ | Reynolds number |
| $R^2$ | coefficient of determination |
| $S$ | domain area |
| $St$ | Strouhal numbers |
| $T$ | temperature |
| $t$ | time |
| $u$ | velocity in LBM equilibrium function |
| $U$ | velocity |
| $Wo$ | Womersley numbers |
| $w_i$ | lattice weights |
| Greek symbols | |
| $\alpha$ | thermal diffusivity |
| $\emptyset$ | porosity |
| $\tau$ | dimensionless momentum relaxation time |
| $\tau_g$ | dimensionless thermal relaxation time |
| $\rho$ | density |
| $\nu$ | kinematic viscosity |
| $\varphi$ | phase angle |
| Subscripts | |
| *cold* | cold temperature |
| *hot* | hot temperature |
| *i* | index notation |
| *in* | inlet |
| *j* | index notation |
| *max* | maximum |
| *out* | outlet |
| *solid* | solid phase |
| *x* | x-direction |
| *y* | y-direction |
| Abbreviations | |
| 2D | two dimensional |
| 3D | three dimensional |
| AI | Artificial Intelligence |
| ANN | Artificial Neural Networks |
| BO | Bayesian Optimization |
| C | Critic |
| CART | Classification and Regression Trees |
| CNN | Convolutional Neural Network |
| CV | cross validation |
| *dist* | distance |
| DL | Deep Learning |
| EGS | Enhanced Geothermal Systems |
| G | Generator |
| GANs | Generative Adversarial Networks |
| GP | Gaussian process |
| KNN | K-Nearest Neighbors |
| LBM | Lattice Boltzmann Method |
| MAE | Mean Absolute Error |
| micro-CT | micro-Computed Tomography |
| ML | Machine Learning |
| MPD | Mean Pairwise Distance |
| MLP | Multilayer Perceptron |
| NCI | Network Connectivity Index |
| Nested_CV | nested cross-validation |
| PCA | Principal Component Analysis |
| PINNs | Physics-Informed Neural Networks |
| PolyRidge | Polynomial Ridge Regression |
| RF | Random Forests |
| RMSE | Root Mean Squared Error |
| ROI | Region of Interest |
| SHAP | SHapley Additive exPlanations |
| SSA | Specific Surface Area |
| SSE | Sum of Squared Errors |
| SST | Total Sum of Squares |
| SVM | Support Vector Machines |
| SVR | Support Vector Regressor |
| wGAN-GP | Wasserstein Generative Adversarial Networks with Gradient Penalty |
| XGBoost | eXtreme Gradient Boosting |

**Data availability**

The code of wGAN-GP can be found at: https://github.com/lzhu26/Lichang_NuPrediction. The code and data applied in this work will be available upon reasonable request.

## Appendix A: List of Implemented Machine Learning Models

Table A.1 presents a comprehensive list of the predictive ML models implemented in this study, as discussed in Section 2.3. The table includes each model's classification and the corresponding Python libraries used for implementation.

***Table A.1*** *ML models implemented in this work*

| Classification | ML models | Major Python Library |
|---|---|---|
| Support Vector Machines (SVM) [48] | Support Vector Regressor, kernel = radial (Radial SVR) | `sklearn.svm.SVR` |
| | Support Vector Regressor, kernel = linear (Linear SVR) | `sklearn.linear_model.SGDRegressor` |
| | Support Vector Regressor, kernel = sigmoid (Sigmoid SVR) | `sklearn.svm.SVR` |
| Tree-based models | Classification and Regression Trees (CART) [51] | `sklearn.tree.DecisionTreeRegressor` |
| | Random Forest (RF) [50] | `sklearn.ensemble.RandomForestRegressor` |
| | eXtreme Gradient Boosting (XGBoost) [49] | `xgboost` |
| Neural Network models | Artificial Neural Networks - Multi-Layer Perceptron (ANN-MLP) [52] | `sklearn.neural_network.MLPRegressor` |
| \ | k-Nearest Neighbors (k-NN) [53] | `sklearn.neighbors.KNeighborsRegressor` |
| Linear models | ElasticNet [55] | `sklearn.linear_model.ElasticNet` |
| | Ridge Regression [54] | `sklearn.linear_model.Ridge` |

## Appendix B: Typical LBM simulation results

This appendix presents LBM simulation results for two representative porous structures with high thermal performance extracted from the database. The corresponding images are shown in Fig. B.1. Both cases were simulated under the same oscillatory flow conditions, with Re = 120 and *St* = 1000. The topological features and their effects on heat transfer were discussed in Sec. 4.3, and the simulation setup is detailed in Sec. 2.2.2. Please note that the representative visualizations shown in this section are not additional validation cases, but examples extracted from the database for illustrative purposes.

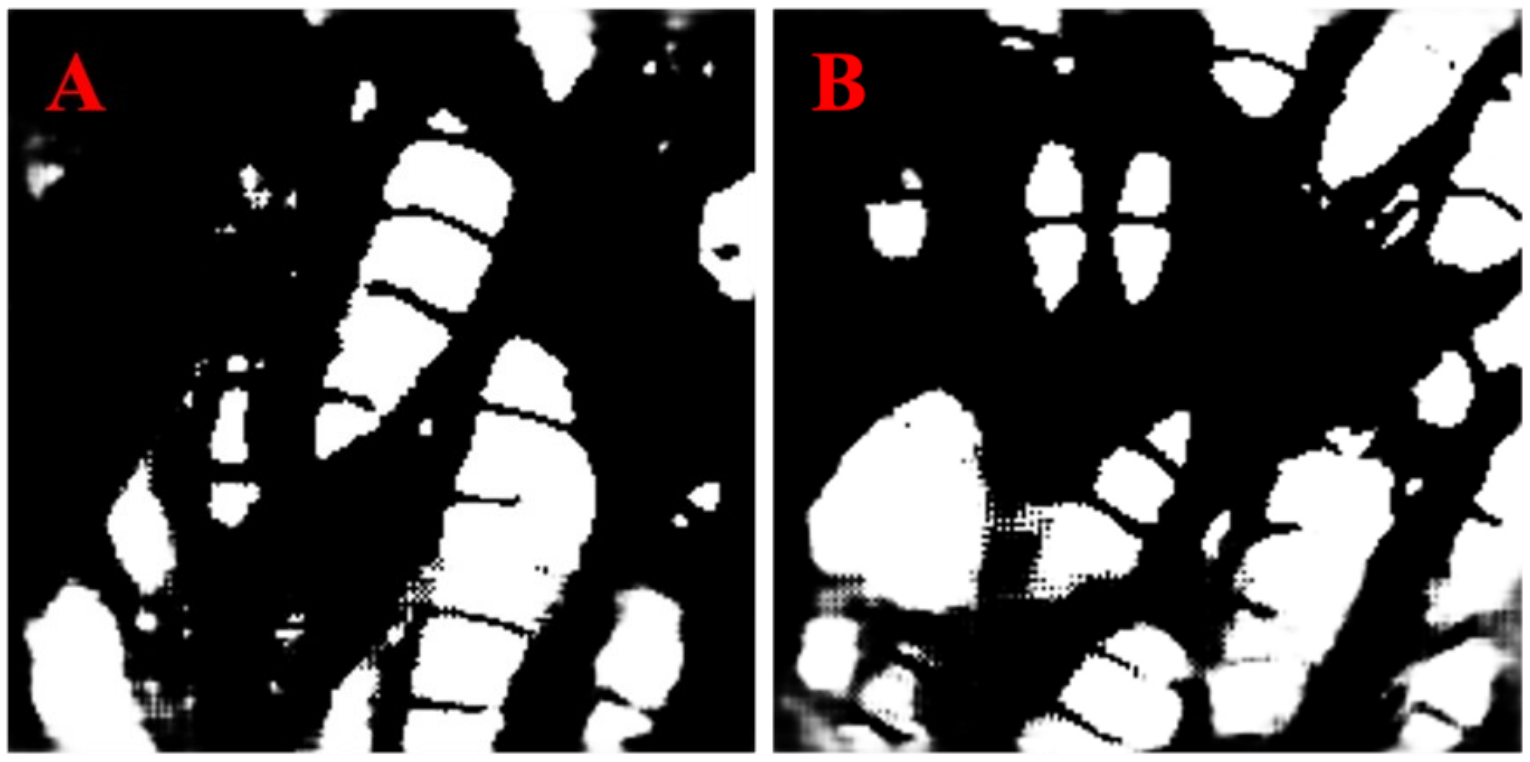

***Fig. B.1*** *Typical high $\overline{Nu}$ porous structures A and B*

For Structure A, velocity vector contours at four different phases within one oscillation cycle are illustrated in Figs. B.3 to B.6. The oscillation cycle period is denoted as *T*, and the selected phases are: $a = 0$, $b = \frac{1}{4}T$, $c = \frac{1}{2}T$, and $d = \frac{3}{4}T$. A schematic of the four selected phases is provided in Fig. B.2.

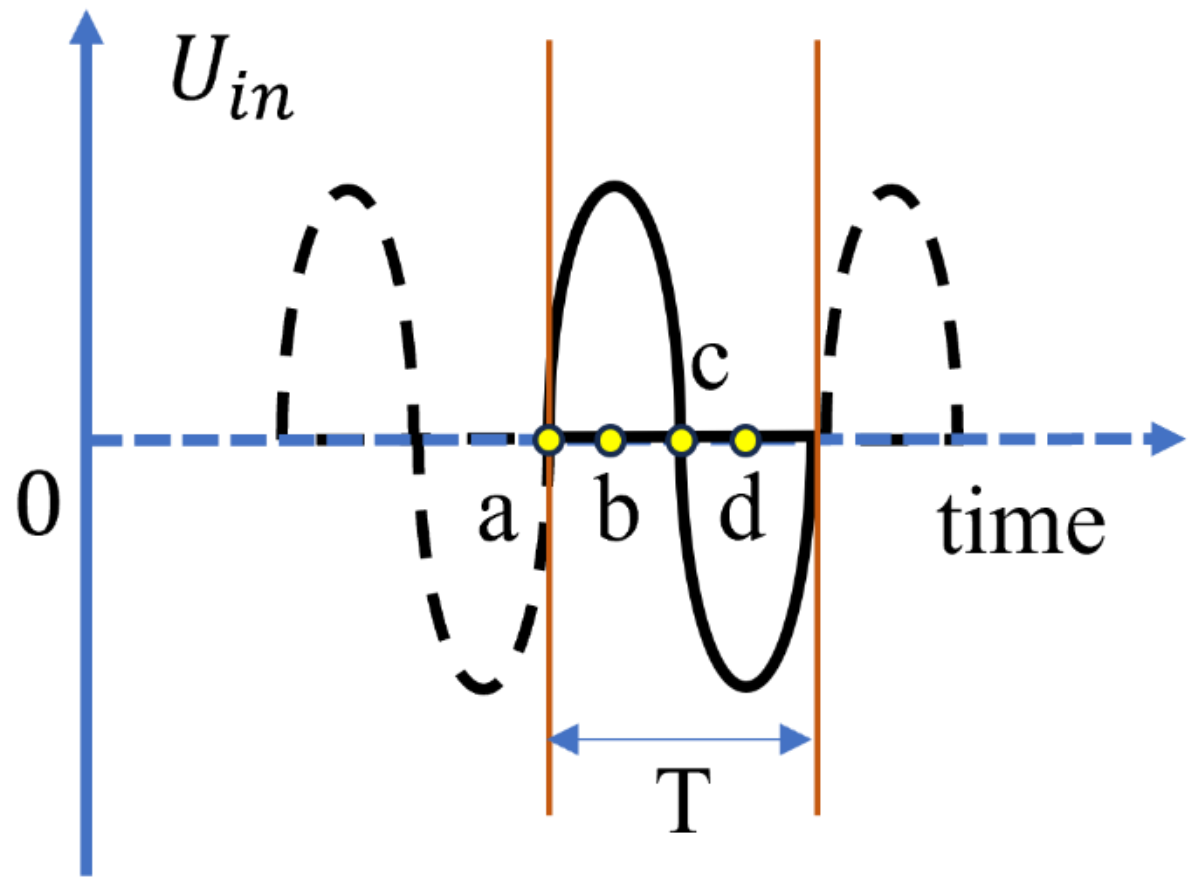

***Fig. B.2*** *Schematic of the location of phase a, b, c, and d*

Figs. B.4 and B.6 show high-velocity magnitudes at phases *b* and *d*, which correspond to the crest and trough of the oscillation cycle. At phase *b*, the inlet flow is directed from left to right (positive), whereas at phase *d*, it reverses direction (right to left, negative).

The porous structure induces significant modulation of the flow pattern. For example, at phase *a* (Fig. B.3), most velocity vectors are negative, consistent with the decaying portion of the oscillation. However, several vectors near the lower-left of the domain point rightward, which is attributed to the inlet velocity transitioning from negative to zero and then to positive. This transient behavior mimics a quasi-non-slip condition at the left boundary, causing partial reversal of the incoming flow and generating reflected velocity vectors. Conversely, at phase *c* (Fig. B.5), the inlet velocity transitions from positive to negative. As a result, most vectors remain positive, but an increasing number begin to orient leftward, reflecting the decaying momentum from the prior phase. This gradual shift may refer to the inertia-driven flow response within the complex porous network.

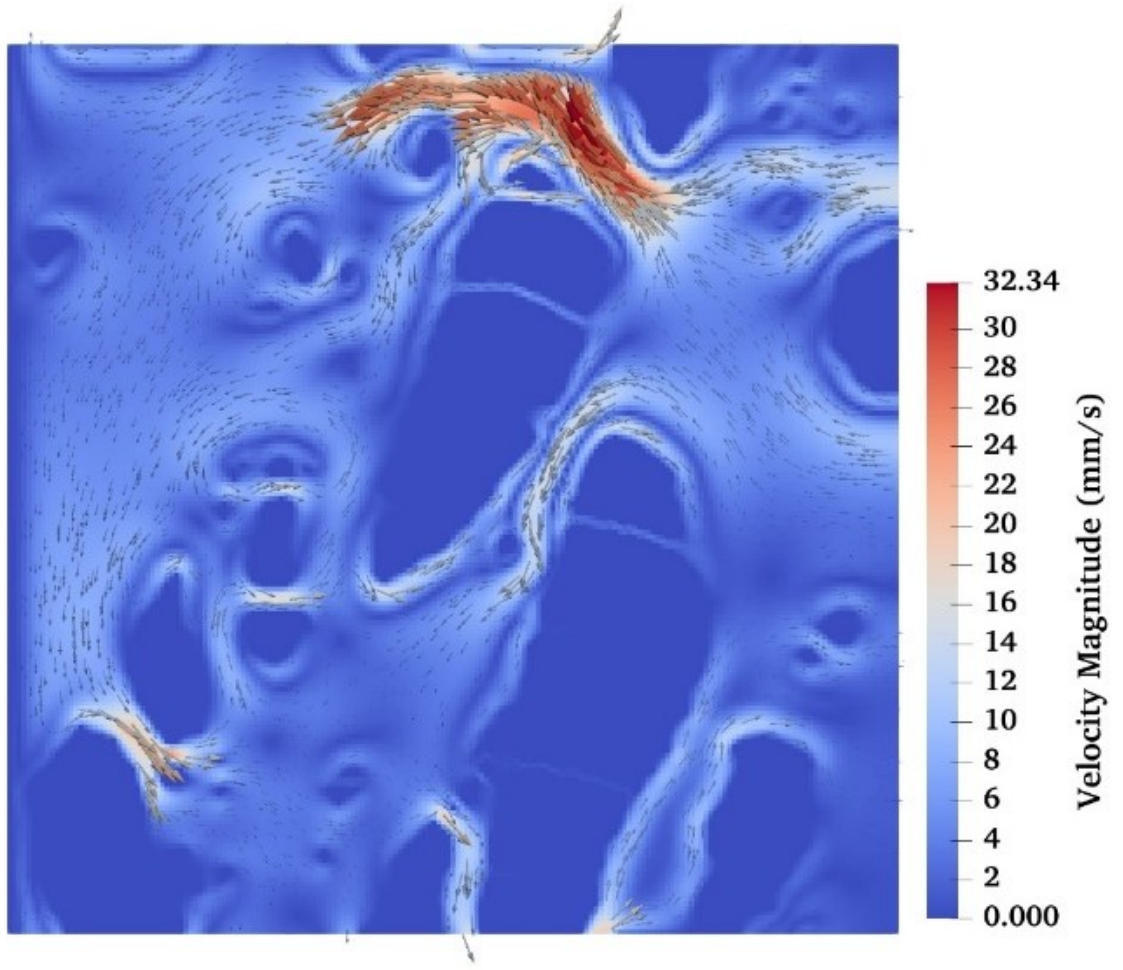

***Fig. B.3*** *Velocity vector contour of structure A at phase a, where* $\alpha = 0$

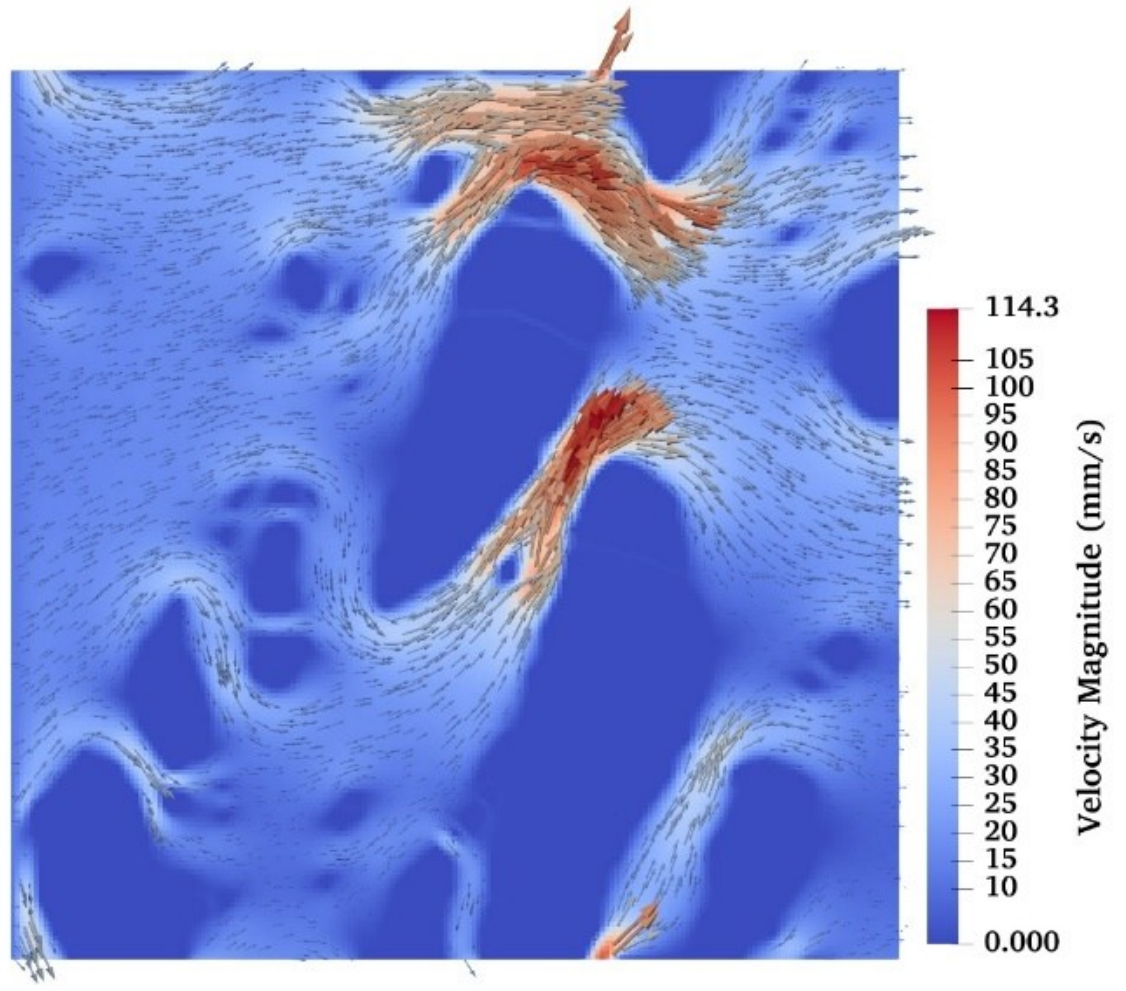

***Fig. B.4*** *Velocity vector contour of structure A at phase b, where* $b = \frac{1}{4}T$

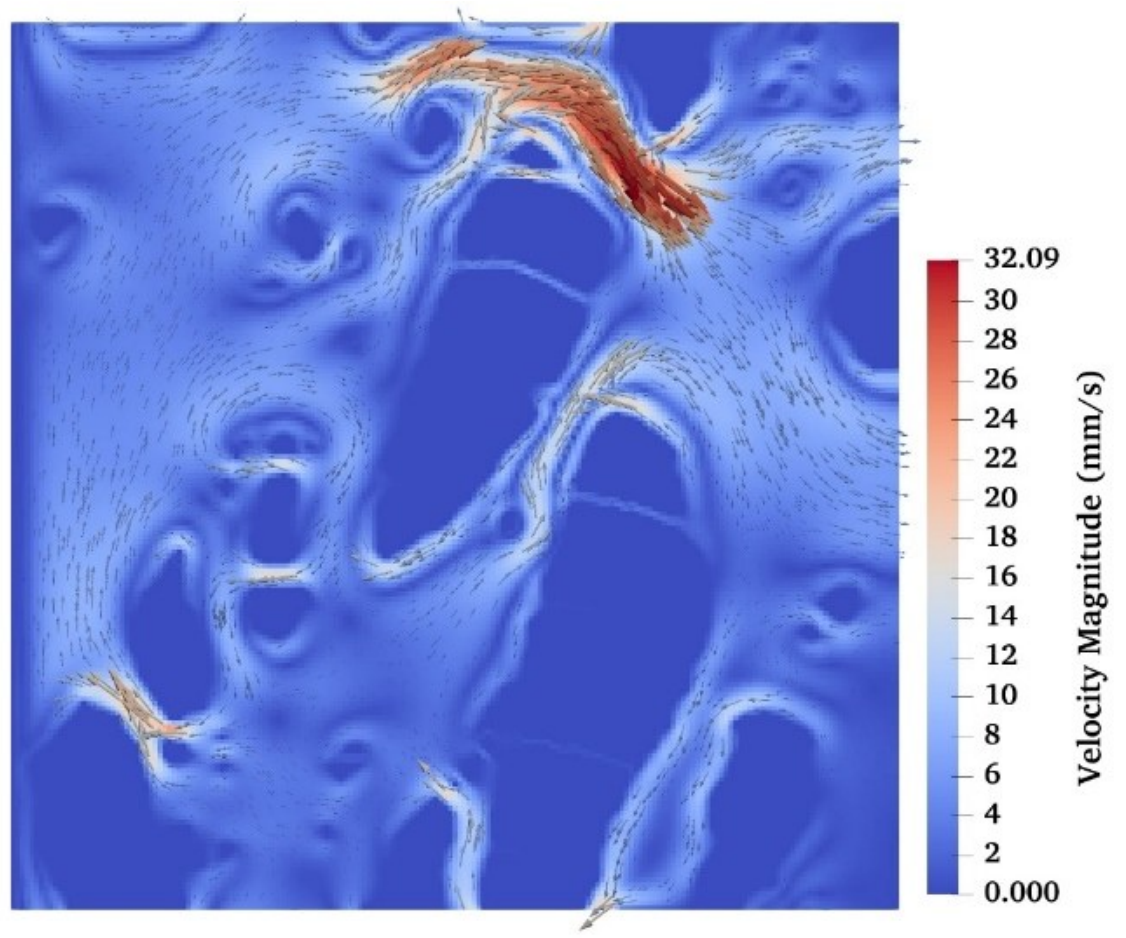

***Fig. B.5*** *Velocity vector contour of structure A at phase c, where* $c = \frac{1}{2}T$

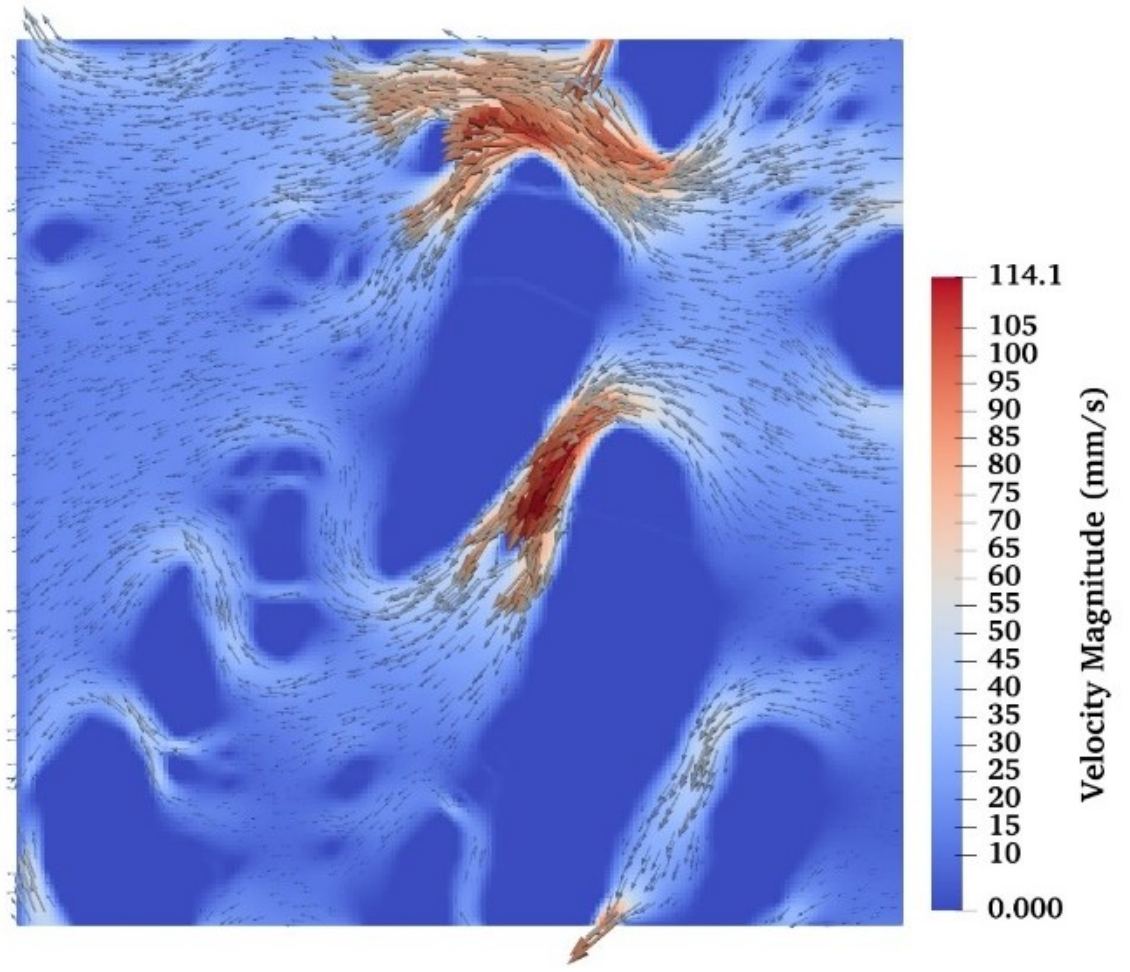

***Fig. B.6*** *Velocity vector contour of structure A at phase d, where* $d = \frac{3}{4}T$

Temperature distributions at the four phases are shown in Figs. B.7 to B.10. The thermal gradients closely follow the velocity field patterns, particularly at phases *b* and *d,* where stronger flow enhances convective heat transfer.

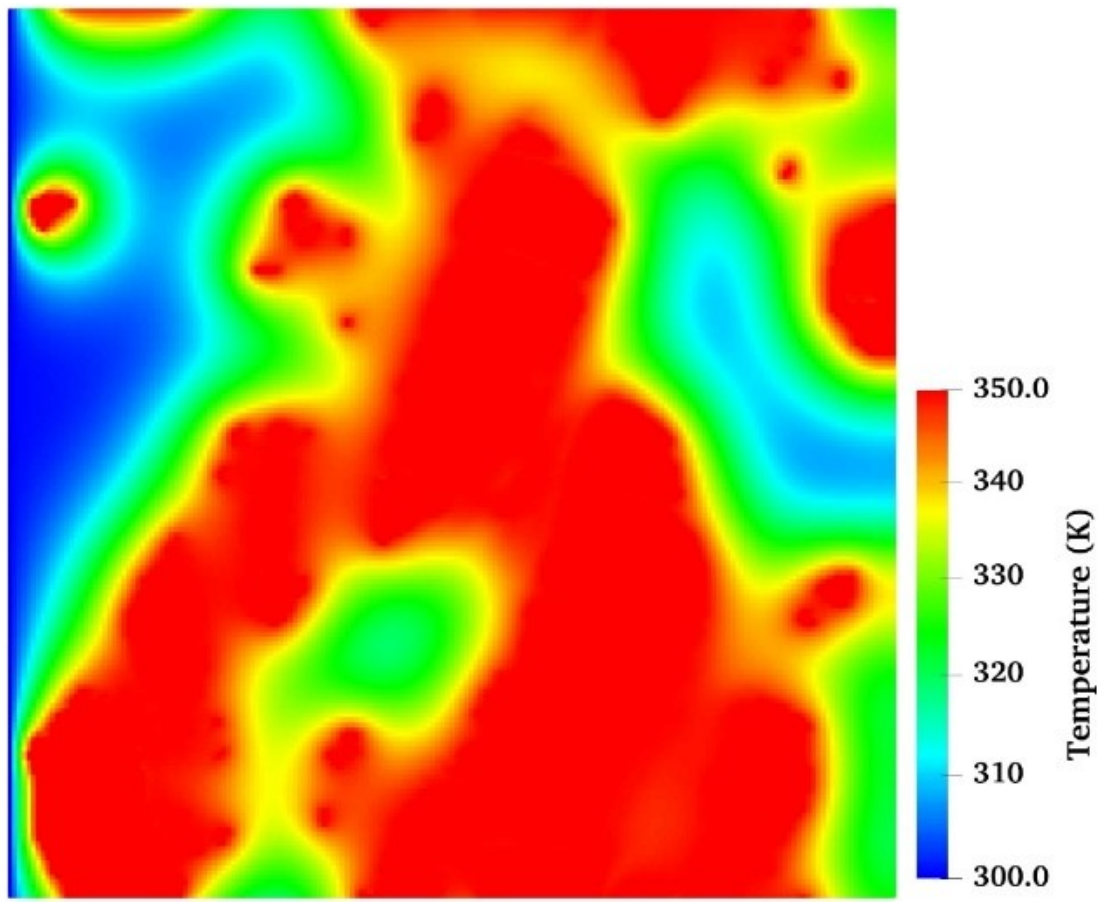

***Fig. B.****7 Temperature contour of structure A at phase a, where* $a = 0$

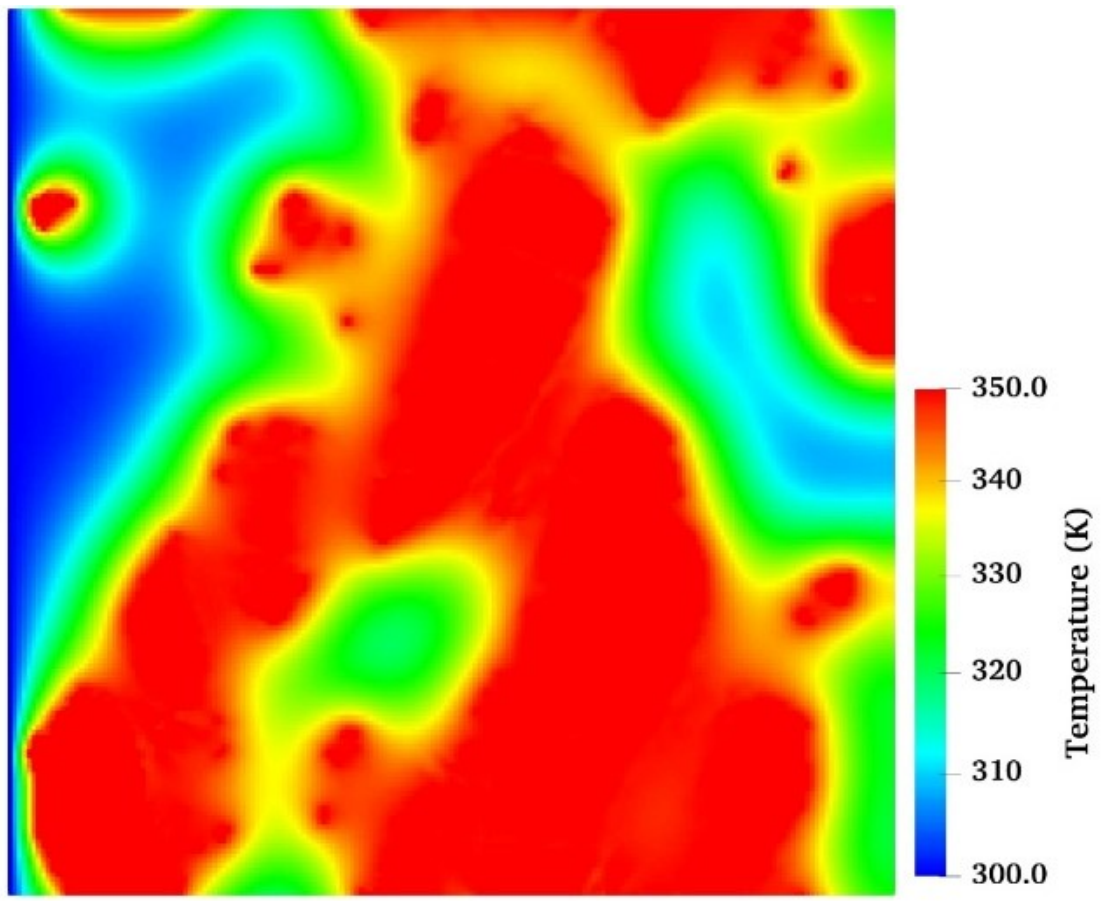

***Fig. B.8*** *Temperature contour of structure A at phase b, where* $b = \frac{1}{4}T$

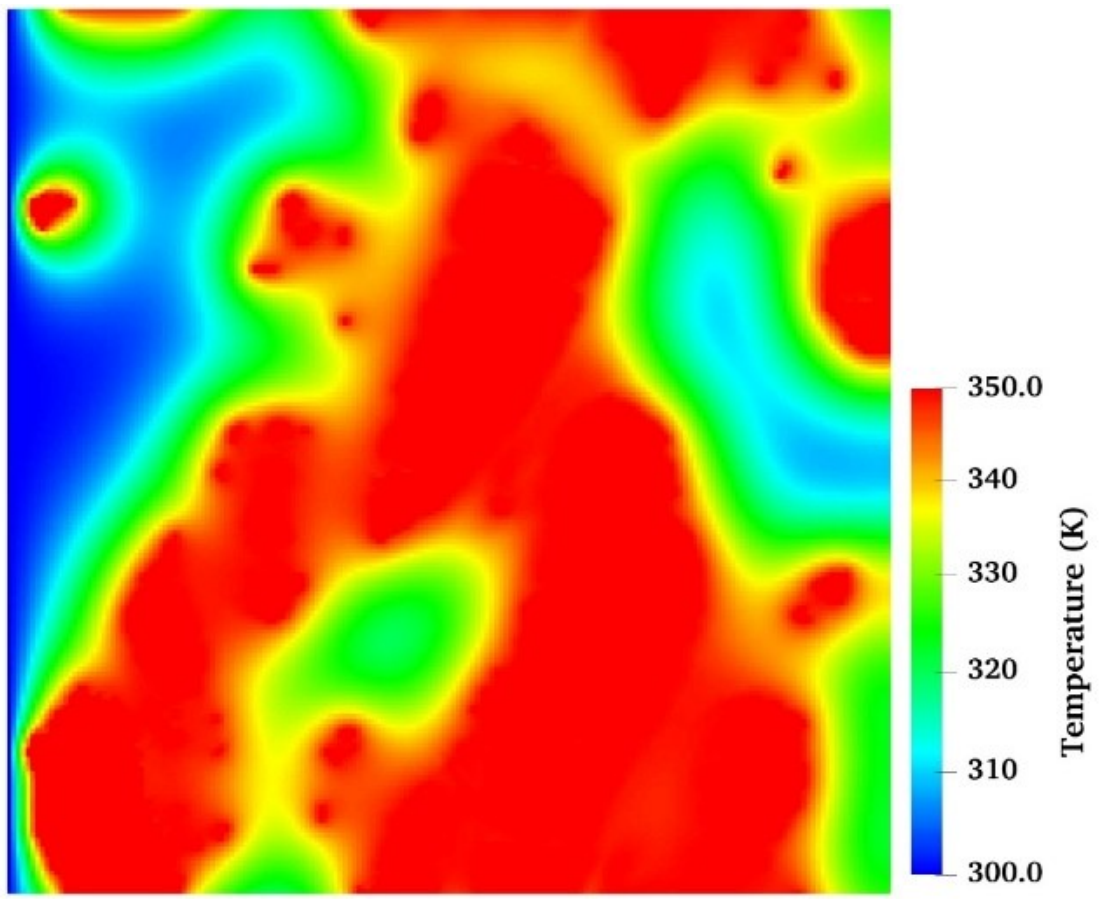

***Fig. B.9*** *Temperature contour of structure A at phase c, where* $c = \frac{1}{2}T$

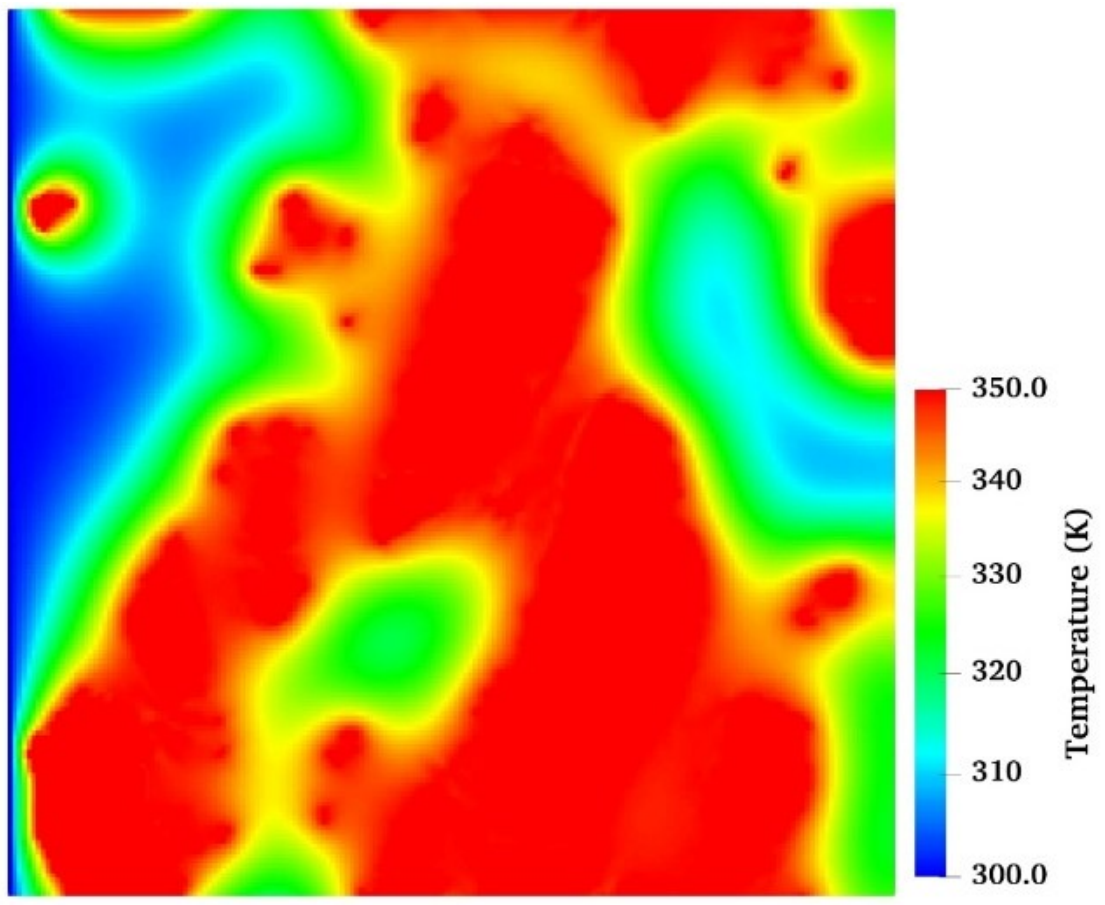

***Fig. B.10*** *Temperature contour of structure A at phase d, where* $d = \frac{3}{4}T$

For Structure B, velocity vector and temperature contours at the same four oscillation phases are presented in Figs. B.11 to B.18. The results reveal similar flow and thermal field behaviors as observed in Structure A.

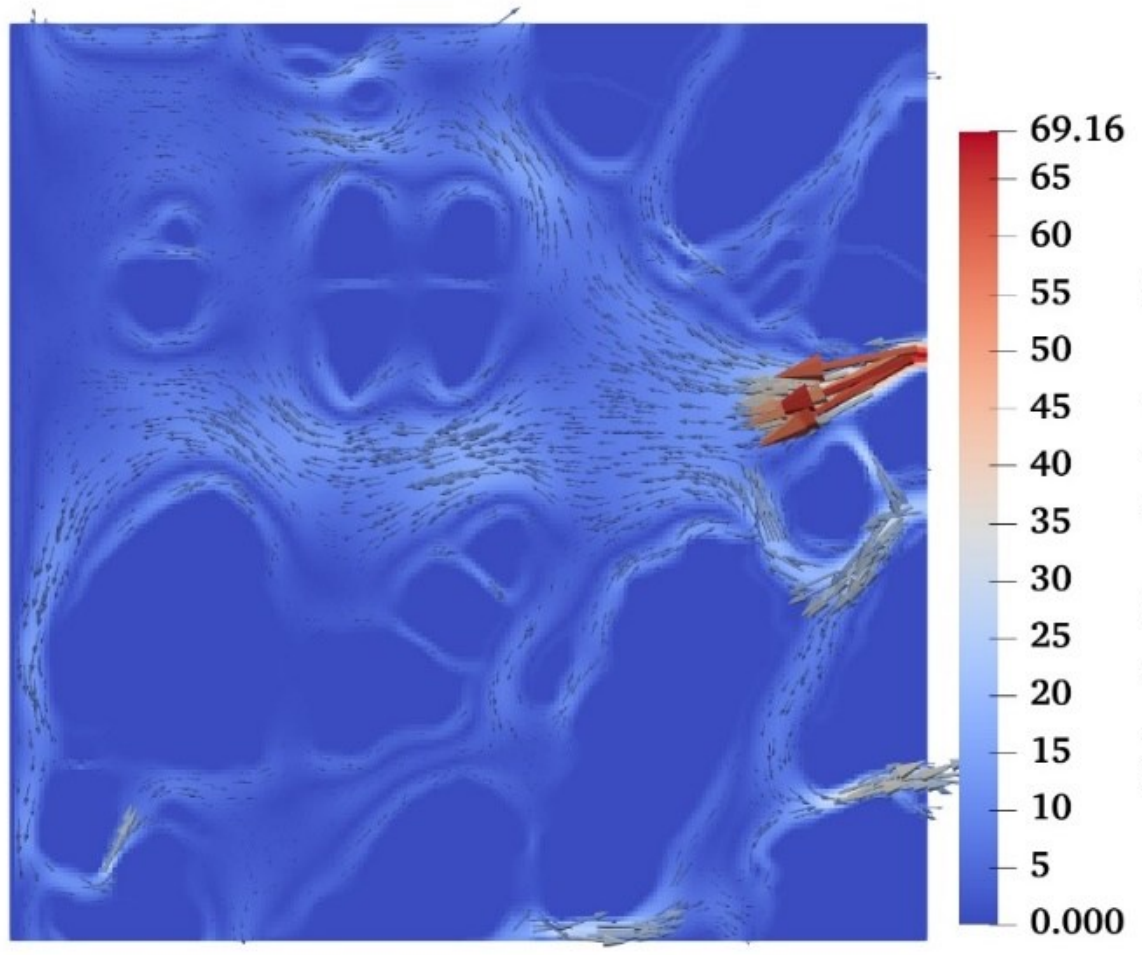

***Fig. B.11*** *Velocity vector contour of structure B at phase a, where* $a = 0$

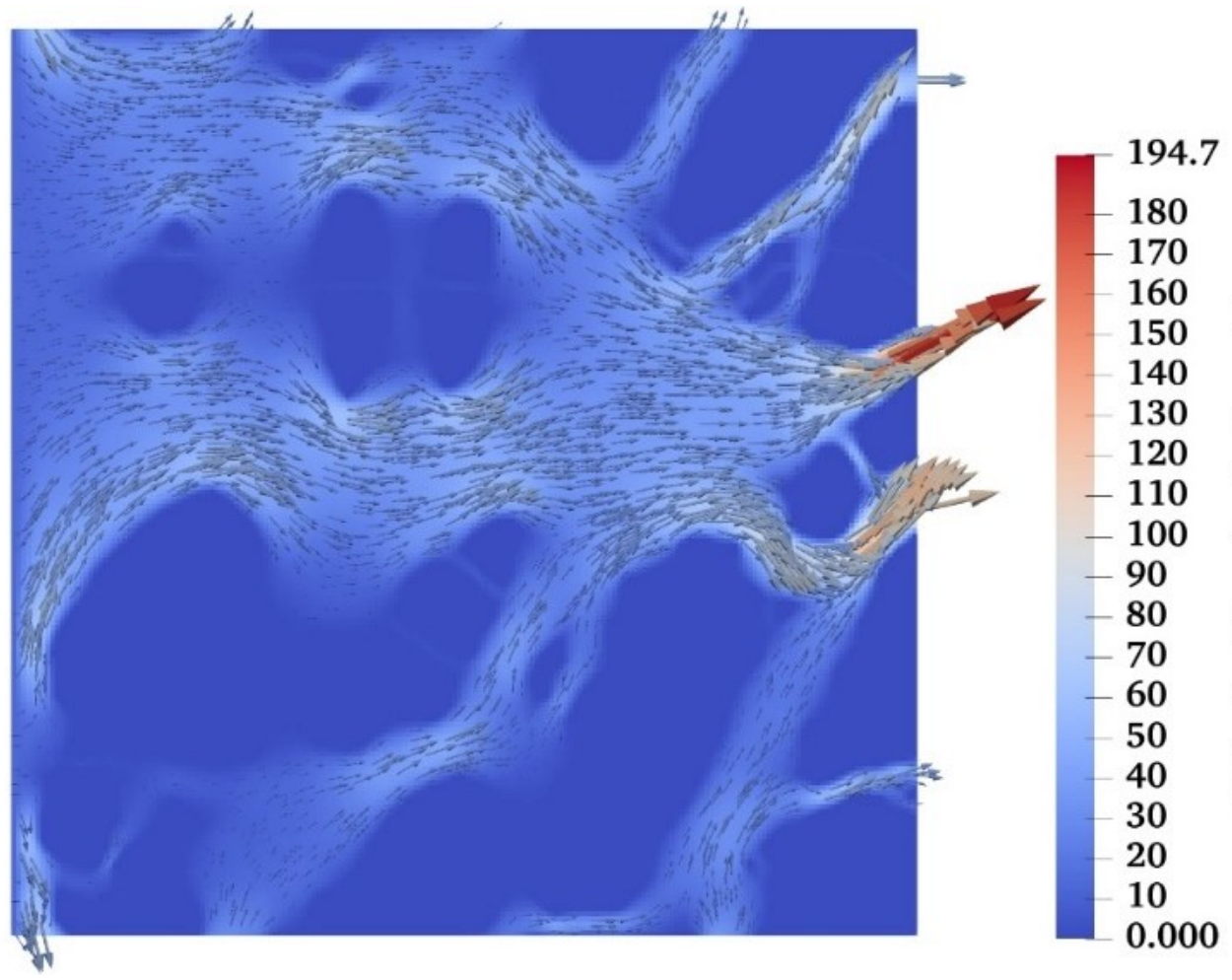

***Fig. B.12*** *Velocity vector contour of structure B at phase b, where* $b = \frac{1}{4}T$

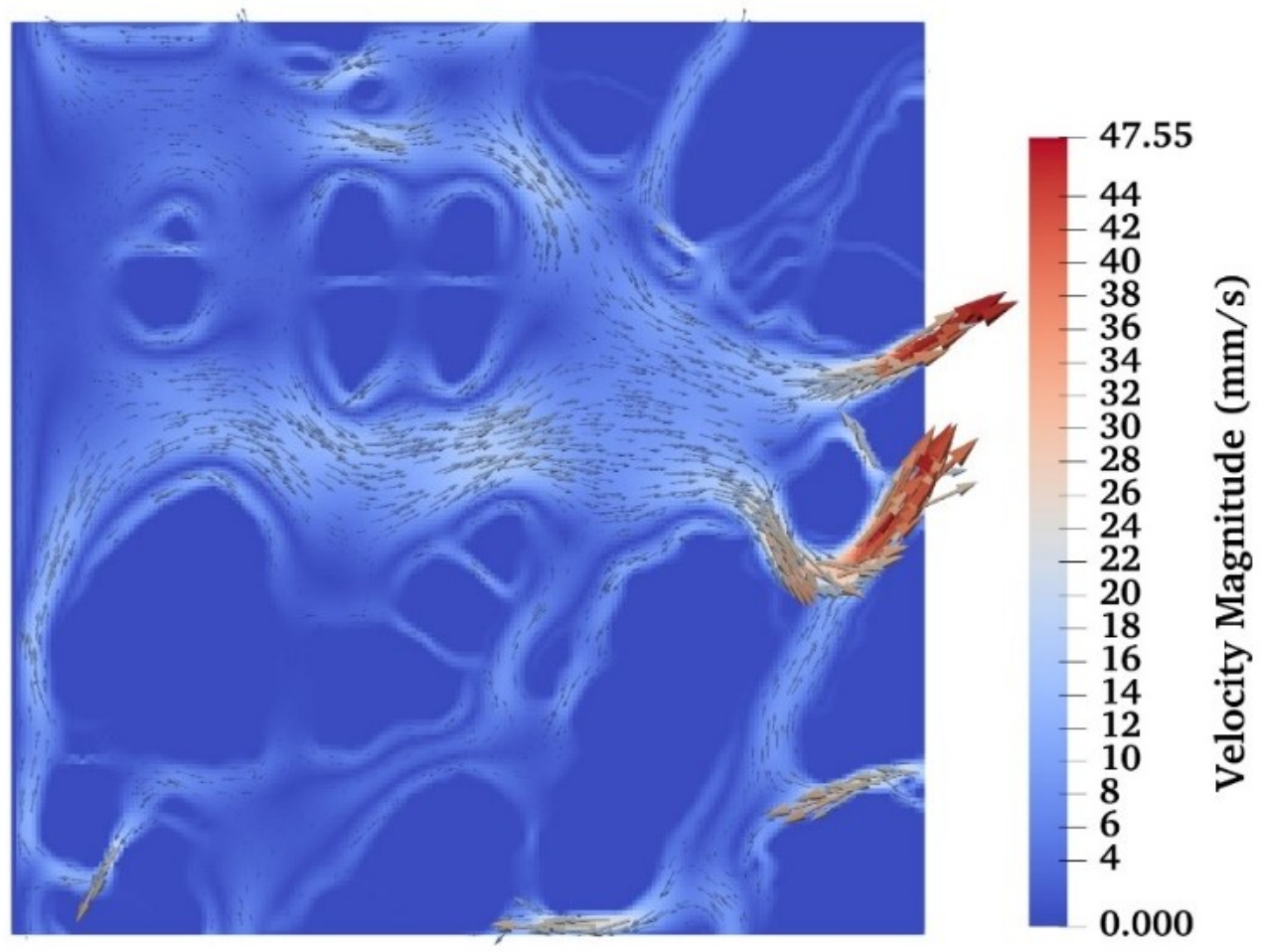

***Fig. B.13*** *Velocity vector contour of structure B at phase c, where* $c = \frac{1}{2}T$

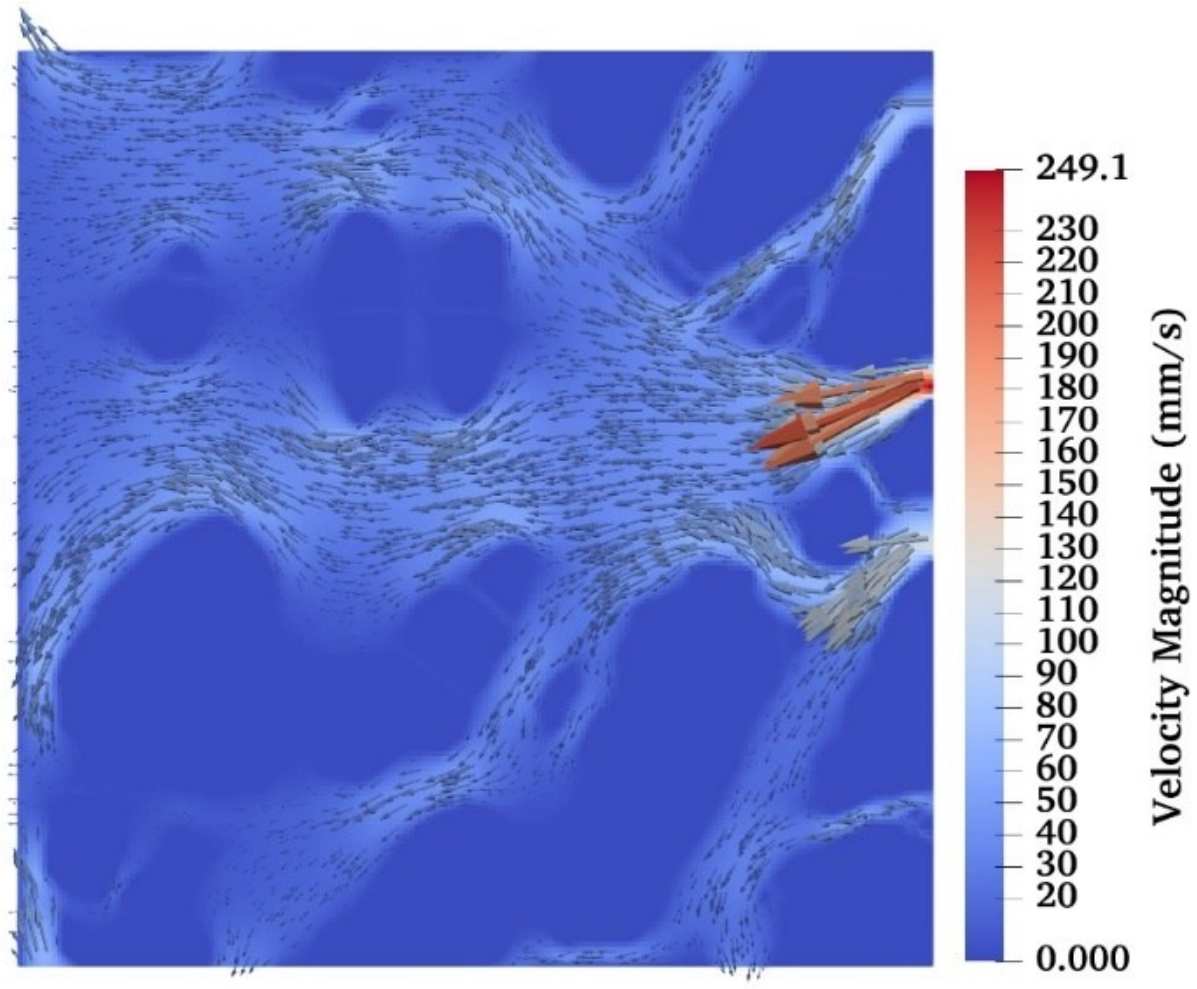

***Fig. B.14*** *Velocity vector contour of structure B at phase d, where* $d = \frac{3}{4}T$

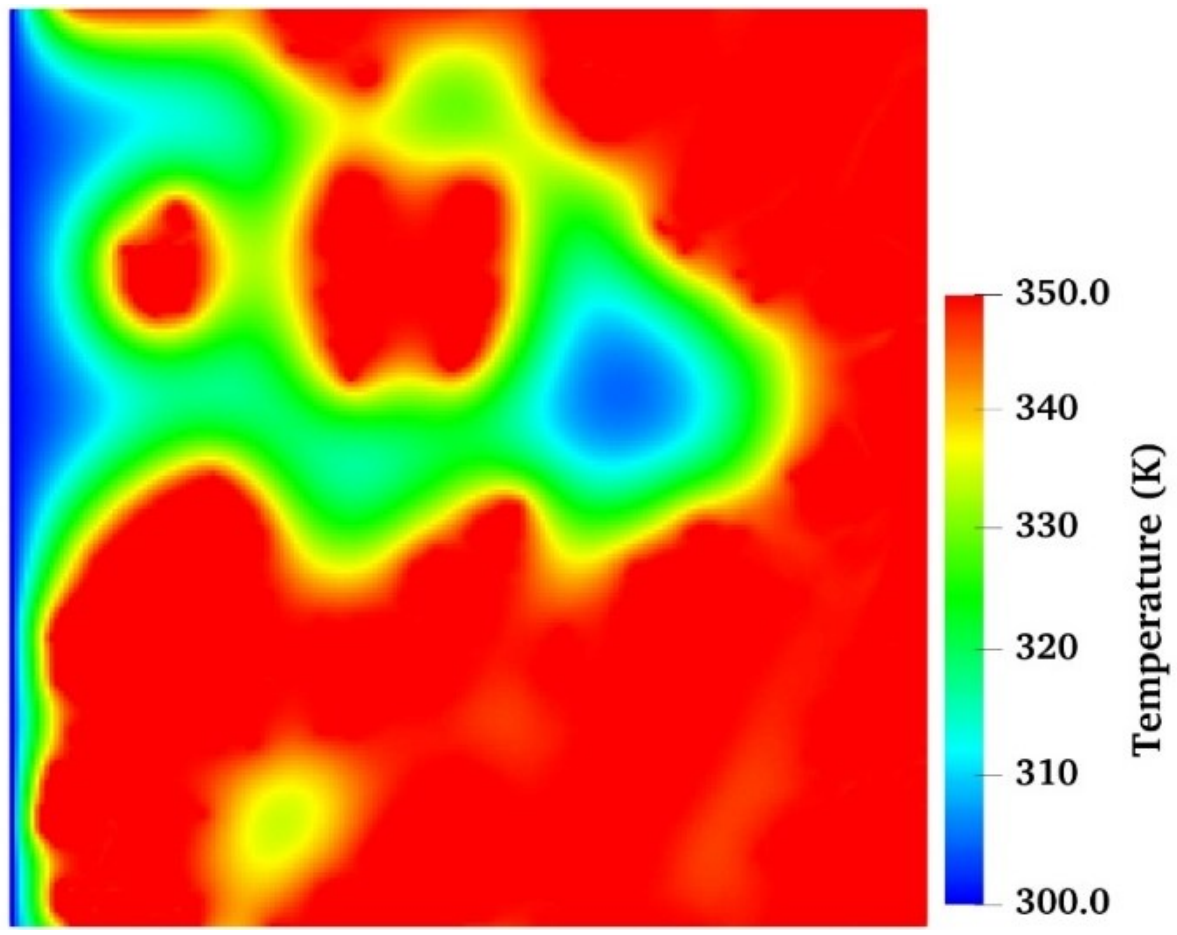

***Fig. B.15*** *Temperature contour of structure B at phase a, where* $a = 0$

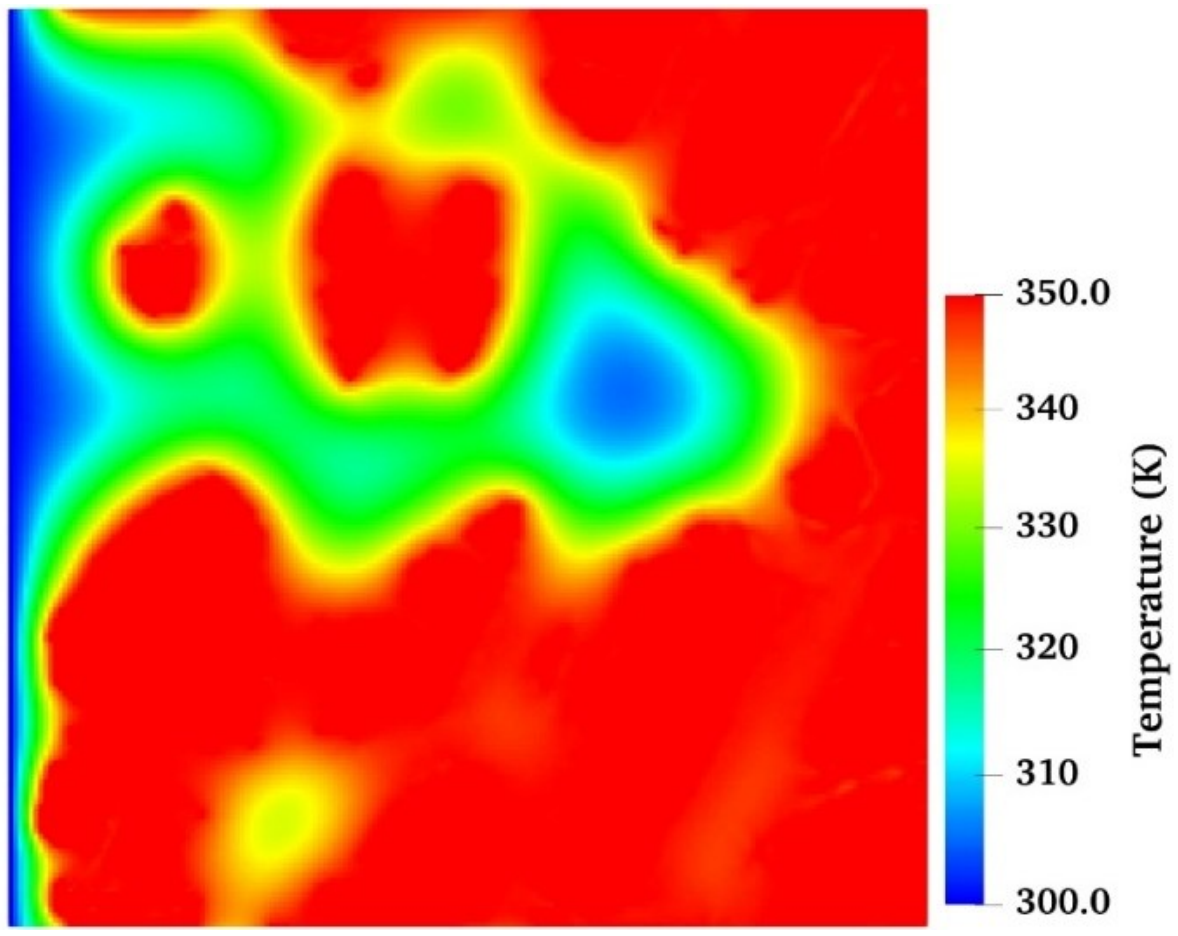

***Fig. B.16*** *Temperature contour of structure B at phase b, where* $b = \frac{1}{4}T$

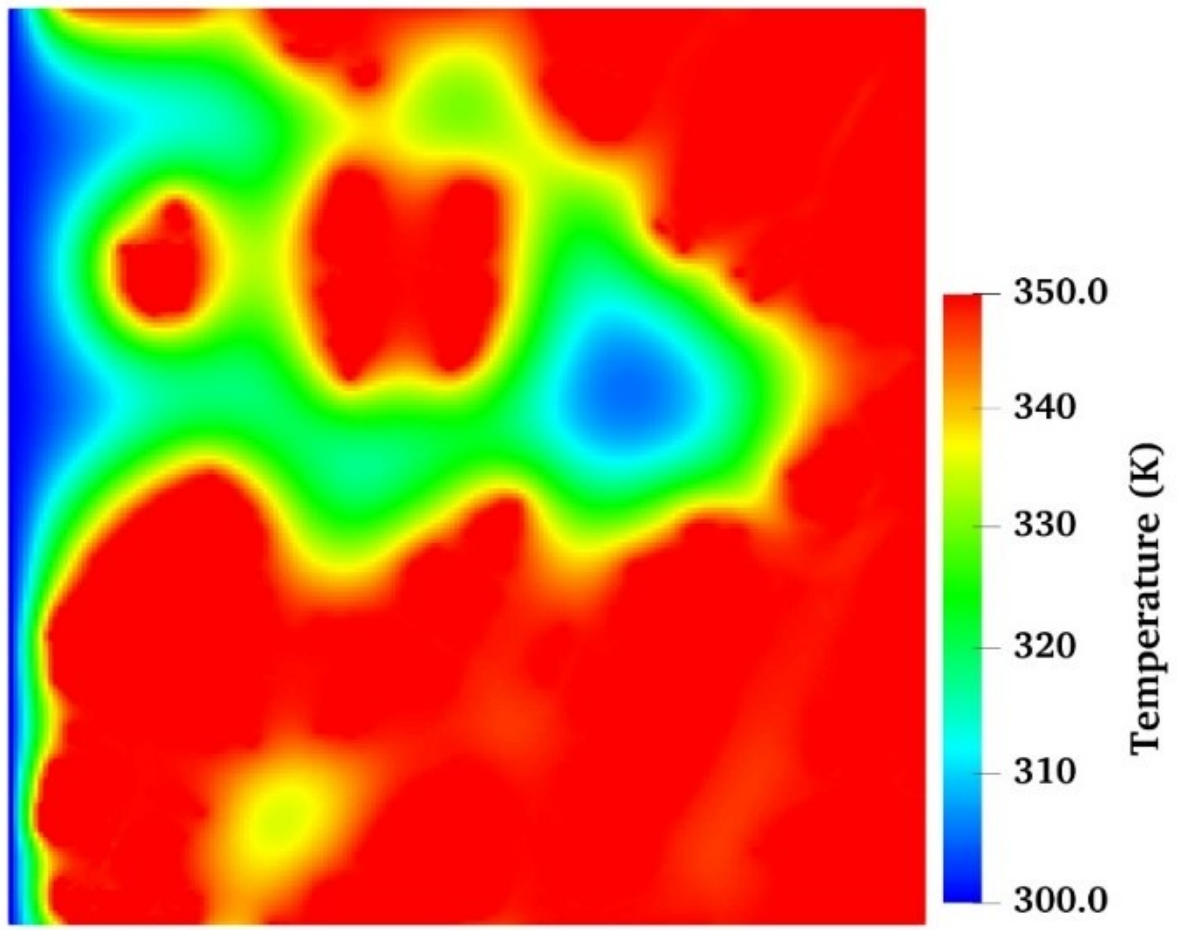

***Fig. B.17*** *Temperature contour of structure B at phase c, where* $c = \frac{1}{2}T$

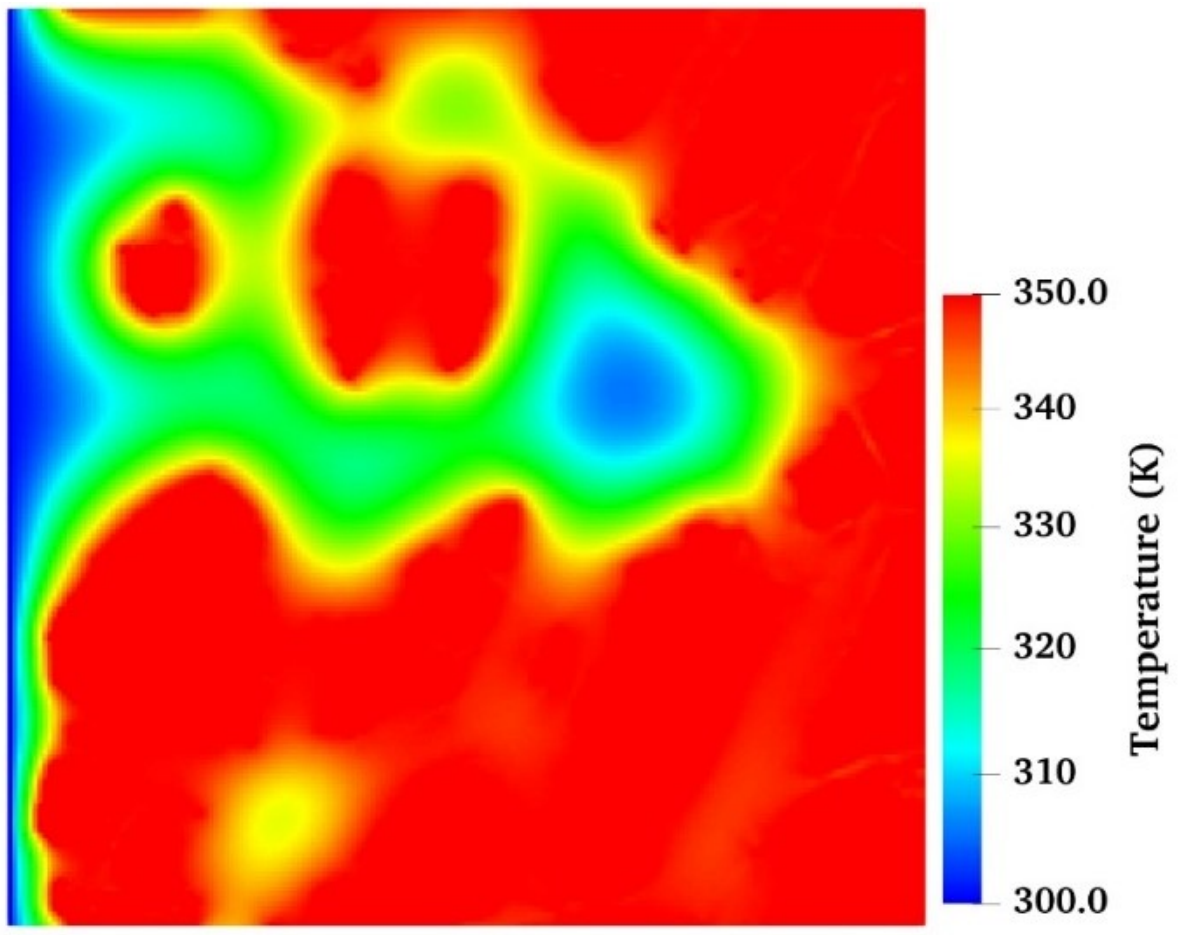

***Fig. B.18*** *Temperature contour of structure B at phase d, where* $d = \frac{3}{4}T$

## Appendix C: SHAP dependence plots for additional topological features

This appendix presents the SHAP dependence plots for the remaining topological features beyond porosity. Detailed analyses and discussions of these features are provided in Sec 4.2.

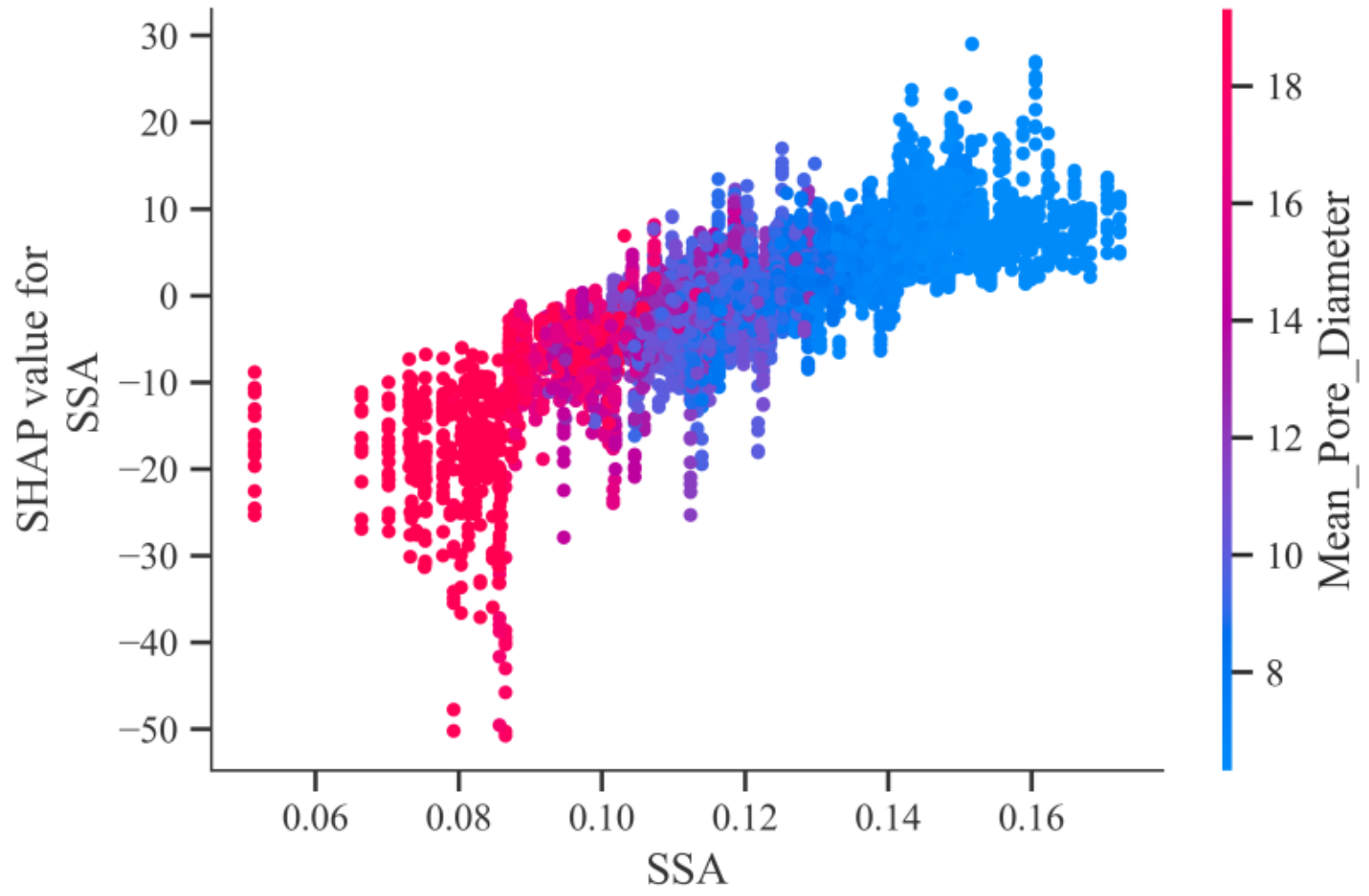

**Fig. C.1** SHAP dependence plots for SSA

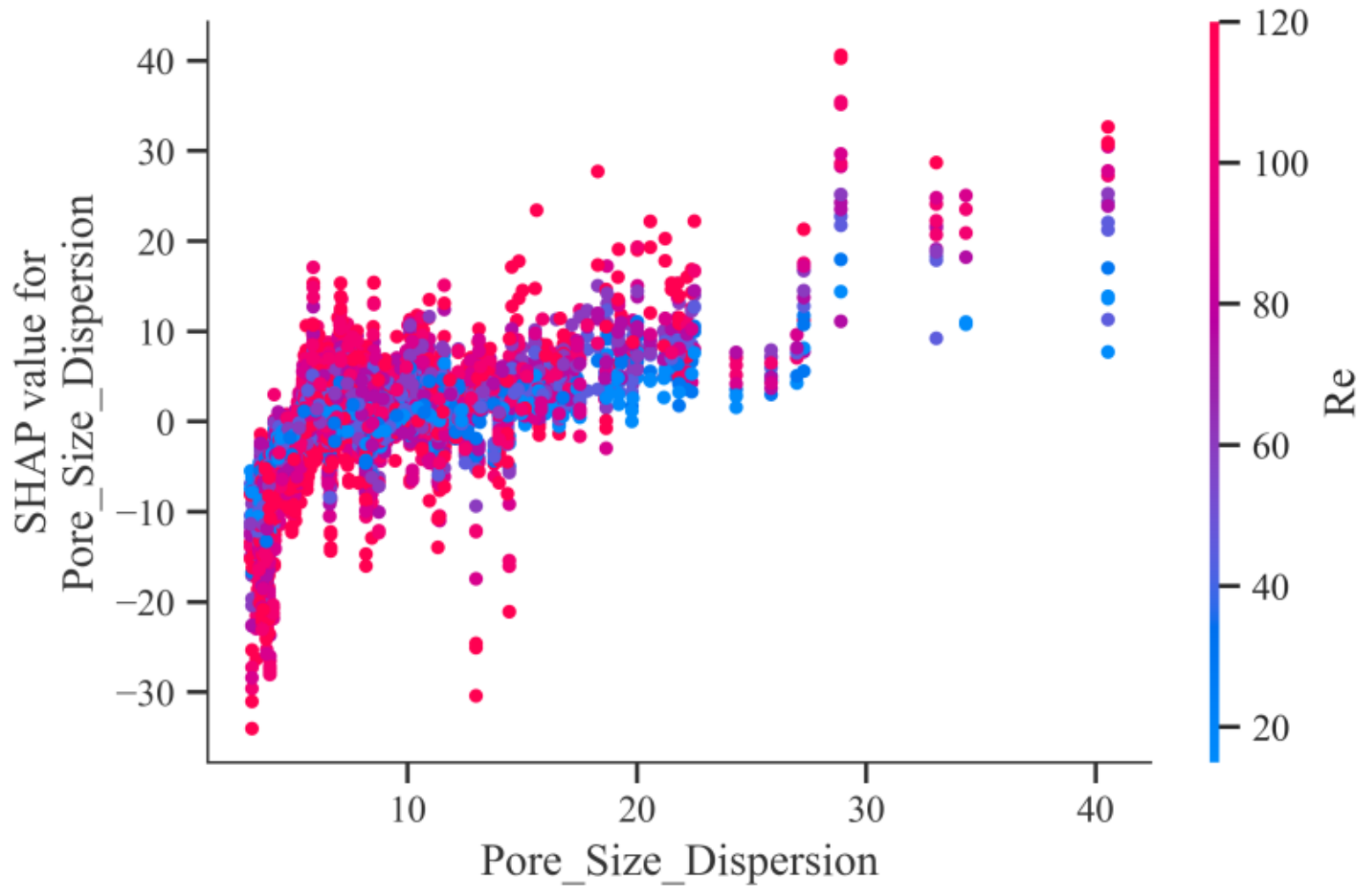

**Fig. C.2** SHAP dependence plots for pore size dispersion

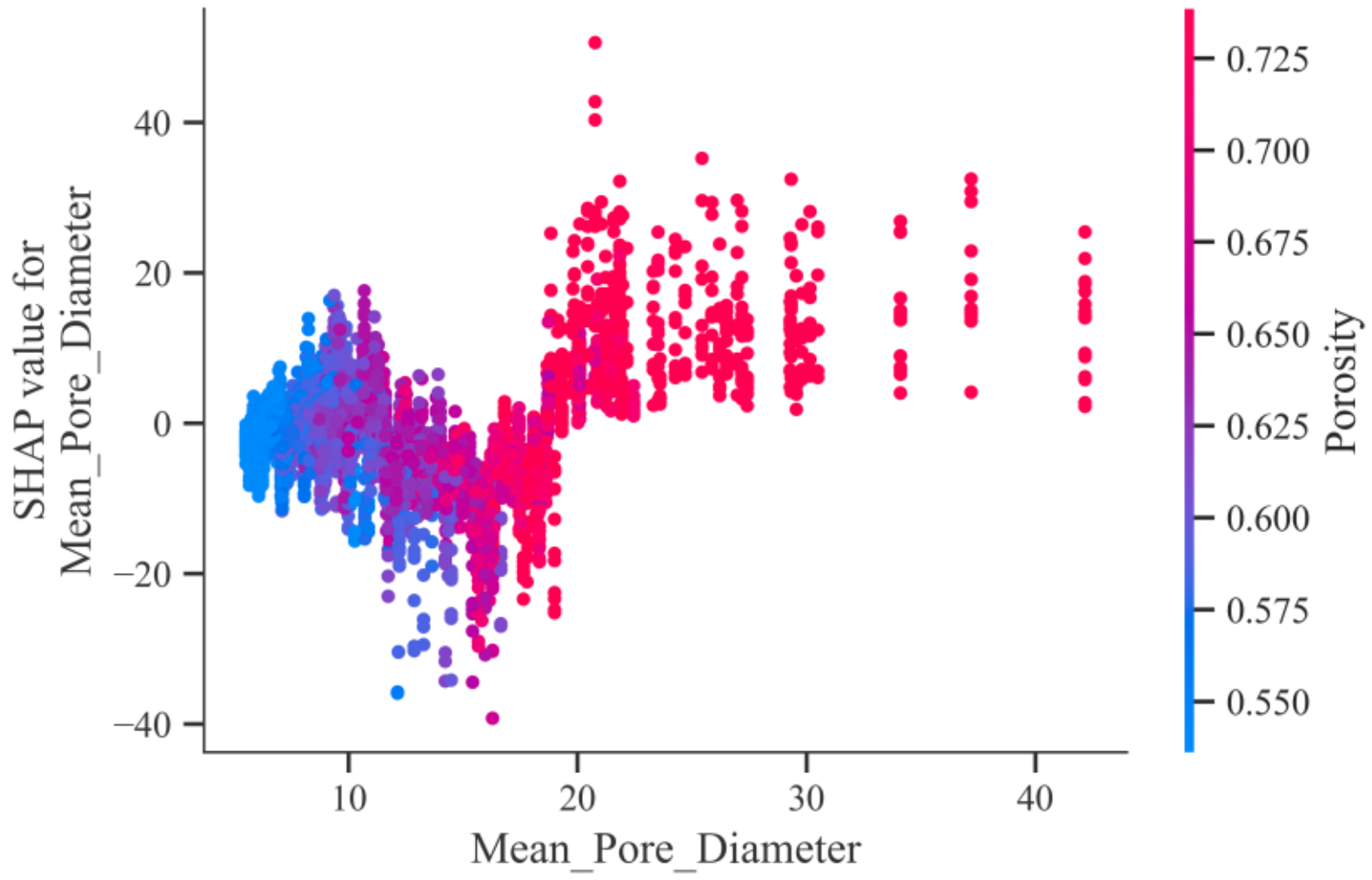

**Fig. C.3** SHAP dependence plots for mean pore diameter

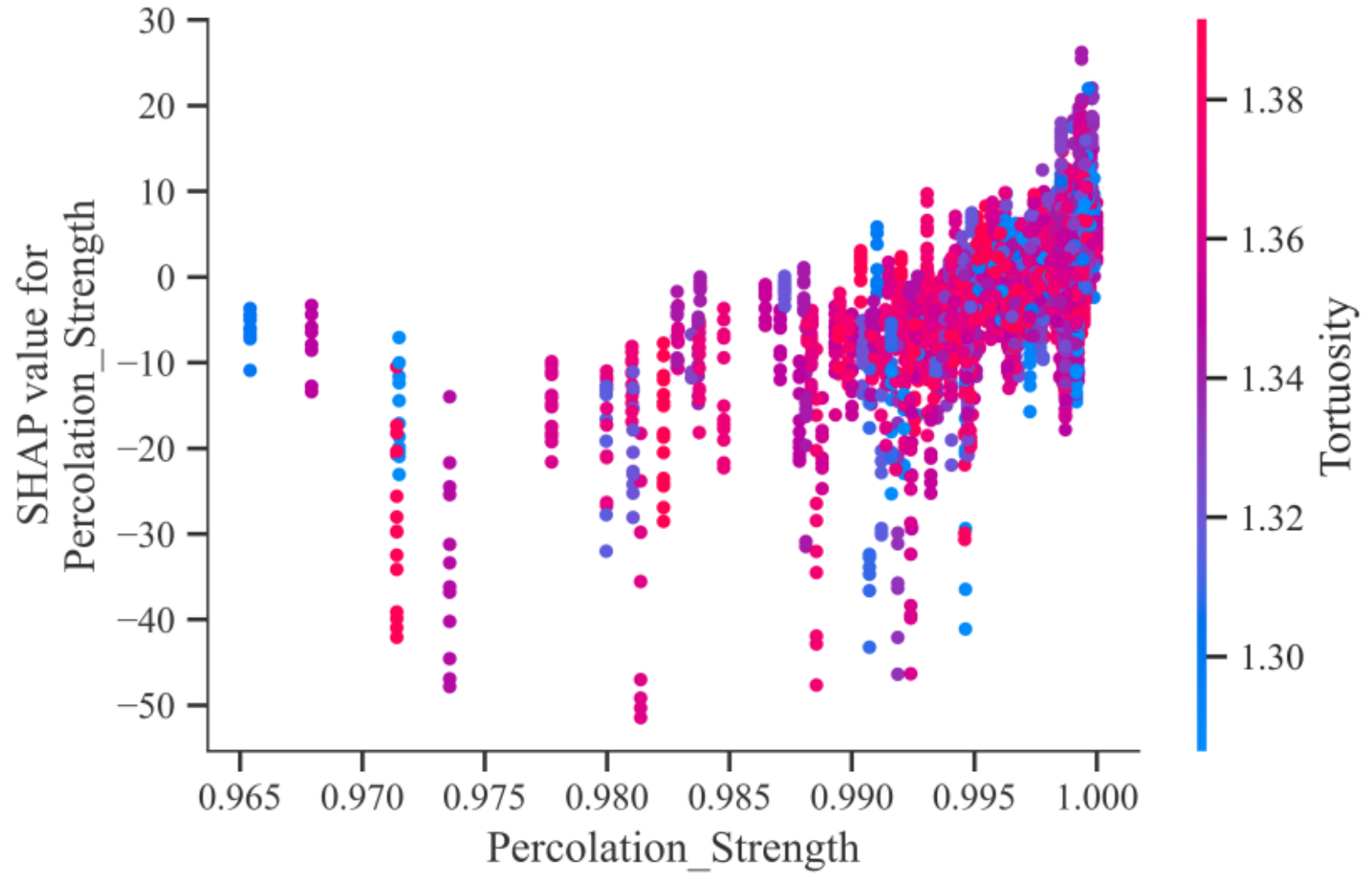

**Fig. C.4** SHAP dependence plots for percolation strength

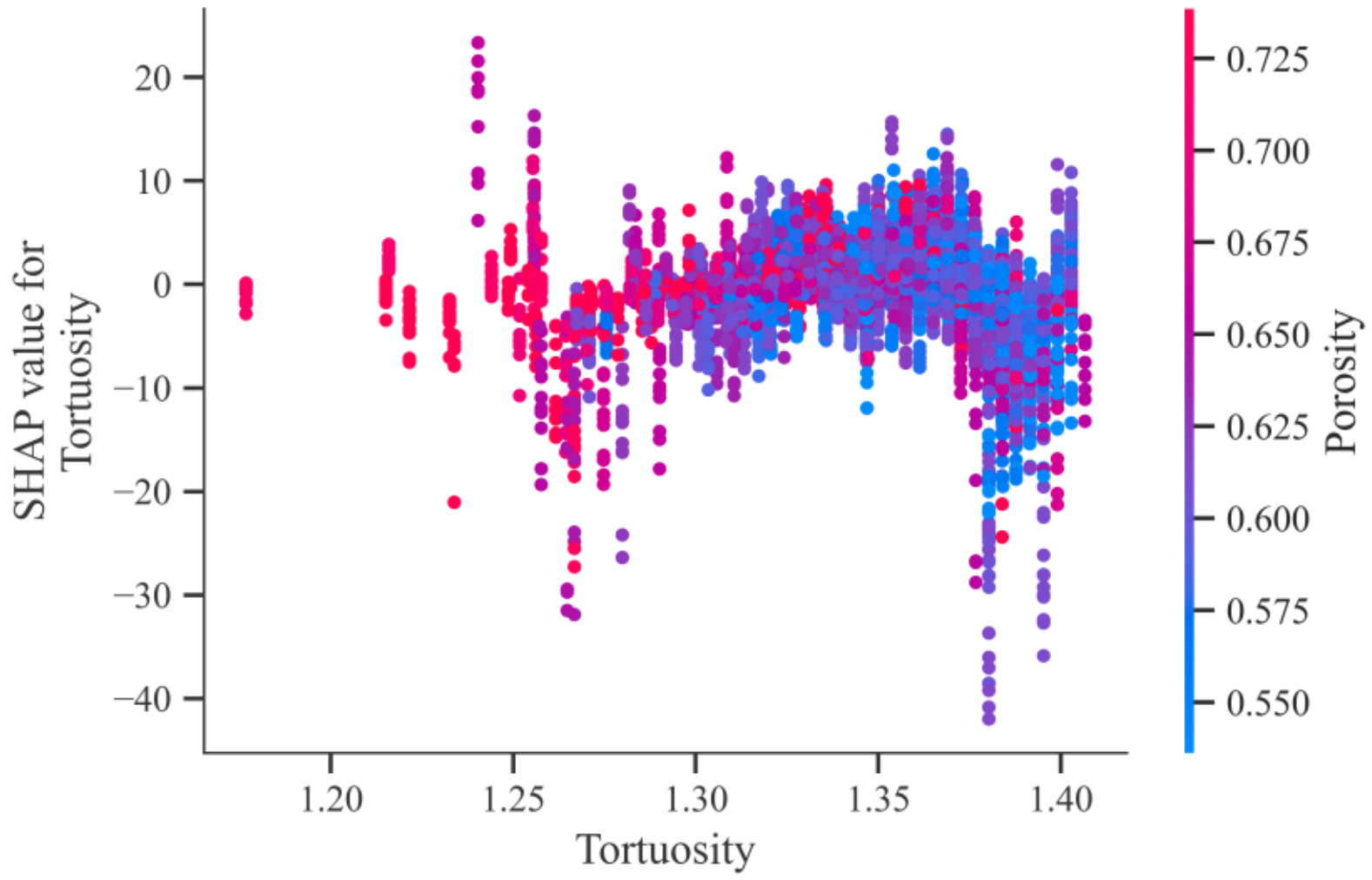

**Fig. C.5** SHAP dependence plots for tortuosity

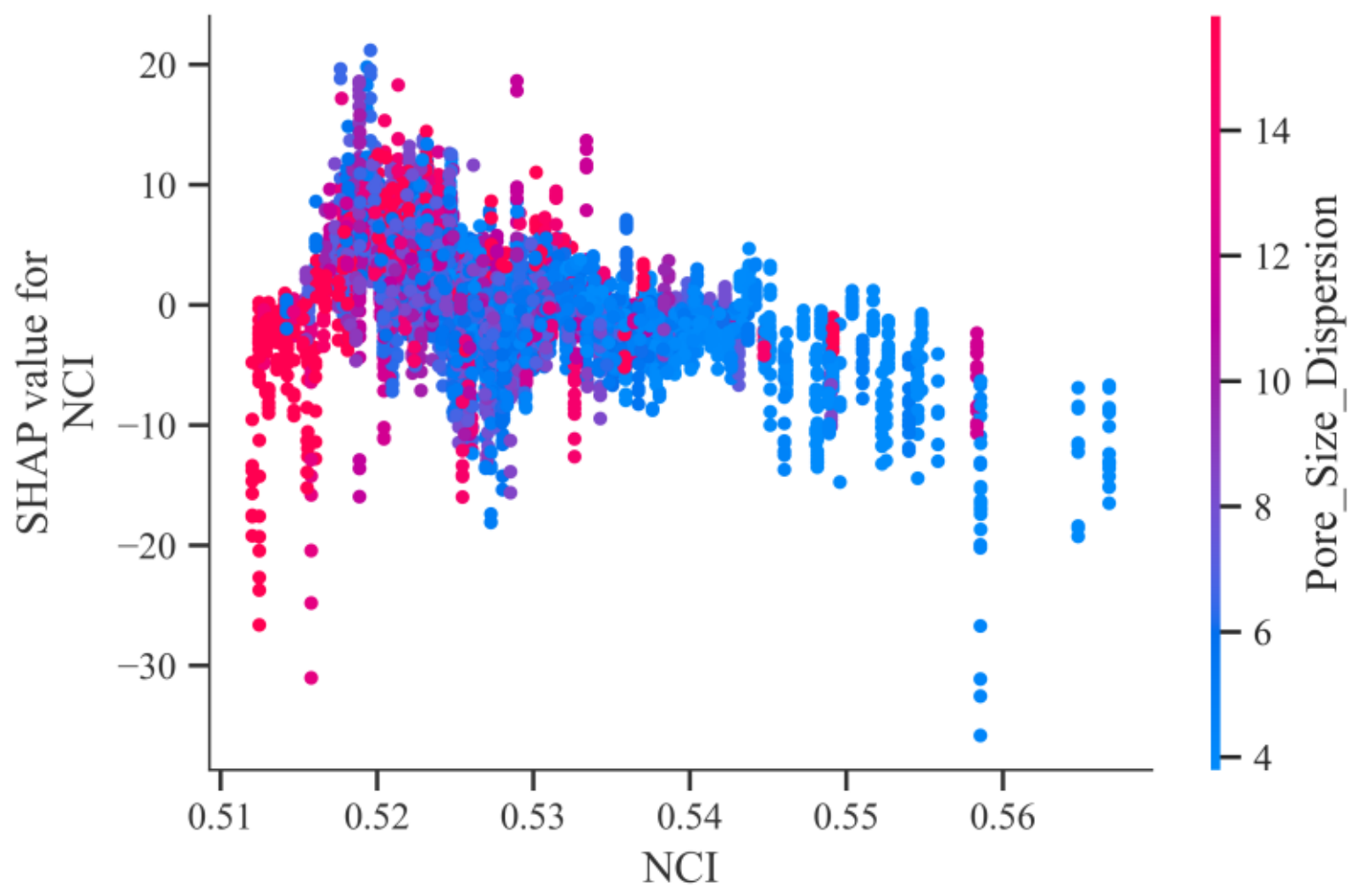

**Fig. C.6** SHAP dependence plots for NCI

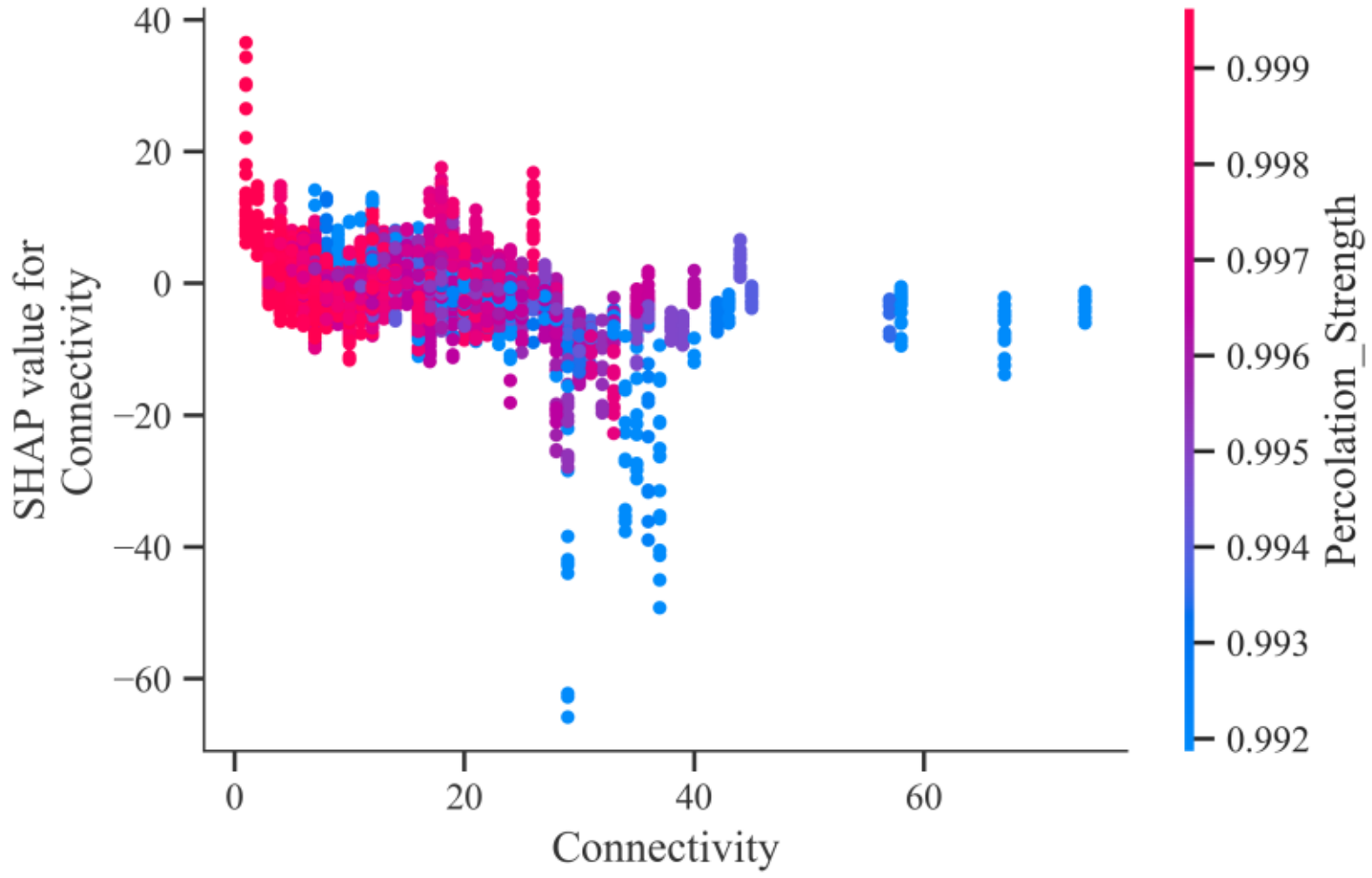

**Fig. C.7** SHAP dependence plots for Connectivity

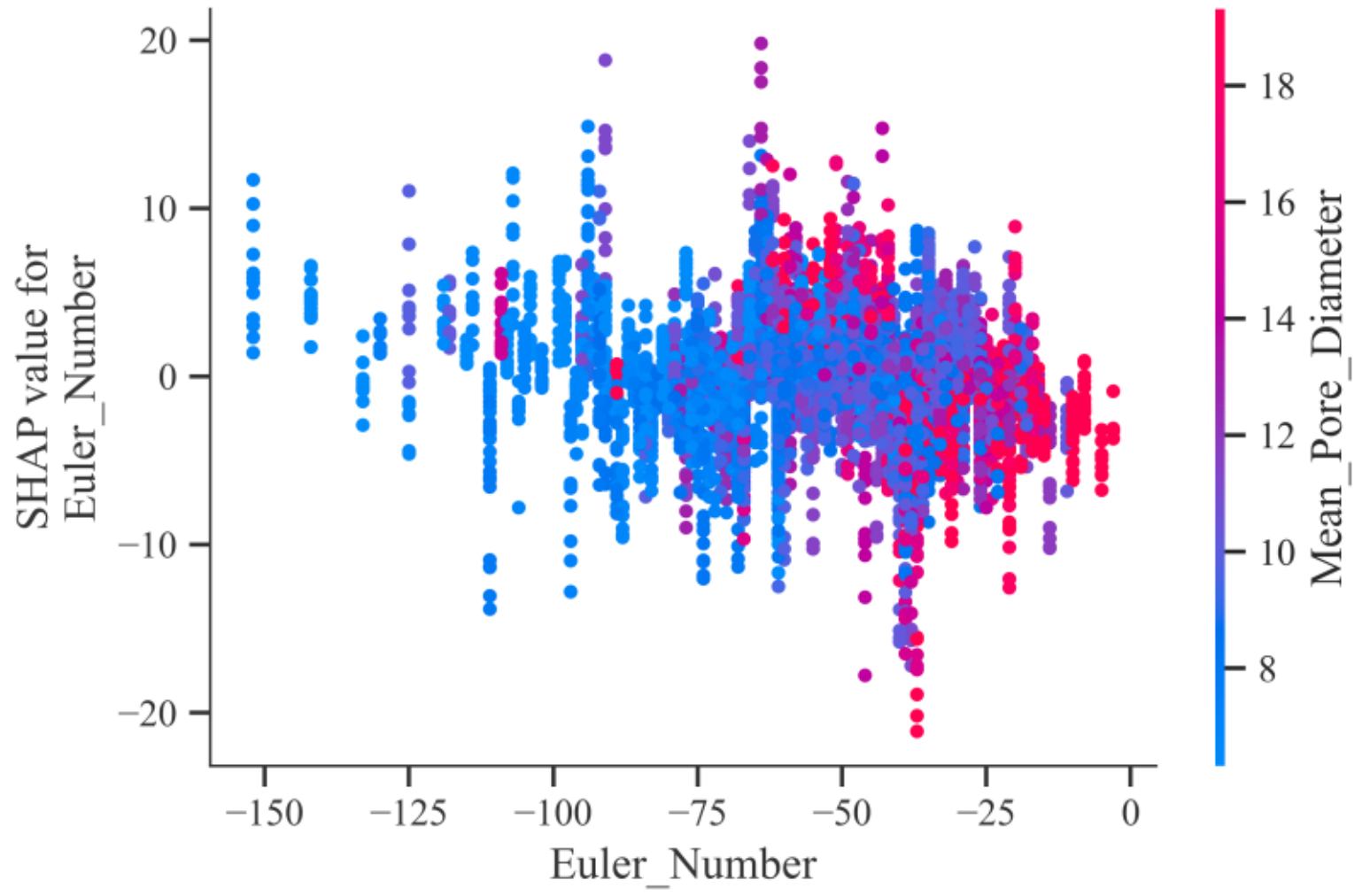

**Fig. C.8** SHAP dependence plots for Euler number